\documentclass[12pt]{article}
\usepackage{a4wide,epsfig,amsmath,amssymb,cite,scalefnt,graphicx}
\numberwithin{equation}{section}
\allowdisplaybreaks
\usepackage{tikz}
\usetikzlibrary{intersections}
\usetikzlibrary{calc}
\usetikzlibrary{shapes}
\newcommand{\vertex}{\node[fill,circle,inner sep=0pt,minimum size=0pt]}
\newcommand{\coord}[1]{({sin(#1)},{cos(#1)})}

\usepackage{tabularx}
\usepackage{graphics}
\usepackage{array}

\def\MOp{{\rm MOM'}}
\def\MO{{\rm MOM}}
\def\mop{{\rm mom'}}
\def\mo{{\rm mom}}

\def\be{\begin{equation}}
\def\ee{\end{equation}}
\def\nn{\nonumber\\}

\def\msbar{{\overline{\rm MS}}}
\def\msba{{\overline{\rm ms}}}

\def\Rbar{\overline R}
\def\Xtil{\tilde X}
\def\Xhat{\hat X}
\def\chat{\hat c}
\def\ehat{{\hat \epsilon}}
\def\Scal{{\cal S}}

\def\Khat{\hat K}

\def\Gtil{\tilde G}
\def\zhat{\hat\zeta}

\def\fbar{\overline f}

\def\fbar{\overline f}

\def\frakk[#1#2{{{#1}\over{#2}}}

\def\bhat{\hat \beta}
\def\tbeta{\tilde \beta}

\def\Ghat{\hat G}

\def\Ftil{\tilde F}

\def\ctil{\tilde c}

\def\pa{\partial}

\def\zbar{\overline \zeta}

\def\tbeta{\tilde \beta}

\input amssym.def
\input amssym
\def\pa{\partial}
\def\be{\begin{equation}}
\def\ee{\end{equation}}
\def\nn{\nonumber\\}

\def \S{{\cal S}}

\def \x{{\rm x}}
\def \y{{\rm y}}

\def\btil{\tilde \beta}
\def\cirk{\,{\raise1pt \hbox{${\scriptscriptstyle \circ}$}}\,}

\def \olr{{\raise6.5pt\hbox{$\leftrightarrow  \! \! \! \! \!$}}}

\def\hbar{\bar h}

\font\ninerm=cmr9 \font\ninesy=cmsy9
\font\eightrm=cmr8 \font\sixrm=cmr6
\font\eighti=cmmi8 \font\sixi=cmmi6
\font\eightsy=cmsy8 \font\sixsy=cmsy6
\font\eightbf=cmbx8 \font\sixbf=cmbx6
\font\eightit=cmti8
\def\eightpoint{\def\rm{\fam0\eightrm}
  \textfont0=\eightrm \scriptfont0=\sixrm \scriptscriptfont0=\fiverm
  \textfont1=\eighti  \scriptfont1=\sixi  \scriptscriptfont1=\fivei
  \textfont2=\eightsy \scriptfont2=\sixsy \scriptscriptfont2=\fivesy
  \textfont3=\tenex   \scriptfont3=\tenex \scriptscriptfont3=\tenex
  \textfont\itfam=\eightit  \def\it{\fam\itfam\eightit}%
  \textfont\bffam=\eightbf  \scriptfont\bffam=\sixbf
  \scriptscriptfont\bffam=\fivebf  \def\bf{\fam\bffam\eightbf}%
  \normalbaselineskip=9pt
  \setbox\strutbox=\hbox{\vrule height7pt depth2pt width0pt}%
  \let\big=\eightbig  \normalbaselines\rm}
\catcode`@=11 %
\def\eightbig#1{{\hbox{$\textfont0=\ninerm\textfont2=\ninesy
  \left#1\vbox to6.5pt{}\right.\n@@space$}}}
\def\vfootnote#1{\insert\footins\bgroup\eightpoint
  \interlinepenalty=\interfootnotelinepenalty
  \splittopskip=\ht\strutbox %
  \splitmaxdepth=\dp\strutbox %
  \leftskip=0pt \rightskip=0pt \spaceskip=0pt \xspaceskip=0pt
  \textindent{#1}\footstrut\futurelet\next\fo@t}
\catcode`@=12 %
\def\today{\number\day\ \ifcase\month\or January\or February\or March\or
April\or May\or June\or July\or
August\or September\or October\or November\or December\fi, \number\year}

\input epsf

\begin{document}

\begin{titlepage}
\begin{flushright}
LTH1356
\end{flushright}
\date{}
\vspace*{3mm}

\begin{center}
{\Huge No-$\pi$ schemes for multi-coupling theories}\\[12mm]
{\bf I.~Jack}\\

\vspace{5mm}
Dept. of Mathematical Sciences,
University of Liverpool, Liverpool L69 3BX, UK\\

\end{center}

\vspace{3mm}

\begin{abstract}
We show that even $\zeta$-functions may be removed from the $\beta$-functions of general multi-coupling theories up to high loop order by means of coupling redefinitions. For theories whose $\beta$-function is determined by the anomalous dimensions of the fields, such as supersymmetric theories, this corresponds to a renormalisation scheme change to a momentum subtraction scheme.
\end{abstract}

\vfill

\end{titlepage}
\section{Introduction}
The vanishing of $\zeta$-functions of even argument for RG functions in certain schemes, and for certain physical quantities, has been known for some time. The first observation was in Ref.~\cite{GKL}, where it was shown that $\zeta_4$ contributions cancelled in the Adler function in QCD at $O(\alpha_s^3)$. Subsequently it was demonstrated that 
the Adler function displayed this property at $O(\alpha_s^4)$ too\cite{baikov6}\cite{BCK1}. However, it was remarked in Refs.~\cite{baikov5}, \cite{herz} that the $\zeta_4$-dependence in the scalar quark and scalar gluonium correlators respectively did not cancel at $O(\alpha_s^4)$. In Ref.~\cite{BCK2} the cancellation of even $\zeta$s in specific physical quantities was explained by a dependence of relevant Feynman integrals only on certain combinations of $\zeta$-functions (first noticed in Ref.~\cite{baikov4}), together with properties of scale invariance. In Ref.~\cite{kotikov} an explanation of this phenomenon was presented in terms of the Landau-Khalatnikov-Fradkin transformation.   In Ref.~\cite{jamin} it was shown that there was a scheme (the ``$C$-scheme", first defined in Ref.~\cite{boito}) in which $\zeta_4$ and $\zeta_6$ terms vanished up to $O(\alpha_s^4)$ for the scalar quark and scalar gluonium correlators as well. It was suggested that the Adler function would display a similar behaviour at next order; i.e. there would be even-$\zeta$ terms in $\msbar$ at $O(\alpha_s^5)$ which would disappear on transformation to the $C$-scheme. Soon afterwards, it was shown in Refs.~\cite{baikov1,baikov2,baikov3} that the explanation of Ref.~\cite{BCK2} extended to a very high loop order and resulted in the even-$\zeta$ parts of $\beta$-functions being related to lower-order $\beta$-functions. Furthermore, this property explained the  
``no-$\pi$'' features of the $C$-scheme and it was conjectured that this ``no-$\pi$'' property would hold to all orders. Further empirical support was provided for instance in Ref.~\cite{davies}.

Meanwhile there has been continuing interest in various ``momentum subtraction'' renormalisation schemes.  In such schemes certain Green functions are set to particular values (often their tree values) for specific values of the external momenta, often as a function of $\mu$, the dimensional regularisation mass parameter. From our point of view these schemes effectively correspond to removing finite contributions to renormalisation constants in addition to the pole terms which would be removed in minimal subtraction. We shall use the term MOM\cite{celgon} to refer to all these schemes, being more precise when appropriate. In any case, most MOM schemes (with the exception of our $\MOp$ scheme, as we shall see later) agree on the treatment of two-point functions which will be our main concern, but differ on three and four point vertices. It has been noticed that in these $\MO$ schemes there are cancellations of even $\zeta$s in individual RG functions. For instance, in Refs.~\cite{chet2}, \cite{verma} the quark mass anomalous dimension was computed in $\msbar$ up to four loops for $SU(N)$, and for a general gauge group within a general covariant gauge respectively; these $\msbar$ results contain $\zeta_4$ at the four-loop level. However, in Ref.~\cite{chetret1}, it was shown that the $\zeta_4$ terms were absent up to four loops in the anomalous dimensions of both the quark mass and quark field in the so-called ``regularisation-invariant'' (RI) scheme, despite being present in $\msbar$. In Ref.~\cite{chetret2}, the same authors evaluated the QCD $\beta$-function up to four loops in a variety of MOM schemes originally introduced in Ref.~\cite{bralev}, once again noting an absence of $\zeta_4$ terms; though in this case, the $\zeta_4$ terms are absent even in $\msbar$ up to four loops\cite{vanrit}. On the other hand, in Ref.~\cite{chet1}, the gluon, ghost and ghost-ghost-gluon anomalous dimensions for QCD were computed up to four loops in $\msbar$ within a general covariant gauge and seen to contain $\zeta_4$.  In Ref.~\cite{vonsmek}, the ``mini-MOM'' scheme was defined as a variant of $\MO$ especially suitable for QCD. In Ref.\cite{gracey1} various anomalous dimensions were computed in QCD up to four loops in this scheme, at the same time extending to general gauges; once again the $\zeta_4$ terms are absent in all covariant gauges (note that in Ref.~\cite{gracey1}, the correct result for the quark mass anomalous dimension is given in an erratum where a sign error is corrected). Turning now to five loops, in Ref.~\cite{BCKR} it was shown that $\zeta_4$ terms appeared in the $\msbar$ QED $\beta$-function but were absent in $\MO$. An interesting feature visible in the later computation of the QCD $\beta$-function at five loops in $\msbar$\cite{york1, york2} is that the gluon and ghost field and ghost-gluon vertex anomalous dimensions all contain also $\zeta_6$ (and indeed $\zeta_7$ and $(\zeta_3)^2$) but these $\zeta$-functions cancel in the combination of anomalous dimensions which gives the QCD $\beta$-function (though a dependence on $\zeta_4$, and of course on $\zeta_3$ and $\zeta_5$, remains). This (and the similar cancellation of $\zeta_4$ in the four-loop gauge $\beta$-function) seems therefore to be a consequence of gauge invariance rather than a property of Feynman diagrams, in contrast to the more general cancellations of even $\zeta$s we are concerned with here. Finally in
Ref.~\cite{gracey2} a range of anomalous dimensions were computed in QCD up to five loops in the ``mini-MOM'' scheme, once again at the same time extending to general gauges; interestingly, now the potential $\zeta_4$ and $\zeta_6$ terms which are visible in $\msbar$\cite{BCK3, york3} are absent, but now only in the Landau gauge. To put it another way, there are gauge-parameter dependent even-$\zeta$ terms in the five-loop mini-$\MO$ results.\footnote{Note that in the case of the quark mass anomalous dimension there is no gauge parameter dependence in the $\msbar$ scheme result; it can only appear in mass-dependent schemes such as $\MO$. However the other anomalous dimension results do depend on the gauge parameter in $\msbar$} It seems surprising that dependence on even $\zeta$s may appear in QCD anomalous dimensions in the mini-$\MO$ scheme, but only at five loops, and even then, only in terms proportional to the gauge parameter.

Turning now to supersymmetric theories, in Ref.~\cite{gracey}, the five-loop anomalous dimension was computed for the supersymmetric Wess-Zumino model. The results were given in both $\msbar$ and MOM, and it was remarked that the $\zeta_4$ and $\zeta_6$ terms in the anomalous dimension (and hence the $\beta$-function) cancelled out in the case of $\MO$. In fact, it had already been noticed in Ref.~\cite{JO} that the $\zeta_4$ terms at four loops could be removed by a coupling redefinition, but the significance in terms of MOM had not been pointed out. It should be stressed that these supersymmetric results are valid for multi-coupling theories. We should also mention that in Ref.~\cite{bellon}, results were given for a subset of diagrams (essentially a bubble sum) using a Hopf algebra approach to a very high order (200 loops), within a MOM scheme. The absence of even $\zeta$s here was an immediate consequence of the method used.


It would appear that all the general results on the absence of even $\zeta$s in RG functions which have been derived hitherto apply to single-coupling theories; however there is empirical evidence, especially in Ref.~\cite{gracey}, that the phenomenon extends to multi-coupling theories. In this article we report on progress towards extending the general results to multi-coupling theories. We focus on what we call AD theories where the $\beta$-function is determined by the anomalous dimensions. This is because in such theories one can define the $\MO$ scheme in an unambiguous fashion, and most if not all variants of $\MO$ agree on the treatment of two-point functions. For our purposes, $\MO$ is defined by subtracting all finite parts in the two-point function with external momentum $p$, after setting $p^2=\mu^2$. The centrepiece of this paper is a demonstration that a general multi-coupling theory is free of $\zeta_4$s up to five-loop level
and free of $\zeta_6$ up to six-loop level within a simpler variant of $\MO$ which we call $\MOp$, where we only subtract finite terms containing even $\zeta$s from the bare coupling. In each case this is one loop beyond the loop order at which the $\zeta$-function first appears in RG quantities. We believe that the same applies to $\MO$, and we give a proof for $\zeta_4$ up to five loops. However our methods are manifestly capable of straightforward extension to higher orders; furthermore they may be immediately applied to other even $\zeta$s, and combinations of even $\zeta$s, at the first two loop orders at which they appear. It will also be clear that we may remove even $\zeta$s by coupling redefinitions in non-AD theories, even though in this case there is no clear interpretation in terms of a change to a well-defined scheme such as $\MO$.

The outline of the paper is as follows: In Section~\ref{basic} we give a brief explanation of the properties of Feyman integrals elucidated in  Refs.~\cite{baikov1,baikov2,baikov3} which underlie the behaviour of the even $\zeta$s. In Section \ref{single} we show that there are simple coupling redefinitions which remove the even-$\zeta$ terms in the RG functions wherever the Feynman integrals have the properties proposed in Refs.~\cite{baikov1,baikov2,baikov3}, at least up to seven loops and for a single-coupling AD theory. We shall see that these redefinitions correspond to changing the RG scheme to $\MOp$. In particular, these redefinitions eradicate the even-$\zeta$ terms in the $\beta$-functions for a single-coupling scalar $\phi^4$ theory computed explicitly at four\cite{kaz}, five\cite{kleina}, six\cite{loopsix} and seven\cite{loopseven} loops. In Section~\ref{susysec} we turn to the Wess-Zumino model, i.e. the supersymmetric scalar-fermion theory. As mentioned earlier, in Ref.~\cite{gracey} Gracey already demonstrated the cancellation of even-$\zeta$ terms in the MOM scheme. Here we show the similar cancellation for the $\MOp$ scheme. 
In Section~\ref{genproof} we make some steps towards a general proof that both MOM and $\MOp$ have the no-$\pi$ property for completely general theories, based on the properties of Feynman integrals proposed in Refs.~\cite{baikov1,baikov2,baikov3}. We show that $\MOp$ has no even-$\zeta$ terms through six loops. Due to the greater calculational complexity, we have only so far showed this property for MOM through five loops; however, for neither scheme is there any apparent obstacle to continuing to higher loops. We offer some pointers for future directions in the Conclusion. In Appendix~\ref{genscal} we show that the even-$\zeta$ terms in the $\beta$-functions may be removed by a coupling redefinition to five loops in a scalar $\phi^4$ theory with general tensor couplings.  (The $\beta$-functions for the general theory were obtained at four loops in Ref.~\cite{JP} using the diagram-by-diagram results helpfully provided in Ref.~\cite{kaz}, and likewise may be obtained at five loops from Ref.~\cite{klein}, though for our purposes here we have focussed solely on the even-$\zeta$ terms.) This redefinition is similar in form to the the one used for the WZ model, but since the general scalar theory is not an AD theory, there is no obvious interpretation as a scheme change.   In Appendix~\ref{betres} we summarise some relations between $\beta$-functions derived in Ref.~\cite{baikov3} and used in Section~\ref{single}.  In Appendix~\ref{feynres} we list results for the divergences and finite parts of Feynman diagrams which were useful in our calculations and which also exemplify the results of Refs.~\cite{baikov1,baikov2,baikov3}. In Appendix~\ref{wzsingle} we record the results for the five-loop $\beta$-function for the single-coupling Wess-Zumino model computed within $\msbar$, $\MO$ and $\MOp$. In the case of $\msbar$ and $\MO$, these results are given in Ref.~\cite{gracey}, but we have largely rederived them as a check on our own calculations. In Appendix~\ref{alt} we describe an alternative computation of the five-loop $\MO$ $\beta$-function for the single-coupling Wess-Zumino model, where we work within $\MO$ from the outset. In Appendix~\ref{susyvar} we give results for the scheme variations of anomalous dimension coefficients in the Wess-Zumino model up to five loops, as required in Section~\ref{susysec}. In Appendix~\ref{mstrans} we discuss the transformations from $\msbar$ to $\MOp$ and $\MO$ and show in general at five loops that the even $\zeta$s are removed by this transformation. We also derive some general results which are useful in Section~\ref{single}.

\section{Basic ideas}\label{basic}



This section is mostly devoted to summarising the results of Ref.~\cite{baikov3} regarding the way in which even $\zeta$s appear in  Feynman integrals. We start with some basic definitions. We use dimensional regularisation with $d=4-\epsilon$ and (for the time being) minimal subtraction. We write
\be
F=\sum F(G), \quad F(G)=-\Rbar(G)
\label{rbar}
\ee
where the sum is over all the relevant graphs $G$ and where the $\Rbar$ operation denotes the inclusion of appropriate subtraction of diagrams with counterterm insertions corresponding to divergent subgraphs. The final minus sign is of course to implement the subtraction of the overall pole. We shall denote by $F^{\zeta_n}_{L,m}$ the $L$-loop, order $\epsilon^{-m}$ $\zeta_n$-dependent contribution to $F$, and by $F_{L,m}$ the $L$-loop, order $\epsilon^{-m}$ non-$\zeta$-dependent contribution to $F$. 
we may write the bare coupling $g_B$ as
\be
g_B=\mu^{\epsilon}\left(g+\frac{1}{\epsilon}\left\{F_{1,1}+F_{2,1}+F^{\zeta_3}_{3,1}\zeta_3+F^{\zeta_4}_{4,1}\zeta_4\right\}+\frac{1}{\epsilon^2}\left\{F_{2,2}+F^{\zeta_3}_{4,2}\zeta_3\right\}\right)+\ldots
\label{gb}
\ee
where $\mu$ is the usual dimensional regularisation mass parameter, so that 
\be
\mu\frac{d}{d\mu} g_B=0.
\label{dgzero}
\ee
Here we are taking as our underlying example a $\phi^4$ theory (other examples may result in different powers of $\mu$).
The $\beta$-function is then defined by
\be
\hat\beta(g)=\mu\frac{d}{d\mu}g=-\epsilon g+\beta(g)
\label{betdefa}
\ee 
where $\beta(g)$ is $\epsilon$-independent in the minimal subtraction scheme, though not in others; see Section \ref{genproof} and Appendix \ref{alt}.
Using Eqs.~\eqref{gb}, \eqref{betdefa}, we may write 
\be
\beta_L^{\zeta_n}=LF^{\zeta_n}_{L,1},
\label{betdef}
\ee
where $\beta_L^{\zeta_n}$ is the $\zeta_n$-dependent contribution to the $L$-loop $\beta$-function (again assuming minimal subtraction). 

It was shown in Ref.~\cite{baikov3} that in a large class of Feynman diagrams, termed ``$\pi$-safe'' the dependence on even $\zeta$s is solely through the combinations
\begin{align}
\zhat_3=&\zeta_3+\frac{3\epsilon}{4}\zeta_4-\frac{5\epsilon^3}{16}\zeta_6+\frac{21\epsilon^5}{64}\zeta_8-\frac{153\epsilon^7}{256}\zeta_{10}+\ldots\nn
\zhat_5=&\zeta_5+\frac{5\epsilon}{4}\zeta_6-\frac{35\epsilon^3}{32}\zeta_8+\frac{63\epsilon^5}{32}\zeta_{10}+\ldots\nn
\zhat_7=&\zeta_7+\frac{7\epsilon}{4}\zeta_8-\frac{21\epsilon^3}{8}\zeta_{10}+\ldots\nn
\zhat_9=&\zeta_9+\frac{9\epsilon}{4}\zeta_{10}+\ldots,\nn
\zhat_{5,3}=&\zbar_{5,3}-\frac{15\epsilon}{4}\zeta_4\zeta_5-\frac{2905\epsilon^2}{1504}\zeta_{10}+\frac{25\epsilon^3}{16}\zeta_5\zeta_6+\ldots,
\label{zcomb}
\end{align}
where
\be
\zbar_{5,3}=\zeta_{5,3}-\frac{29}{12}\zeta_8.
\ee 
Note that in writing Eq.~\eqref{zcomb} we have taken account of a factor of 2 difference in our definition of $\epsilon$ compared with Ref.~\cite{baikov3}.
Following Ref.~\cite{baikov3}, we would like to use the fact that the even $\zeta$s are packaged as in Eq.~\eqref{zcomb} to derive relations between contributions to Eqs.~\eqref{gb}, \eqref{betdef} for different $\zeta$s. However we must be careful, since 
an $L$-loop term $F^{\zeta_n}_{L,m}$ in $g_B$ has contributions not only from the $L$-loop diagrams, but also incorporates subtractions involving lower-loop $(L-p)$-loop diagrams with insertions of $p$th-order counterterms. Since these lower-order counterterms may also involve $\zeta$-functions, the $L$-loop subtractions will not necessarily be in accord with the form Eq.~\eqref{zcomb}.
We therefore write  $G^{\zeta_n}_{L,m}$ for the value of $F^{\zeta_n}_{L,m}$ after we omit these $\zeta_n$-dependent contributions to the lower-order counterterms involved in the subtractions. We should emphasise that the $G^{\zeta_n}_{L,m}$ are not necessarily the ``raw'' diagram contributions, since they include the non-$\zeta_n$-dependent counterterm contributions. We could have defined them to be exactly the diagrammatic contributions, but this would complicate the relations between the $G^{\zeta_n}_{L,m}$ and the $F^{\zeta_n}_{L,m}$.
At the lowest order at which a given $\zeta_n$ appears, there is no difference between the $F^{\zeta_n}$ and the $G^{\zeta_n}$, and we have for instance
\be
F^{\zeta_4}_{3,0}=G^{\zeta_4}_{3,0}, \quad F^{\zeta_3}_{3,1}=G^{\zeta_3}_{3,1}.
\label{bet3a}
\ee
However, at four loops there is a difference between  $G^{\zeta_3}_{4,1}$ and $F^{\zeta_3}_{4,1}$. We must omit the $\zeta_3$-dependent part of the three-loop counterterm, which has the coefficient $F^{\zeta_3}_{3,1}$. Since this counterterm is an insertion on the basic one-loop diagram, we find
\be
G^{\zeta_3}_{4,1}=F^{\zeta_3}_{4,1}-F^{\zeta_3}_{3,1}\cdot F_{1,0}.
\label{fgid}
\ee
where $\cdot$ is a shorthand for $\frac{\pa}{\pa g}$. (Again, for simplicity we are assuming the use of minimal subtraction; this relation will be modified in other schemes such $\MO$, as we shall see in Section 5.)

Now we {\it can} use Eq.~\eqref{zcomb} to derive relations for the $G^{\zeta_n}_{L,m}$. 
We obtain
\begin{subequations}\label{Grels:main}
\begin{align}
G^{\zeta_4}_{L,m}=&\frac34G_{L,m+1}^{\zeta_3},\label{Grels:a}\\
G_{L,m}^{\zeta_6}=&\frac54\left(G_{L,m+1}^{\zeta_5}-\frac13G_{L,m+2}^{\zeta_4}\right),\label{Grels:b}\\
G_{L,m}^{\zeta_8}=&\frac74\left(G_{L,m+1}^{\zeta_7}-\frac12G_{L,m+2}^{\zeta_6}+\frac{1}{24}G_{L,m+4}^{\zeta_4}\right),\label{Grels:c}\\
G_{L,m}^{\zeta_{10}}=&\frac94\left(G_{L,m+1}^{\zeta_9}-\frac23G_{L,m+2}^{\zeta_8}+\frac{7}{60}G_{L,m+4}^{\zeta_6}-\frac{1}{72}G_{L,m+6}^{\zeta_4}\right),\label{Grels:d}
\end{align}
which link $G^{\zeta_n}_{L,m}$ for even $n$ to $G^{\zeta_{n'}}_{L,m'}$ for $n'<n$ (and $m'>m$). 
\end{subequations}
We also have
\begin{subequations}\label{Grelsa:main}
\begin{align}
G_{L,m}^{\zeta_4\zeta_3}=&\frac32G_{L,m+1}^{(\zeta_3)^2},\label{Grelsa:a}\\
G_{L,m}^{\zeta_4^2}=&\frac38G_{L,m+1}^{\zeta_4\zeta_3},\label{Grelsa:b}\\
G_{L,m}^{\zeta_6\zeta_3}=&\frac54G_{L,m+1}^{\zeta_5\zeta_3}-\frac58G_{L,m+3}^{(\zeta_3)^2},\label{Grelsa:c}\\
G_{L,m}^{\zeta_4\zeta_5}=&\frac34G_{L,m+1}^{\zeta_3\zeta_5}-\frac{15}{4}G_{L,m+1}^{\zbar_{5,3}},\label{Grelsa:d}\\
G_{L,m}^{(\zeta_3)^2\zeta_4}=&\frac94G_{L,m+1}^{(\zeta_3)^3}.\label{Grelsa:e}
\end{align}
\end{subequations}
It was shown in Ref.~\cite{baikov3} that the properties of the class of $\pi$-safe diagrams extend to their finite parts, and so Eqs.~\eqref{Grels:main}, \eqref{Grelsa:main} are valid for $m\ge0$. It was also shown that Eq.~\eqref{zcomb} resulted in a large number of relations between the even-$\zeta$ parts of $\beta$ functions, and the odd-$\zeta$ parts of lower-order $\beta$-functions, which were given explicitly in the case of single-coupling theories up to eight loops. For ease of reference we have repeated these results up to seven loops in Appendix A (with some factors of two to account for our own conventions). We shall show here that all the even-$\zeta$ $\beta$-function contributions expressed in this way may be removed by a coupling redefinition, at least up to seven loops. This is already implicit in Ref.~\cite{baikov3}, but here we shall extend to the multi-coupling case. We shall also focus particular attention on theories where the $\beta$-functions depend only on the anomalous dimensions, which we term AD theories. Obvious examples are supersymmetric theories and QCD coupled only to fermions. We shall show that for AD theories, this redefinition corresponds to a switch to a well-defined renormalisation scheme, the one we have dubbed $\MOp$, where we remove the finite even-$\zeta$ contributions to the bare coupling; these finite contributions have a unique well-defined physical meaning in the case of AD theories where setting the external momentum in two-point diagrams equal to the $\msbar$ mass scale $\mu$  gives a unique numerical result for the finite part of the integral. Again, similar ideas have already been presented in Refs.~\cite{baikov1}, \cite{baikov3}, where the $\Ghat$ scheme appears to be equivalent to our $\MOp$ scheme; but here we give an explicit treatment suitable to extension to the multiple coupling case, which we cover in detail in later sections. For the case of an AD theory, we also have at our disposal the well-defined and quite well-known $\MO$ scheme, where we remove {\it all} the finite contributions to the bare coupling. Rather remarkably, this scheme also appears to be free of even $\zeta$s, at least to the order we have been able to investigate.

We finish this section with a brief discussion of scheme changes, which in fact is easily extended to multi-coupling theories. A change of renormalisation scheme results in a redefinition $g\rightarrow g'$ of the coupling $g$ which in turn leads to a variation $\beta(g)\rightarrow \beta'(g')$ of the $\beta$-function given by 
\be
\beta'(g')=\beta(g)\cdot  g' 
\label{bdef}
\ee
A variation $g'=g+\delta g$ then results in a variation of the $\beta$-function given by 
\be
\delta\beta=[\beta,\delta g]-\delta g\cdot[\beta,\delta g]-\frac12 \delta g\cdot(|\delta g|\cdot \beta)+\ldots
\label{blin}
\ee
where
\be
[X,Y]=X\cdot  Y-X\cdot  Y,
\label{comdef}
\ee
and where in the last term, the vertical lines indicate that $\delta g$ is not acted on by the first derivative.

\section{The single-coupling case}\label{single}
In this section we show how the relations between $\beta$-functions derived in Ref.~\cite{baikov3} for single-coupling theories may be used to construct a simple scheme change which removes even-$\zeta$-terms from the $\beta$-function. We show this explicitly up to seven loops, but we expect that the procedure can be continued with the eight-loop results given in Ref.~\cite{baikov3}. We have made some progress towards deriving similar relations for a general multi-coupling theory, and we have derived other useful results for this general case. These general results are collected in Appendix \ref{mstrans} and referred to where necessary or appropriate in the main text; but in this Section we confine ourselves to the single-coupling case. We also mostly consider the case of an AD theory where the scheme change which removes the even $\zeta$s has a manifest physical meaning; however, as we shall explain, the even $\zeta$s may also be removed in the general case by a coupling redefinition, though this redefinition has no obvious interpretation here as a renormalisation scheme. In this case of the general single coupling, we are essentially recapitulating the ideas of Ref.~\cite{baikov3}, but we believe it is useful as a preliminary to the multi-coupling case. 

We start with the 4-loop $\zeta_4$ contribution to $\beta$, $\beta_4^{\zeta_4}$, given in Eq.~\eqref{z4:a}. For definiteness we assume we are working in the $\msbar$ scheme and unless otherwise stated, RG functions are assumed to be expressed in terms of the $\msbar$ coupling, $g^{\msbar}$. $\beta_4^{\zeta_4}$ is the first non-trivial even-$\zeta$ contribution as we increase the loop order.
From Eqs.~\eqref{betdef}, \eqref{bet3a}, \eqref{Grels:a}, we have
\be
F^{\zeta_4}_{3,0}=G^{\zeta_4}_{3,0}=\frac34G^{\zeta_3}_{3,1}=\frac34F^{\zeta_3}_{3,1}=\frac{1}{4}\beta^{\zeta_3}_3,
\label{bet3}
\ee
and bearing in mind that an $L$-loop RG quantity in scalar $\phi^4$ theory is $O(g^{L+1})$, we may rewrite Eq.~\eqref{z4:a} as
\be
\beta_4^{\zeta_4}=\frac14[\beta_1,\beta_3^{\zeta_3}]=[\beta_1,F_{3,0}^{\zeta_4}].
\label{bet4a}
\ee
Comparing with Eq.~\eqref{blin}, we see that  $\beta_4^{\zeta_4}$ is exactly of the form to be removed by the four-loop variation induced at first order by 
\be
(\delta g)^{(3)}=-\frac14\beta^{\zeta_3}_3\zeta_4=-F_{3,0}^{\zeta_4}\zeta_4.
\label{delone}
\ee
This result on its own is somewhat trivial, but it is in line with expectations since it is exactly the redefinition required to remove the lowest-order finite $\zeta_4$ term, as required by the $\MOp$ scheme; furthermore we shall now go on to show that all the even-$\zeta$ contributions to the $\beta$-function may be removed by an extension of this redefinition to higher loop orders (after also taking into account higher-order effects in $\delta g$ where necessary). However, a couple of subtleties immediately present themselves. The first is that for loop order $L>3$, it is no longer true that $F^{\zeta_3}_{L,1}=G^{\zeta_3}_{L,1}$ in $\msbar$; for instance, at four loops we have
\begin{subequations}\label{FGrel:main}
\begin{align}
G^{\zeta_3\msbar}_{4,1}=&F^{\zeta_3\msbar}_{4,1}-F^{\zeta_3\msba}_{3,1}\cdot F^{\msba}_{1,0},\label{FGrel:a}\\
G^{\zeta_4\msbar}_{4,0}=&F^{\zeta_4\msbar}_{4,0}.\label{FGrel:b}
\end{align}
\end{subequations}
Here we introduce some new notation. Quantities denoted $F^{\zeta_3\msbar}_{4,1}$ (with upper-case $\msbar$) are scheme-dependent and in addition are assumed (unless otherwise stated) to be functions of the coupling appropriate to that scheme, i.e. $g^{\msbar}$ in this case; quantities denoted $F^{\zeta_3\msba}_{3,1}$ (with lower-case $\msba$)have a scheme-independent functional form but are expressed in terms of the particular coupling, again $g^{\msbar}$ in this case, so that explicitly
\begin{align}
F^{\zeta_3\msbar}_{4,1}\equiv& F^{\zeta_3\msbar}_{4,1}(g^{\msbar}),\nn
F^{\zeta_3\msba}_{3,1}\equiv& F^{\zeta_3}_{3,1}(g^{\msbar}).
\end{align}
We shall make similar distinctions later in Section~\ref{genproof} between $\MO$ and $\mo$ superscripts, and between $\MO'$ and $\mo'$ superscripts.
 For a particular $\zeta_n$, quantities $F^{\zeta_n}_{L,p}$ are scheme-independent at the loop order $L_{\zeta_n}$ where they first appear. We also have 
\be
F^{\zeta_n}_{L_{\zeta_n},p}=G^{\zeta_n}_{L_{\zeta_n},p}
\label{Lmin}
\ee
 at this lowest loop-order $L_{\zeta_n}$. 

The second subtlety is here we are trying to transform from the $\msbar$ scheme to $\MOp$, but it is more natural to define the reverse transformation. The $\MOp$ scheme is defined by subtracting finite even-$\zeta$ contributions order by order within the scheme itself. The $\msbar$ scheme is then defined by subsuming all these finite parts into the $\msbar$ coupling, thereby giving $g^{\msbar}$ as a function of $g^{\MOp}$; afterwards we may invert to write $g^{\MOp}$ in terms of $g^{\msbar}$. This is all explained in some detail in Appendix \ref{mstrans}; but the upshot is that it is natural to write the coupling redefinition taking us from $\msbar$ to $\MOp$ in terms of $F^{\zeta_4\MOp}_{L,0}(g^{\msbar})$, i.e. $F^{\zeta_4\MOp}_{L,0}$ with its natural argument $g^{\MOp}$ replaced by $g^{\msbar}$; as we see for instance in Eq.~\eqref{MS10}. Providentially, we find that for $L>3$, we have an analogous result to \eqref{Grels:a}
\be
F^{\zeta_4\MOp}_{L,0}(g^{\msbar})=\frac34F^{\zeta_4\msbar}_{L,1},
\label{FMOM}
\ee
which turns out to be exactly what we need. Again, see Appendix \ref{mstrans} for further details. In the rest of this section we return to suppressing explicit mention of the $\msbar$ scheme since as explained, this is the default here. We now apply to Eqs.~\eqref{z4:b}-\eqref{z4:d} the same chain of reasoning which led to Eq.~\eqref{bet4a}, but using \eqref{FMOM} where appropriate. 
We find, using Eqs.~\eqref{comdef}, \eqref{bet3}, 
\begin{subequations}\label{z4betas:main}
\begin{align}
\beta_5^{\zeta_4}=&[\beta_1,F_{4,0}^{\zeta_4\MOp}]+[\beta_2,F_{3,0}^{\zeta_4\msbar}],\label{z4betas:a}\\
\beta_6^{\zeta_4}=&[\beta_1,F_{5,0}^{\zeta_4\MOp}]+[\beta_2,F_{4,0}^{\zeta_4\MOp}]+[\beta_3,F_{3,0}^{\zeta_4\msbar}],\label{z4betas:b}\\
\beta_7^{\zeta_4}=&[\beta_1,F_{6,0}^{\zeta_4\MOp}]+[\beta_2,F_{5,0}^{\zeta_4\MOp}]+[\beta_3,F_{4,0}^{\zeta_4\MOp}]+[\beta_4,F_{3,0}^{\zeta_4\msbar}]\label{z4betas:c}.
\end{align}
\end{subequations}
Just to be absolutely clear here since we are suppressing $g^{\msbar}$: in Eq.~\eqref{z4betas:main} and elsewhere in this section, scheme-dependent quantities like $F_{4,0}^{\zeta_4\MOp}$ have their natural arguments of $g^{\MOp}$ replaced by $g^{\msbar}$, while  scheme-independent quantities like $F_{3,0}^{\zeta_4\msbar}$ are written in terms of $g^{\msbar}$.
The results in Eq.~\eqref{z4betas:main} are of the expected form and the appearance of, for instance, $F_{3,0}^{\zeta_4}$ throughout with the same coefficient is highly non-trivial. The term $[\beta_3,F_{3,0}^{\zeta_4}]$ is zero according to Eq.~\eqref{comdef}, since both $\beta_3$, $F_{3,0}^{\zeta_4}$ are the same order in $g$; but we have included it to emphasise the developing pattern.

We now turn to the $\zeta_6$ contributions. As shown in Appendix \ref{mstrans}, just as Eq.~\eqref{Grels:a} leads to Eq.~\eqref{FMOM}, Eq.~\eqref{Grels:b} leads to similar relations between $F^{\zeta_6\MOp}_{L,0}(g^{\msbar})$ and $ F^{\zeta_5\msbar}_{L,1}$, and hence $\beta^{\zeta_5}_{L}$. We start with $\beta^{\zeta_6}_5$.  At four loops, using Eqs.~\eqref{MS20a:a} and \eqref{betdef}, 
and observing that $F^{\zeta_4}_{4,2}=0$, leads immediately to 
\be
F_{4,0}^{\zeta_6\MOp}=\frac{5}{16}\beta^{\zeta_5}_4.
\label{bet4}
\ee
We may then rewrite Eq.~\eqref{z6:a} as
\be
\beta^{\zeta_6}_5=\frac{5}{16}[\beta_1,\beta^{\zeta_5\msbar}_{4}]=[\beta_1,F_{4,0}^{\zeta_6}].
\label{MS23a}
\ee
We remark here that since only the first terms on the RHS of Eqs.~\eqref{Grels:b}  appear at the lowest level, the same line of reasoning which led to Eq.~\eqref{MS3a} may be used to prove Eq.~\eqref{MS23a} in the general multi-coupling case.

Now for $\beta^{\zeta_6}_6$. At five loops, Eq.~\eqref{MS20a:a} and \eqref{betdef} lead to
\be
F_{5,0}^{\zeta_6\MOp}=\frac14\beta^{\zeta_5}_5-\frac{5}{12}G^{\zeta_4}_{5,2}.
\label{MS22a}
\ee
The double pole term $G^{\zeta_4}_{4,2}$ is related via 
\be
G^{\zeta_4}_{5,2}=F^{\zeta_4}_{5,2}-F^{\zeta_4}_{4,1}\cdot F_{1,1},
\label{G2def}
\ee
to the double pole contribution $F^{\zeta_4}_{5,2}$, which in turn is given by standard RG arguments applied to Eq.~\eqref{gb} as
\be
F^{\zeta_4}_{5,2}=\frac{1}{5}\left(F_{1,1}\cdot F^{\zeta_4}_{4,1}+\beta^{\zeta_4}_{4,1}\cdot F_{1,1}\right)
\ee
We therefore find
\be
G^{\zeta_4}_{5,2}=\frac{1}{20}\left[\beta_1,\beta^{\zeta_4}_4\right].
\label{Gdouble}
\ee
This relation is valid in the multi-coupling case; but specialising to the single-coupling case, we find using \eqref{z4:a}:
\be
G^{\zeta_4}_{5,2}=\frac{3}{40}(\beta_1)^2\beta^{\zeta_3}_3
\label{G52a}
\ee
and hence from Eqs.~\eqref{z6:b}, \eqref{bet4} and \eqref{MS22a}
\be
\beta_6^{\zeta_6}=[\beta_1,F_{5,0}^{\zeta_6\MOp}]+[\beta_2,F_{4,0}^{\zeta_6}].
\ee

We now turn to $\beta^{\zeta_6}_7$. At six loops,
\be
F_{6,0}^{\zeta_6\MOp}=\frac{5}{24}\beta^{\zeta_5}_6-\frac{5}{12}(G^{\zeta_4}_{6,2}+G^{\zeta_4}_{5,2}\cdot F_{1,0}).
\label{MS22b}
\ee
We now derive a useful identity. We readily find using
\be
G^{\zeta_{2m}}_{6,2}=F^{\zeta_{2m}}_{6,2}-F^{\zeta_{2m}}_{5,1}\cdot F_{1,1}-F^{\zeta_{2m}}_{4,1}\cdot F_{2,1}-G^{\zeta_{2m}}_{5,2}\cdot F_{1,0},\quad m=2,3
\ee
combined with standard RG results for $F^{\zeta_{2m}}_{6,2}$ that
\be
G^{\zeta_{2m}}_{6,2}+G^{\zeta_{2m}}_{5,2}\cdot F_{1,0}=\frac{1}{30}\left[\beta_1,\beta^{\zeta_{2m}}_5\right]
+\frac{1}{24}\left[\beta_2,\beta^{\zeta_{2m}}_4\right]. \quad m=2,3
\label{Gdoublea}
\ee
A very similar computation to that for $\beta^{\zeta_6}_6$, using Eqs.~\eqref{z6:c}, \eqref{MS22b}, \eqref{Gdoublea} (for $m=2$), \eqref{bet4}, \eqref{MS22a}, \eqref{z4:a}, \eqref{z4:b}, leads to
\be
\beta_7^{\zeta_6}=[\beta_1,F_{6,0}^{\zeta_6\MOp}]+[\beta_2,F_{5,0}^{\zeta_6\MOp}]+[\beta_3,F_{4,0}^{\zeta_6}].
\ee

Turning now to the $\zeta_8$ and $\zeta_{10}$ cases: We have from Eq.~\eqref{MS20a:b}
\begin{subequations}\label{MS23a:main}
\begin{align}
F^{\zeta_8\MOp}_{5,0}=&\frac{7}{20}\beta^{\zeta_7}_{5},\label{MS23a:a}\\
F_{6,0}^{\zeta_8\MOp}=&\frac{7}{24}\beta^{\zeta_7}_6-\frac{7}{8}(G^{\zeta_6}_{6,2}+G^{\zeta_6}_{5,2}\cdot F_{1,1}).\label{MS23a:b}
\end{align}
\end{subequations}
and similarly from Eq.~\eqref{MS20a:c}
\be
F^{\zeta_{10}\MOp}_{6,0}=\frac{3}{8}\beta^{\zeta_9}_{6}.
\label{MS24}
\ee
Then we may show using Eqs.~\eqref{MS23a:a}, \eqref{MS24}, \eqref{z8:a}, \eqref{z10} that
\begin{align}
\beta^{\zeta_8}_6=&\frac{7}{20}[\beta_1,\beta^{\zeta_7}_{5}]=[\beta_1,F_{5,0}^{\zeta_8},]\nn
\beta^{\zeta_{10}}_7=&\frac{3}{8}[\beta_1,\beta^{\zeta_9}_{6}]=[\beta_1,F_{6,0}^{\zeta_{10}}].
\label{MS24a}
\end{align}
As in the case of $\beta^{\zeta_6}_5$, since only the first terms on the RHS of Eqs.~\eqref{Grels:c}, \eqref{Grels:d} appear at the lowest level, we can use the same line of reasoning which led to Eq.~\eqref{MS3a} to show that Eq.~\eqref{MS24a} is valid in the multi-coupling case. 
We shall postpone consideration of $\beta^{\zeta_8}_{6}$ for a short while, since there are a couple of points which require further discussion.

In the case of products of $\zeta$s more work is potentially required to obtain the required generalisation of Eq.~\eqref{Grelsa:main} analogous to Eq.~\eqref{FMOM}; however, up to the level we are working ($L=6$), it is only Eqs.~\eqref{Grelsa:a}, \eqref{Grelsa:b} which are non-trivial.
We find, as noted in Eqs.~\eqref{MS25}, \eqref{MS26}, 
\be
F^{\zeta_4\zeta_3}_{5,0}=\frac{3}{10}\beta^{(\zeta_3)^2}_5,\quad
\quad F^{\zeta_4\zeta_3\MOp}_{6,0}=\frac14\beta^{(\zeta_3)^2}_6.
\label{z43six}
\ee
We also readily find from Eqs.~\eqref{Grelsa:c}-\eqref{Grelsa:e}
\be
 F^{\zeta_6\zeta_3\MOp}_{6,0}=\frac{5}{24}\beta^{\zeta_5\zeta_3}_6,\quad 
F^{\zeta_4\zeta_5\MOp}_{6,0}=\frac18\beta^{\zeta_3\zeta_5}_6-\frac{5}{8}\beta_6^{\zbar_{5,3}},\quad
F_{6,0}^{(\zeta_3)^2\zeta_4\MOp}=\frac38\beta_6^{(\zeta_3)^3}.
\label{zmixed}
\ee

We then find from Eqs.~\eqref{z43six}, \eqref{zmixed}, \eqref{z43:a}, \eqref{z43:b}, \eqref{z45:a}, \eqref{z45:b}, \eqref{z433},
\begin{subequations}\label{mixed:main}
\begin{align}
\beta_6^{\zeta_4\zeta_3}=&[\beta_1,F_{5,0}^{\zeta_4\zeta_3}]\label{mixed:a}\\
\beta_7^{\zeta_4\zeta_3}=&[\beta_1,F_{6,0}^{\zeta_4\zeta_3\MOp}]+[\beta_2,F_{5,0}^{\zeta_4\zeta_3}]+[\beta_3^{\zeta_3},F_{4,0}^{\zeta_4\MOp}]+[\beta_4^{\zeta_3},F_{3,0}^{\zeta_4}],\label{mixed:b}\\
\beta_7^{\zeta_6\zeta_3}=&[\beta_1,F_{6,0}^{\zeta_6\zeta_3}]+[\beta^{\zeta_3}_3,F_{4,0}^{\zeta_6}],\label{mixed:c}\\
\beta_7^{\zeta_4\zeta_5}=&[\beta_1,F_{5,0}^{\zbar_{5,3}}]+[\beta^{\zeta_5}_4,F_{3,0}^{\zeta_4}],\label{mixed:d}\\
\beta^{\zeta_4(\zeta_3)^2}_7=&[\beta_1,F_{6,0}^{\zeta_4(\zeta_3)^2}].\label{mixed:e}
\end{align}
\end{subequations}
In the case of Eqs.~\eqref{mixed:a}, \eqref{mixed:b}, since there is only one term on the right-hand side of Eq.~\eqref{Grelsa:a}, the same arguments leading to Eqs.~\eqref{MS3a}, \eqref{MS7} can be used to show that these equations are valid in the general multi-coupling case. Furthermore, at lowest order (seven loops) only the first term on the right-hand side of Eqs.~\eqref{Grels:c}, \eqref{Grels:d} appears, and so for similar reasons
 Eqs.~\eqref{mixed:c}, \eqref{mixed:d}, \eqref{mixed:e} are valid in the general case. 

Now for the postponed discussion of $\beta^{\zeta_8}_{7}$. Firstly, in view of the identity $\zeta_8=\frac67(\zeta_4)^2$, one can combine the $\zeta_8$ and $(\zeta_4)^2$ terms. It is not clear that one is obliged to do this, at least at this level, but for comparison purposes we have followed Ref.~\cite{baikov1} in doing so. The anticipated variation of $g^{\msbar}$ corresponding to $(\zeta_4)^2$ may be read off from Eq.~\eqref{MS28}. Accordingly, consider
\be
\delta g^{\zeta_8}_6=-F^{\zeta_8\MOp}_{6,0}-\frac76\left(F^{(\zeta_4)^2\MOp}_{6,0}-F^{\zeta_4}_{3,0}\cdot F^{\zeta_4}_{3,0}\right);
\label{delz8}
\ee
the factor of $\frac76$ being inserted to convert $(\zeta_4)^2$ to $\zeta_8$. We then find, using Eqs.~\eqref{betdef}, \eqref{bet3}, \eqref{Gdoublea}, 
\eqref{MS23a:b}, \eqref{MS27}, \eqref{z6:a}, \eqref{z43:a},
\be
\delta g^{\zeta_8}_6=-\frac{7}{24}\beta^{\zeta_7}_6+\frac{7}{64}\beta_1^2\beta^{\zeta_5}_4-\frac{7}{80}\beta_1\beta^{(\zeta_3)^2}_5+\frac{7}{48}(\beta_3^{\zeta_3})^2.
\ee
The second, somewhat related subtlety is the need to incorporate effects of lower-order variations to second-order; related since the $\zeta_4^2$ term is the first to appear in a second order variation. We find using Eqs.~\eqref{Grels:a}, \eqref{bet3}, \eqref{MS23a:a}, \eqref{z4:a} and \eqref{z8:b} that 
\be
\beta^{\zeta_8}_7=-[\beta_1,\delta g^{\zeta_8}_6]+[\beta_2,F^{\zeta_8}_{5,0}]+\frac76([\beta^{\zeta_4}_4,F^{\zeta_4\MOp}_{3,0}]-\delta^{(2)}_8),
\ee
where
\be
\delta^{(2)}_8=-\frac{11}{16}\beta_1(\beta_3^{\zeta_3})^2.
\ee
This may be rewritten in the form
\be
\delta^{(2)}_8=-(\delta g)^{(3)}\cdot[\beta_1,(\delta g)^{(3)}]-\frac12(\delta g)^{(3)}\cdot (|(\delta g)^{(3)}|\cdot \beta_1),
\ee
where again, in the last term, the vertical lines indicate that $(\delta g)^{(3)}$ is not acted on by the first derivative;
this is the second-order variation induced by $\delta g^{(3)}$ in Eq.~\eqref{delone}.
We now see that all the even-$\zeta$ $\beta$-function contributions listed so far are of the form which may be removed by the first-order and (where relevant) second-order changes resulting from a coupling redefinition 
\begin{align}
\delta g=& -\Bigl(\left[F_{3,0}^{\zeta_4}+F_{4,0}^{\zeta_4\MOp}+F_{5,0}^{\zeta_4\MOp}+F_{6,0}^{\zeta_4\MOp}\right]\zeta_4\nn
&+\left[F_{4,0}^{\zeta_6}+F_{5,0}^{\zeta_6\MOp}+F_{6,0}^{\zeta_6\MOp}\right]\zeta_6\nn
&+\left[F_{5,0}^{\zeta_8}+F_{6,0}^{\zeta_8\MOp}\right]\zeta_8+F_{6,0}^{\zeta_{10}}\zeta_{10}\nn
&+\left[F_{5,0}^{\zeta_4\zeta_3}+F_{6,0}^{\zeta_4\zeta_3\MOp}\right]\zeta_4\zeta_3+F_{6,0}^{\zeta_6\zeta_3}\zeta_6\zeta_3\nn
&+F_{6,0}^{\zeta_4\zeta_5}\zeta_4\zeta_5+F_{6,0}^{(\zeta_3)^2\zeta_4}(\zeta_3)^2\zeta_4\nn
&+\left(F^{(\zeta_4)^2}_{6,0}-F^{\zeta_4}_{3,0}\cdot F^{\zeta_4}_{3,0}\right)(\zeta_4)^2\Bigr)-\ldots
\label{MOM'}
\end{align}

The coupling redefinitions in Eq.~\eqref{MOM'} are exactly those required to remove the finite even-$\zeta$ terms in the bare coupling, and no more, as specified by the definition of $\MOp$. One can also consider performing the redefinitions required to remove the other finite terms, starting at one loop, as required by the MOM scheme. There are then second-order effects on the $\beta$-functions starting at five loops, but also the form of the coupling redefinition is modified at second order; we show in general in Appendix \ref{mstrans} that the combination of these two effects still results in the cancellation of $\zeta_4$ terms in the five-loop $\beta$-functions. It seems plausible that this feature continues at higher orders, in Eqs.~\eqref{z4betas:b}, \eqref{z4betas:c} etc.

Finally, as emphasised earlier, we have been thinking throughout this section of the example of an AD theory. The packaging of even $\zeta$s described in Eq.~\eqref{zcomb} was derived in Refs. \cite{baikov1},  \cite{baikov3} for the case of $p$-integrals which are essentially the integrals which appear in two-point functions and depend on a single momentum $p$. In such integrals, the relations in Eqs.~\eqref{Grels:main}, \eqref{Grelsa:main} govern both the poles and the finite parts of the integrals after setting $p^2=\mu^2$. For an AD theory, the resulting finite parts are unique and have a well-defined physical meaning in terms of picking a renormalisation point. In other cases such as scalar $\phi^4$ theory, one can reduce the computation to $p$-integrals by ``infra-red rearrangement'', i.e. by judiciously setting momenta to zero. The simple poles thereby obtained are well-defined irrespective of which momenta are nullified, and since the $\pi$-safe integrals satisfy Eqs.~\eqref{Grels:main}, \eqref{Grelsa:main} for $m\ge0$, it seems that such quantities as $F^{\zeta_4}_{L,0}$ which appear in the scheme change to Eq.~\eqref{MOM'} will also be well-defined. As in the AD case, one can thus express the scheme definition in terms of simple pole residues such as $F^{\zeta_3}_{L,1}$ or in terms of finite quantities such as $F^{\zeta_4}_{L,0}$; as we see in Eqs.~\eqref{bet3}, \eqref{bet4a} for instance. However, nevertheless it seems unlikely that the scheme defined by this coupling redefinition can be characterised in any physically meaningful way, in contrast to the $\MOp$ and $\MO$ schemes.

\section{The supersymmetric case}\label{susysec}
In this section we consider the supersymmetric Wess-Zumino model. The non-renormalisation theorem implies that the $\beta$-function is completely determined by the anomalous dimension of the field(s). In Ref.~\cite{abbott}, the anomalous dimension was computed up to three loops in $\msbar$. It was also computed in the MOM scheme, with a clear explanation of the definition of the scheme and details of the computation. The anomalous dimension was computed at four loops for a single coupling theory in Ref.~\cite{sen}, but some errors in both Refs.~\cite{abbott} and Ref.~\cite{sen} were identified and corrected in Ref.~\cite{avdeev}. This result was generalised to a multi-coupling theory in Ref.~\cite{FJJ}.
In Ref.~\cite{gracey}, the computation of the anomalous dimension (and hence the $\beta$-function) for the supersymmetric Wess-Zumino model was extended to five-loop order. Results were given for the single-coupling case and also for the case of a general tensor coupling in the multi-field case. Furthermore, in each of these cases, results were given for the MOM scheme in addition to the minimal subtraction scheme, and the cancellation of even $\zeta$s in the MOM scheme was remarked upon. Again, many useful explanations and details were included. The main results have therefore already been obtained, but what we would like to do here is to set them in the context of the previous sections, and also show the cancellation of the even $\zeta$s in the more easily defined $\MOp$ scheme as well as in $\MO$. For this we need a detailed discussion of scheme variations. A useful formalism for this was introduced in Ref.~\cite{scheme} with particular reference to the Wess-Zumino model, and further developed in Refs.~\cite{jot} and \cite{JP}. The main points are summarised in Appendix \ref{susyvar} and extended to the five-loop case as required here; we shall make frequent reference to these results in this section.

The general multi-field Wess-Zumino model is described by couplings $y^{ijk}$ (where $i$ labels the fields) and their complex conjugates $y_{ijk}$. According to the non-renormalisation theorem, the $\beta$ function is then given by 
\be
\beta^{ijk}=y^{l(ij}\gamma^{k)}{}_l, \quad \beta_{ijk}=y_{l(ij}\gamma^l{}_{k)},
\label{gambet}
\ee
where $\gamma^i{}_j$ is the anomalous dimension\footnote{Ref.~\cite{gracey} uses the notation $g^{ijk}$ for the Yukawa couplings; but we prefer to retain $g$ for a scalar-type coupling. The main practical difference from our point of view is the replacement of $\mu^{\epsilon}$ by $\mu^{\frac12\epsilon}$ in Eq.~\eqref{gb} for a Yukawa coupling.}.
The anomalous dimension results up to three loops are given by
\begin{align}
\gamma&{} = c_{11} \, 
\tikz[baseline=(vert_cent.base)]{
  \node (vert_cent) {\hspace{-13pt}$\phantom{-}$};
  \draw (-0.4,0)--(0.1,0)
        (0.7,0) ++(0:0.6cm) arc (0:360:0.6cm and 0.4cm)
        (1.3,0)--(1.8,0);
        \filldraw [gray] (0.1,0) circle [radius=1.5pt];
}
+ c_{21}\, 
\tikz[baseline=(vert_cent.base)]{
  \node (vert_cent) {\hspace{-13pt}$\phantom{-}$};
  \draw (-0.4,0)--(0.1,0)
       (0.7,0) ++(0:0.6cm and 0.4cm) arc (0:50:0.6cm and 0.4cm) node (n1)
        {}
       (0.7,0) ++(50:0.6cm and 0.4cm) arc (50:130:0.6cm and 0.4cm) node (n2) 
       {}
       (0.7,0) ++(130:0.6cm and 0.4cm) arc (130:360:0.6cm and 0.4cm)
        (n1.base) to[out=215,in=325] (n2.base)
                (1.3,0)--(1.8,0);
                  \filldraw [gray] (n1) circle [radius=1.5pt]; 
          \filldraw [gray]  (0.1,0) circle [radius=1.5pt]; 
          }
\nn 
&{} + c_{31} \,
\tikz[baseline=(vert_cent.base)]{
  \node (vert_cent) {\hspace{-13pt}$\phantom{-}$};
  \draw (-0.4,0)--(0.1,0)
     (0.7,0) ++(0:0.6cm and 0.4cm) arc (0:50:0.6cm and 0.4cm) node (n1)
            {}
     (0.7,0) ++(50:0.6cm and 0.4cm) arc (50:130:0.6cm and 0.4cm) node(n2)
       {}
        (0.7,0) ++(130:0.6cm and 0.4cm) arc (130:230:0.6cm and 0.4cm) node(n3){}
          (0.7,0) ++(230:0.6cm and 0.4cm) arc (230:310:0.6cm and 0.4cm)  node(n4) {} 
          (0.7,0) ++(310:0.6cm and 0.4cm) arc (310:360:0.6cm and 0.4cm) 
           (n1.base) to (0.75,0.05)
         (n2.base) to  (n4.base) 
         (0.65,-0.05) to (n3.base)
         (1.3,0)--(1.8,0);
                   \filldraw [gray] (n1) circle [radius=1.5pt]; 
          \filldraw [gray]  (0.1,0) circle [radius=1.5pt];
          \filldraw [gray] (n4) circle [radius=1.5pt]; 
          }
+ c_{32} \,
\tikz[baseline=(vert_cent.base)]{[scale=1.5]
  \node (vert_cent) {\hspace{-13pt}$\phantom{-}$};
  \draw (-0.4,0)--(0.1,0)
     (0.7,0) ++(0:0.6cm and 0.4cm) arc (0:50:0.6cm and 0.4cm) node (n1)
            {}
     (0.7,0) ++(50:0.6cm and 0.4cm) arc (50:130:0.6cm and 0.4cm) node(n2)
       {}
        (0.7,0) ++(130:0.6cm and 0.4cm) arc (130:230:0.6cm and 0.4cm) node(n3){}
          (n1.base) to[out=215,in=325] (n2.base) 
          (0.7,0) ++(230:0.6cm and 0.4cm) arc (230:310:0.6cm and 0.4cm)  node(n4) {} 
          (0.7,0) ++(310:0.6cm and 0.4cm) arc (310:360:0.6cm and 0.4cm) 
         (n3.base) to[out=-325,in=-215] (n4.base) 
         (1.3,0)--(1.8,0);
                   \filldraw [gray] (n1) circle [radius=1.5pt]; 
          \filldraw [gray]  (0.1,0) circle [radius=1.5pt];
          \filldraw [gray] (n4) circle [radius=1.5pt]; 
          } 
+ c_{33} \,
\tikz[baseline=(vert_cent.base)]{
  \node (vert_cent) {\hspace{-13pt}$\phantom{-}$};
  \draw (-0.4,0)--(0.1,0)
     (0.7,0) ++(0:0.6cm and 0.4cm) arc (0:35:0.6cm and 0.4cm) node (n1)
            {}
     (0.7,0) ++(35:0.6cm and 0.4cm) arc (35:75:0.6cm and 0.4cm) node(n2)
       {}
        (0.7,0) ++(75:0.6cm and 0.4cm) arc (75:105:0.6cm and 0.4cm) node(n3){}
      (n1.base) to  [out=270,in=240] (n2.base) 
          (0.7,0) ++(105:0.6cm and 0.4cm) arc (105:145:0.6cm and 0.4cm)  node(n4) {} 
          (0.7,0) ++(145:0.6cm and 0.4cm) arc (145:360:0.6cm and 0.4cm) 
       (n3.base) to  [out= 330,in= 300] (n4.base) 
         (1.3,0)--(1.8,0);
                   \filldraw [gray] (n1) circle [radius=1.5pt]; 
          \filldraw [gray]  (0.1,0) circle [radius=1.5pt];
          \filldraw [gray] (n3) circle [radius=1.5pt]; 
          }
{}+ c_{34}  \,
\tikz[baseline=(vert_cent.base)]{
  \node (vert_cent) {\hspace{-13pt}$\phantom{-}$};
  \draw (-0.4,0)--(0.1,0)
     (0.7,0) ++(0:0.6cm and 0.4cm) arc (0:35:0.6cm and 0.4cm) node (n1)
            {}
     (0.7,0) ++(35:0.6cm and 0.4cm) arc (35:65:0.6cm and 0.4cm) node(n2)
       {}
        (0.7,0) ++(65:0.6cm and 0.4cm) arc (65:115:0.6cm and 0.4cm) node(n3){}
          (0.7,0) ++(115:0.6cm and 0.4cm) arc (115:145:0.6cm and 0.4cm)  node(n4) {} 
          (0.7,0) ++(145:0.6cm and 0.4cm) arc (145:360:0.6cm and 0.4cm) 
       (n2.base) to  [out= 220,in= 320] (n3.base) 
       (n1.base) to  [out=220,in=320] (n4.base) 
         (1.3,0)--(1.8,0);
                   \filldraw [gray] (n1) circle [radius=1.5pt]; 
          \filldraw [gray]  (0.1,0) circle [radius=1.5pt];
          \filldraw [gray] (n3) circle [radius=1.5pt]; 
          }+\ldots,
          \label{gam3}
\end{align}
where for ease of comparison we follow the diagram labelling of Ref.~\cite{gracey}. Each vertex represents a tensor coupling, though we suppress indices as far as possible. We distinguish the conjugate vertices by dots; the propagators link conjugate to unconjugated vertices. We assume that within a particular renormalisation scheme, the renormalisation functions are expressed in terms of the coupling appropriate to that scheme. In this Section we are working in $\msbar$ and therefore quantities are expressed in terms of $y^{\msbar}$ unless otherwise stated. The $\msbar$ values of the coefficients in Eq.~\eqref{gam3} are
\begin{align}
c_1\equiv c_{11}=\tfrac12 \, , \quad& c_{21}= - \tfrac12\, , \nn
c_{31}= \tfrac32 \, \zeta_3,\quad  c^{\msbar}_{32}=-\frac18\, ,\quad &c^{\msbar}_{33}=-\frac14,\quad    c^{\msbar}_{34}=1 
\label{c3WZ}
\end{align}
(the one and two loop values and also $c_{31}$ are actually scheme-independent so there is no need to label them). Here for future convenience we introduce $c_1$ as a shorthand notation for $c_{11}$, since it appears frequently later on. So far there are no even-$\zeta$ terms even in $\msbar$; they appear for the first time at four loops, as of course we have already seen in the single-coupling scalar field case in Section \ref{single}. 
The four-loop anomalous dimension is\cite{avdeev,FJJ}
\begin{align}
\gamma^{(4)}={}& c_{41} \, 
\tikz[baseline=(vert_cent.base)]{
  \node (vert_cent) {\hspace{-13pt}$\phantom{-}$};
  \draw (-0.6,0)--(-0.1,0)
     (0.7,0) ++(0:0.8cm and 0.5cm) arc (0:50:0.8cm and 0.5cm) node (n1)  {}
     (0.7,0) ++(50:0.8cm and 0.5cm) arc (50:130:0.8cm and 0.5cm) node(n2) {}
              (0.7,0) ++(130:0.8cm and 0.5cm) arc (130:230:0.8cm and 0.5cm)  node(n3) {} 
          (0.7,0) ++(230:0.8cm and 0.5cm) arc (230:310:0.8cm and 0.5cm) node(n4) {}
            (0.7,0) ++(310:0.8cm and 0.5cm) arc (310:360:0.8cm and 0.5cm) 
                (n2.base) to [out = 330, in = 90] (0.45,0)
           (0.45,0) to [out = 270, in =30] (n3.base)
            (n1.base) to  [out= 220,in= 90]  (0.95,0)
            (0.95,0)  to [out = 270, in = 330] (n4.base) 
            (0.45,0) to (0.95,0) 
         (1.5,0)--(2,0);
                   \filldraw [gray] (n1) circle [radius=1.5pt]; 
          \filldraw [gray]  (-0.1,0) circle [radius=1.5pt];
         \filldraw [gray] (n4) circle [radius=1.5pt]; 
       \filldraw [gray] (0.45,0) circle [radius=1.5pt]; 
          } 
+ c_{42} \, 
 \tikz[baseline=(vert_cent.base)]{
  \node (vert_cent) {\hspace{-13pt}$\phantom{-}$};
  \draw (-0.6,0)--(-0.1,0)
     (0.7,0) ++(0:0.8cm and 0.5cm) arc (0:45:0.8cm and 0.5cm) node (n1)  {}
     (0.7,0) ++(45:0.8cm and 0.5cm) arc (45:95:0.8cm and 0.5cm) node(n2) {}
              (0.7,0) ++(95:0.8cm and 0.5cm) arc (95:115:0.8cm and 0.5cm)  node(n3) {} 
          (0.7,0) ++(115:0.8cm and 0.5cm) arc (115:155:0.8cm and 0.5cm) node(n4) {}
            (0.7,0) ++(155:0.8cm and 0.5cm) arc (155:265:0.8cm and 0.5cm)  node(n5) {}
             (0.7,0) ++(265:0.8cm and 0.5cm) arc (265:315:0.8cm and 0.5cm) node(n6) {}
              (0.7,0) ++(315:0.8cm and 0.5cm) arc (315:360:0.8cm and 0.5cm)
            (n1.base) to (1.03,0.05)
            (0.95,-0.05) to (n5.base)
           (n2.base) to (n6.base) 
     (n3.base) to  [out= 320,in= 300] (n4.base) 
         (1.5,0)--(2,0);
                   \filldraw [gray] (n1) circle [radius=1.5pt]; 
          \filldraw [gray]  (-0.1,0) circle [radius=1.5pt];
         \filldraw [gray] (n3) circle [radius=1.5pt]; 
         \filldraw [gray] (n6) circle [radius=1.5pt]; 
          }
          + c_{43}
 \tikz[baseline=(vert_cent.base)]{
  \node (vert_cent) {\hspace{-13pt}$\phantom{-}$};
  \draw (-0.6,0)--(-0.1,0)
     (0.7,0) ++(0:0.8cm and 0.5cm) arc (0:25:0.8cm and 0.5cm) node (n1)
            {}
     (0.7,0) ++(25:0.8cm and 0.5cm) arc (25:65:0.8cm and 0.5cm) node(n2)
       {}
        (0.7,0) ++(65:0.8cm and 0.5cm) arc (65:85:0.8cm and 0.5cm) node(n3){}
        (0.7,0) ++(85:0.8cm and 0.5cm) arc (85:135:0.8cm and 0.5cm) node(n4){}
        (0.7,0) ++(135:0.8cm and 0.5cm) arc (135:225:0.8cm and 0.5cm)  node(n5) {} 
        (0.7,0) ++(225:0.8cm and 0.5cm) arc (225:275:0.8cm and 0.5cm) node(n6) {}
            (0.7,0) ++(275:0.8cm and 0.5cm) arc (275:360:0.8cm and 0.5cm)
             (n1.base) to  [out=240,in=240] (n2.base) 
             (n3.base) to  (0.43,0.05)
             (0.37,-0.05) to (n5.base) 
     (n4.base) to  (n6.base) 
         (1.5,0)--(2,0);
                   \filldraw [gray] (n1) circle [radius=1.5pt]; 
          \filldraw [gray]  (-0.1,0) circle [radius=1.5pt];
          \filldraw [gray] (n3) circle [radius=1.5pt]; 
           \filldraw [gray] (n6) circle [radius=1.5pt]; 
  }       
  \nn 
 &{}+ c_{44} \, 
  \tikz[baseline=(vert_cent.base)]{
  \node (vert_cent) {\hspace{-13pt}$\phantom{-}$};
  \draw (-0.6,0)--(-0.1,0)
     (0.7,0) ++(0:0.8cm and 0.5cm) arc (0:45:0.8cm and 0.5cm) node (n1)  {}
     (0.7,0) ++(45:0.8cm and 0.5cm) arc (45:73:0.8cm and 0.5cm) node(n5)  {}
     (0.7,0) ++(73:0.8cm and 0.5cm) arc (73:107:0.8cm and 0.5cm) node(n6)  {}
     (0.7,0) ++(107:0.8cm and 0.5cm) arc (107:135:0.8cm and 0.5cm) node(n2)  {}
        (0.7,0) ++(135:0.8cm and 0.5cm) arc (135:225:0.8cm and 0.5cm) node(n3){}
          (0.7,0) ++(225:0.8cm and 0.5cm) arc (225:315:0.8cm and 0.5cm)  node(n4) {} 
          (0.7,0) ++(315:0.8cm and 0.5cm) arc (315:360:0.8cm and 0.5cm) 
           (n1.base) to (0.76,0.05)
         (n2.base) to  (n4.base) 
         (0.65,-0.05) to (n3.base)
         (n5.base) to [out = 280,in = 260] (n6.base)
         (1.5,0)--(2,0);
                   \filldraw [gray] (n1) circle [radius=1.5pt]; 
          \filldraw [gray]  (-0.1,0) circle [radius=1.5pt];
          \filldraw [gray] (n4) circle [radius=1.5pt]; 
           \filldraw [gray] (n6) circle [radius=1.5pt];
          }
+ c_{45} \, 
\tikz[baseline=(vert_cent.base)]{
  \node (vert_cent) {\hspace{-13pt}$\phantom{-}$};
  \draw (-0.6,0)--(-0.1,0)
     (0.7,0) ++(0:0.8cm and 0.5cm) arc (0:35:0.8cm and 0.5cm) node (n1)
            {}
     (0.7,0) ++(35:0.8cm and 0.5cm) arc (35:75:0.8cm and 0.5cm) node(n2)
       {}
        (0.7,0) ++(75:0.8cm and 0.5cm) arc (75:105:0.8cm and 0.5cm) node(n3){}
      (n1.base) to  [out=270,in=240] (n2.base) 
          (0.7,0) ++(105:0.8cm and 0.5cm) arc (105:145:0.8cm and 0.5cm)  node(n4) {} 
          (0.7,0) ++(145:0.8cm and 0.5cm) arc (145:240:0.8cm and 0.5cm) node(n5){}
          (0.7,0) ++(240:0.8cm and 0.5cm) arc (240:300:0.8cm and 0.5cm) node(n6){}
          (0.7,0) ++(300:0.8cm and 0.5cm) arc (300:360:0.8cm and 0.5cm) 
       (n3.base) to  [out= 280,in= 320] (n4.base) 
       (n5.base) to  [out= 80,in= 100] (n6.base) 
         (1.5,0)--(2,0);
                   \filldraw [gray] (n1) circle [radius=1.5pt]; 
          \filldraw [gray]  (-0.1,0) circle [radius=1.5pt];
          \filldraw [gray] (n3) circle [radius=1.5pt]; 
           \filldraw [gray] (n6) circle [radius=1.5pt];
           }
+ c_{46} \, 
  \tikz[baseline=(vert_cent.base)]{
  \node (vert_cent) {\hspace{-13pt}$\phantom{-}$};
  \draw (-0.6,0)--(-0.1,0)
     (0.7,0) ++(0:0.8cm and 0.5cm) arc (0:35:0.8cm and 0.5cm) node (n1)
            {}
     (0.7,0) ++(35:0.8cm and 0.5cm) arc (35:65:0.8cm and 0.5cm) node(n2)
       {}
        (0.7,0) ++(65:0.8cm and 0.5cm) arc (65:115:0.8cm and 0.5cm) node(n3){}
          (0.7,0) ++(115:0.8cm and 0.5cm) arc (115:145:0.8cm and 0.5cm)  node(n4) {} 
          (0.7,0) ++(145:0.8cm and 0.5cm) arc (145:240:0.8cm and 0.5cm)  node(n5){}
           (0.7,0) ++(240:0.8cm and 0.5cm) arc (240:300:0.8cm and 0.5cm) node(n6){}
          (0.7,0) ++(300:0.8cm and 0.5cm) arc (300:360:0.8cm and 0.5cm) 
       (n2.base) to  [out= 220,in= 320] (n3.base) 
       (n1.base) to  [out=220,in=320] (n4.base) 
         (n5.base) to  [out= 70,in= 110] (n6.base)
         (1.5,0)--(2,0);
                   \filldraw [gray] (n1) circle [radius=1.5pt]; 
          \filldraw [gray]  (-0.1,0) circle [radius=1.5pt];
          \filldraw [gray] (n3) circle [radius=1.5pt]; 
           \filldraw [gray] (n6) circle [radius=1.5pt]; 
          } 
          \nn 
 &+ c_{47} \, 
\tikz[baseline=(vert_cent.base)]{
  \node (vert_cent) {\hspace{-13pt}$\phantom{-}$};
  \draw (-0.6,0)--(-0.1,0)
     (0.7,0) ++(0:0.8cm and 0.5cm) arc (0:45:0.8cm and 0.5cm) node (n1)  {}
            (0.7,0) ++(135:0.8cm and 0.5cm) arc (135:360:0.8cm and 0.5cm) 
                (0.7,0.36) ++(0:0.55cm and 0.35cm) arc (0:50:0.55cm and 0.35cm) node (n2) {}
          (0.7,0.36) ++(50:0.55cm and 0.35cm) arc (50:130:0.55cm and 0.35cm)  node (n3)  {}
           (0.7,0.36) ++(130:0.55cm and 0.35cm) arc (130:230:0.55cm and 0.35cm) node (n4)  {}
            (0.7,0.36) ++(230:0.55cm and 0.35cm) arc (230:310:0.55cm and 0.35cm) node (n5)  {}
             (0.7,0.36) ++(310:0.55cm and 0.35cm) arc (310:360:0.55cm and 0.35cm) node (n6)  {}
             (n2.base) to (0.74,0.4)
             (0.62,0.3) to (n4.base)
             (n3.base) to (n5.base)
         (1.5,0)--(2,0);
                   \filldraw [gray] (n1) circle [radius=1.5pt]; 
         \filldraw [gray]  (-0.1,0) circle [radius=1.5pt];
         \filldraw [gray] (n3) circle [radius=1.5pt]; 
          \filldraw [gray] (n4) circle [radius=1.5pt];
          }
+ c_{48} \,
          \tikz[baseline=(vert_cent.base)]{
  \node (vert_cent) {\hspace{-13pt}  $\phantom{-}$};
  \draw (-0.6,0)--(-0.1,0)
     (0.7,0) ++(0:0.8cm and 0.5cm) arc (0:35:0.8cm and 0.5cm) node (n1)
            {}
     (0.7,0) ++(35:0.8cm and 0.5cm) arc (35:65:0.8cm and 0.5cm) node(n2)
       {}
        (0.7,0) ++(65:0.8cm and 0.5cm) arc (65:115:0.8cm and 0.5cm) node(n3){}
          (0.7,0) ++(115:0.8cm and 0.5cm) arc (115:145:0.8cm and 0.5cm)  node(n4) {} 
          (0.7,0) ++(145:0.8cm and 0.5cm) arc (145:360:0.8cm and 0.5cm) 
       (n2.base) to  [out= 220,in= 320] (n3.base) 
        (0.7,0.7) ++(35:0.8cm and  -0.7cm) arc (35:65:0.8cm and -0.7cm) node(n5) {}
         (0.7,0.7) ++(65:0.8cm and  -0.7cm) arc (65:115:0.8cm and -0.7cm) node(n6) {}
          (0.7,0.7) ++(115:0.8cm and  -0.7cm) arc (115:145:0.8cm and -0.7cm) 
           (n5.base) to  [out= 140,in= 40] (n6.base)
         (1.5,0)--(2,0);
                   \filldraw [gray] (n1) circle [radius=1.5pt]; 
          \filldraw [gray]  (-0.1,0) circle [radius=1.5pt];
          \filldraw [gray] (n3) circle [radius=1.5pt]; 
           \filldraw [gray] (n6) circle [radius=1.5pt]; 
          }\nn
&+ c_{49} 
           \tikz[baseline=(vert_cent.base)]{
  \node (vert_cent) {\hspace{-13pt}$\phantom{-}$};
  \draw (-0.6,0)--(-0.1,0)
     (0.7,0) ++(0:0.8cm and 0.5cm) arc (0:25:0.8cm and 0.5cm) node (n1)
            {}
     (0.7,0) ++(25:0.8cm and 0.5cm) arc (25:50:0.8cm and 0.5cm) node(n2)
       {}
        (0.7,0) ++(50:0.8cm and 0.5cm) arc (50:80:0.8cm and 0.5cm) node(n3){}
        (0.7,0) ++(80:0.8cm and 0.5cm) arc (80:95:0.8cm and 0.5cm) node(n4){}
          (0.7,0) ++(95:0.8cm and 0.5cm) arc (95:120:0.8cm and 0.5cm)  node(n5) {} 
          (0.7,0) ++(120:0.8cm and 0.5cm) arc (120:155:0.8cm and 0.5cm) node(n6) {}
            (0.7,0) ++(155:0.8cm and 0.5cm) arc (155:360:0.8cm and 0.5cm)
             (n1.base) to  [out=240,in=250] (n4.base) 
             (n2.base) to  [out= 300,in= 260] (n3.base) 
     (n5.base) to  [out= 320,in= 300] (n6.base) 
         (1.5,0)--(2,0);
                   \filldraw [gray] (n1) circle [radius=1.5pt]; 
          \filldraw [gray]  (-0.1,0) circle [radius=1.5pt];
          \filldraw [gray] (n3) circle [radius=1.5pt]; 
           \filldraw [gray] (n5) circle [radius=1.5pt]; 
          }
          + c_{410}
 \tikz[baseline=(vert_cent.base)]{
  \node (vert_cent) {\hspace{-13pt}$\phantom{-}$};
  \draw (-0.6,0)--(-0.1,0)
     (0.7,0) ++(0:0.8cm and 0.5cm) arc (0:25:0.8cm and 0.5cm) node (n1)
            {}
     (0.7,0) ++(25:0.8cm and 0.5cm) arc (25:60:0.8cm and 0.5cm) node(n2)
       {}
        (0.7,0) ++(60:0.8cm and 0.5cm) arc (60:85:0.8cm and 0.5cm) node(n3){}
        (0.7,0) ++(85:0.8cm and 0.5cm) arc (85:100:0.8cm and 0.5cm) node(n4){}
          (0.7,0) ++(100:0.8cm and 0.5cm) arc (100:130:0.8cm and 0.5cm)  node(n5) {} 
          (0.7,0) ++(130:0.8cm and 0.5cm) arc (130:155:0.8cm and 0.5cm) node(n6) {}
            (0.7,0) ++(155:0.8cm and 0.5cm) arc (155:360:0.8cm and 0.5cm)
             (n1.base) to  [out=240,in=240] (n2.base) 
             (n3.base) to  [out= 290,in= 300] (n6.base) 
     (n4.base) to  [out= 320,in= 270] (n5.base) 
         (1.5,0)--(2,0);
                   \filldraw [gray] (n1) circle [radius=1.5pt]; 
          \filldraw [gray]  (-0.1,0) circle [radius=1.5pt];
          \filldraw [gray] (n3) circle [radius=1.5pt]; 
           \filldraw [gray] (n5) circle [radius=1.5pt]; 
  }       
  +c_{411} \, 
\tikz[baseline=(vert_cent.base)]{
  \node (vert_cent) {\hspace{-13pt}$\phantom{-}$};
  \draw (-0.6,0)--(-0.1,0)
     (0.7,0) ++(0:0.8cm and 0.5cm) arc (0:25:0.8cm and 0.5cm) node (n1)
            {}
     (0.7,0) ++(25:0.8cm and 0.5cm) arc (25:60:0.8cm and 0.5cm) node(n2)
       {}
        (0.7,0) ++(60:0.8cm and 0.5cm) arc (60:72.5:0.8cm and 0.5cm) node(n3){}
      (n1.base) to  [out=270,in=250] (n2.base) 
        (0.7,0) ++(72.5:0.8cm and 0.5cm) arc (72.5:107.5:0.8cm and 0.5cm) node(n4){}
          (0.7,0) ++(107.5:0.8cm and 0.5cm) arc (107.5:120:0.8cm and 0.5cm)  node(n5) {} 
          (0.7,0) ++(120:0.8cm and 0.5cm) arc (120:155:0.8cm and 0.5cm) node(n6) {}
            (0.7,0) ++(155:0.8cm and 0.5cm) arc (155:360:0.8cm and 0.5cm)
             (n3.base) to  [out= 280,in= 260] (n4.base) 
     (n5.base) to  [out= 320,in= 300] (n6.base) 
         (1.5,0)--(2,0);
                   \filldraw [gray] (n1) circle [radius=1.5pt]; 
          \filldraw [gray]  (-0.1,0) circle [radius=1.5pt];
          \filldraw [gray] (n3) circle [radius=1.5pt]; 
           \filldraw [gray] (n5) circle [radius=1.5pt]; 
          }
\nn 
          &{}+ c_{412} \,
          \tikz[baseline=(vert_cent.base)]{
  \node (vert_cent) {\hspace{-13pt}$\phantom{-}$};
  \draw (-0.6,0)--(-0.1,0)
   (0.7,0) ++(0:0.8cm and 0.5cm) arc (0:25:0.8cm and 0.5cm) node (n1) {}
     (0.7,0) ++(25:0.8cm and 0.5cm) arc (25:42.5:0.8cm and 0.5cm) node (n2)
            {}
     (0.7,0) ++(42.5:0.8cm and 0.5cm) arc (42.5:80:0.8cm and 0.5cm) node(n3)
       {}
        (0.7,0) ++(80:0.8cm and 0.5cm) arc (80:100:0.8cm and 0.5cm) node(n4){}
      (n2.base) to  [out=240,in=240] (n3.base) 
          (0.7,0) ++(100:0.8cm and 0.5cm) arc (100:137.5:0.8cm and 0.5cm)  node(n5){}
           (0.7,0) ++(137.5:0.8cm and 0.5cm) arc (137.5:155:0.8cm and 0.5cm)  node(n6) {} 
            (0.7,0) ++(155:0.8cm and 0.5cm) arc (155:360:0.8cm and 0.5cm) 
         (n4.base) to  [out= 290,in= 320] (n5.base) 
         (n1.base) to  [out= 220,in= 320] (n6.base) 
         (1.5,0)--(2,0);
                   \filldraw [gray] (n1) circle [radius=1.5pt]; 
          \filldraw [gray]  (-0.1,0) circle [radius=1.5pt];
          \filldraw [gray] (n3) circle [radius=1.5pt]; 
           \filldraw [gray] (n5) circle [radius=1.5pt];
          }
          + c_{413} \, 
          \tikz[baseline=(vert_cent.base)]{
  \node (vert_cent) {\hspace{-13pt}$\phantom{-}$};
  \draw (-0.6,0)--(-0.1,0)
     (0.7,0) ++(0:0.8cm and 0.5cm) arc (0:30:0.8cm and 0.5cm) node (n1)
            {}
     (0.7,0) ++(30:0.8cm and 0.5cm) arc (30:55:0.8cm and 0.5cm) node(n2)
       {}
        (0.7,0) ++(55:0.8cm and 0.5cm) arc (55:72.5:0.8cm and 0.5cm) node(n3){}
        (0.7,0) ++(72.5:0.8cm and 0.5cm) arc (72.5:107.5:0.8cm and 0.5cm) node(n4){}
          (0.7,0) ++(107.5:0.8cm and 0.5cm) arc (107.5:125:0.8cm and 0.5cm)  node(n5) {} 
          (0.7,0) ++(125:0.8cm and 0.5cm) arc (125:150:0.8cm and 0.5cm) node(n6) {}
            (0.7,0) ++(150:0.8cm and 0.5cm) arc (150:360:0.8cm and 0.5cm)
            (n1.base) to  [out=230,in=310] (n6.base) 
             (n2.base) to  [out= 280,in= 260] (n5.base) 
     (n3.base) to  [out= 280,in= 260] (n4.base) 
         (1.5,0)--(2,0);
                   \filldraw [gray] (n1) circle [radius=1.5pt]; 
          \filldraw [gray]  (-0.1,0) circle [radius=1.5pt];
          \filldraw [gray] (n3) circle [radius=1.5pt]; 
           \filldraw [gray] (n5) circle [radius=1.5pt]; 
          }
\label{gam4}
\end{align}
where the $\msbar$ values of the coefficients are 
\begin{align}
c^{\msbar}_{41}=&-10\zeta_5,\nn
c^{\msbar}_{42} = c^{\msbar}_{43}=\frac12c^{\msbar}_{44}=&- \frac14(6 \zeta_3 - 3 \zeta_4) ,\nn
c^{\msbar}_{45}=c^{\msbar}_{411}=& \tfrac18(2\zeta_3-1), \nn
c^{\msbar}_{46}=c^{\msbar}_{49}= c^{\msbar}_{410}=& \tfrac13, \nn
c^{\msbar}_{47}=& - \tfrac14( 6 \zeta_3+ 3 \zeta_4),\nn
c^{\msbar}_{48}=\tfrac{5}{24},\quad c^{\msbar}_{412}=& \tfrac{1}{12} ( 5 - 6 \zeta_3), \quad c^{\msbar}_{413}=-\tfrac52.
\label{c4WZ}
\end{align}
Here  we indeed see for the first time in this theory the appearance of $\zeta_4$ terms. We now therefore seek a transformation to a different scheme in which the $\zeta_4$ terms are absent. Based on our experience in Section \ref{single}, we expect to find that they cancel in $\MOp$; on the other hand it has been shown in Ref.~\cite{gracey} that they also cancel in $\MO$. We shall discuss these two schemes in turn. As we show in Appendix \ref{mstrans}, the transformation from $\msbar$ to $\MOp$ is effected by a change of coupling corresponding to the part of the finite term proportional to even $\zeta$s, as evaluated in the $\MOp$ scheme itself; while the transformation from $\msbar$ to $\MO$ is effected by a change of coupling corresponding to the full finite term, as evaluated in the $\msbar$ scheme. We shall denote quantities associated with these transformations by superscripts $A$ and $B$ respectively.

We may express variations of the coupling $g$ in an analogous fashion to Eq.~\eqref{gambet}, by
\be
\delta y^{ijk}=y^{l(ij}h^{k)}{}_l, \quad \delta y_{ijk}=y_{l(ij}h^l_{k)}.
\label{delgam}
\ee
We parametrise the quantities $h^i{}_j$  in a similar manner to Eqs.~\eqref{gam3}, \eqref{gam4} but with $c_{21}$ etc replaced by $\epsilon_{21}$, etc, and again denoting $\epsilon_1\equiv \epsilon_{11}$, so we shall use superscripts $A$ and $B$ for the $\epsilon$ and quantities constructed from them.
As in Eq.~\eqref{delone}, we expect the four-loop $\zeta_4$ terms in Eqs.~\eqref{gam4}, \eqref{c4WZ} to be cancelled by $(\delta y)^{(3)}$ as in Eq.~\eqref{delgam}, with
\be
h^{(3)}=-\frac14\gamma^{\zeta_3}_3\zeta_4=f_{3,0}^{\zeta_4}\zeta_4
=-\frac38\zeta_4\tikz[baseline=(vert_cent.base)]{
  \node (vert_cent) {\hspace{-13pt}$\phantom{-}$};
  \draw (-0.4,0)--(0.1,0)
     (0.7,0) ++(0:0.6cm and 0.4cm) arc (0:50:0.6cm and 0.4cm) node (n1)
            {}
     (0.7,0) ++(50:0.6cm and 0.4cm) arc (50:130:0.6cm and 0.4cm) node(n2)
       {}
        (0.7,0) ++(130:0.6cm and 0.4cm) arc (130:230:0.6cm and 0.4cm) node(n3){}
          (0.7,0) ++(230:0.6cm and 0.4cm) arc (230:310:0.6cm and 0.4cm)  node(n4) {} 
          (0.7,0) ++(310:0.6cm and 0.4cm) arc (310:360:0.6cm and 0.4cm) 
           (n1.base) to (0.75,0.05)
         (n2.base) to  (n4.base) 
         (0.65,-0.05) to (n3.base)
         (1.3,0)--(1.8,0);
                   \filldraw [gray] (n1) circle [radius=1.5pt]; 
          \filldraw [gray]  (0.1,0) circle [radius=1.5pt];
          \filldraw [gray] (n4) circle [radius=1.5pt]; 
          };
\label{dg3susy}
\ee
which corresponds to taking 
\be
\epsilon^A_{31}=-\frac14c^{\zeta_3}_{31}\zeta_4=-\frac38\zeta_4
\label{e31def}
\ee
(where $c^{\zeta_3}_{31}$ is the coefficient of $\zeta_3$ in $c_{31}$) and all other $\epsilon$s through three loops to be zero. Note the appearance of $f$ (defined in Eq.~\eqref{ffdef}) and $\gamma$, replacing $F$ and $\beta$, since we are now dealing with two-point functions. Also note the different relative sign in Eq.~\eqref{dg3susy} compared with Eq.~\eqref{delone}, due to the minus sign in Eq.~\eqref{gamdefb}. This transformation where we do the minimum necessary to remove the even-$\zeta$ terms corresponds to the $\MOp$ scheme and so we use the $A$ superscript. As indicated in Eqs.~\eqref{dg3susy}, \eqref{e31def}, we can extract the value of $\epsilon^A_{31}$ either from the three-loop $\beta$-function values in Eq.~\eqref{c3WZ}, or from the finite $\zeta_4$ coefficient in $\fbar_{31}^{\msbar}$ in Eq.~\eqref{zms3}, multiplied by $(\frac12)$. This factor of $(\frac12)$ is required by our definition of $\gamma$ in Eq.~\eqref{gamdefa}, which in turn ensures the standard expression for the supersymmetric $\beta$-function in Eq.~\eqref{gambet}.
It is easy to check using Eq.~\eqref{blin}, where now in this multi-coupling case
\be
B\cdot  \equiv B^{ijk}\frac{\pa}{\pa y^{ijk}},
\ee
that this transformation does indeed work (with the one-loop $\beta$ function from Eqs.~\eqref{gam3}, \eqref{c3WZ}); more explicitly, in Eq.~\eqref{susyvars4} the variations of the four-loop coefficients are expressed in terms of a small set of quantities $X_{1,31}$, $\Xhat_{1,32}-\Xhat_{1,34}$ (the hats represent the presence of potential second-order variations described in Appendix \ref{susyvar}, which are relevant for the discussion of $\MO$ but not for the current case of $\MOp$). From Eqs.~\eqref{X31def}, \eqref{X32def}, since the only non-zero $\epsilon$ in the $A$ scheme at this level is $\epsilon^A_{31}$ in Eq.~\eqref{e31def}, we readily see
\be
X_{1,31}^{A}=-\frac{3}{16}\zeta_4,\quad \Xhat_{1,32}^{A}=\Xhat_{1,33}^{A}=\Xhat_{1,34}^{A}=0.
\ee
This results in variations $\delta c_{42}$, $\delta c_{43}$, $\delta c_{44}$, and $\delta c_{47}$ in Eq.~\eqref{susyvars4} which cancel the $\zeta_4$ terms in Eq.~\eqref{c4WZ}. 
 
The full MOM ($B$) scheme corresponds to including {\it all} the finite parts in our coupling redefinitions, not just those proportional to $\zeta_4$. In particular we need to use the value $\epsilon^B_1=-\frac12$ as again is easily computed directly from Eq.~\eqref{zms12}. However this makes no change to the $\zeta_4$ dependence of  the $X$, $\Xhat$ quantities in Eq.~\eqref{susyvars4}, since according to Eq.~\eqref{X31def}, in $X_{1,31}$, $\epsilon_1$ multiplies $c_{31}$ which has no $\zeta_4$-dependence, and similarly for $\Xhat_{1,32}-\Xhat_{1,34}$ (as one may see explicitly in Eq.~\eqref{xthree}); so we easily find
\be
X_{1,31}^{\zeta_4B}\zeta_4=X^A_{1,31}=-\frac{3}{16}\zeta_4,\quad \Xhat_{1,32}^{\zeta_4B}=\Xhat_{1,33}^{\zeta_4B}=\Xhat_{1,34}^{\zeta_4B}=0.
\ee
in other words the $\zeta_4$ dependence of the $X$ and $\Xhat$ quantities is unchanged. Hence at this level, the $\zeta_4$ independence of MOM follows immediately from that of $\rm MOM'$.
 
Similarly, we expect that the five-loop even $\zeta$s will be cancelled by taking the four-loop contribution to the $\MOp$ coupling redefinition in Eq.~\eqref{MOM'}, again using $(\delta y)^{(4)}$ as defined by Eq.~\eqref{delgam} and with 
\be
h^{(4)}=-\frac{3}{16}\gamma^{\zeta_3\msbar}_4\zeta_4-\frac{5}{16}\gamma^{\zeta_5\msbar}_4\zeta_6=f^{\zeta_4\MOp}_{4,0}(y^{\msbar})\zeta_4+f^{\zeta_6\MOp}_{4,0}(y^{\msbar})\zeta_6
\label{delg4def}
\ee
where we have used Eqs.~\eqref{MS18}, \eqref{bet4} to give an alternative form for the coupling redefinition, analogous to that in Eq.~\eqref{dg3susy}, and once again with a different relative sign to that in Eq.~\eqref{delone}. 
Parametrising the $\epsilon$ coefficients in a similar manner to Eq.~\eqref{c4WZ}, we immediately see (from the second expression on the right-hand side of Eq.~\eqref{delg4def})
\begin{align}
\epsilon^A_{41}=&c^{\zeta_5\msbar}_{41}Z_6,\nn
\epsilon^A_{4J}=&c^{\zeta_3\msbar}_{4J}Z_4,\quad J=2\ldots13
\label{epsvals}
\end{align}
where $Z_4$, $Z_6$ are given by
\be
Z_4=-\frac{3}{16}\zeta_4,\quad Z_6=-\frac{5}{16}\zeta_6,
\label{Zdefs}
\ee
so that 
\be
\epsilon^A_{46}=\epsilon^A_{48}=\epsilon^A_{49}=\epsilon^A_{410}=\epsilon^A_{413}=0
\label{zeroeps}
\ee
 and we may write
\begin{align}
h^{(4)}=&\epsilon^{A}_{41} \, 
\tikz[baseline=(vert_cent.base)]{
  \node (vert_cent) {\hspace{-13pt}$\phantom{-}$};
  \draw (-0.6,0)--(-0.1,0)
     (0.7,0) ++(0:0.8cm and 0.5cm) arc (0:50:0.8cm and 0.5cm) node (n1)  {}
     (0.7,0) ++(50:0.8cm and 0.5cm) arc (50:130:0.8cm and 0.5cm) node(n2) {}
              (0.7,0) ++(130:0.8cm and 0.5cm) arc (130:230:0.8cm and 0.5cm)  node(n3) {} 
          (0.7,0) ++(230:0.8cm and 0.5cm) arc (230:310:0.8cm and 0.5cm) node(n4) {}
            (0.7,0) ++(310:0.8cm and 0.5cm) arc (310:360:0.8cm and 0.5cm) 
                (n2.base) to [out = 330, in = 90] (0.45,0)
           (0.45,0) to [out = 270, in =30] (n3.base)
            (n1.base) to  [out= 220,in= 90]  (0.95,0)
            (0.95,0)  to [out = 270, in = 330] (n4.base) 
            (0.45,0) to (0.95,0) 
         (1.5,0)--(2,0);
                   \filldraw [gray] (n1) circle [radius=1.5pt]; 
          \filldraw [gray]  (-0.1,0) circle [radius=1.5pt];
         \filldraw [gray] (n4) circle [radius=1.5pt]; 
       \filldraw [gray] (0.45,0) circle [radius=1.5pt]; 
          } 
+ \epsilon^{A}_{42} \, 
 \tikz[baseline=(vert_cent.base)]{
  \node (vert_cent) {\hspace{-13pt}$\phantom{-}$};
  \draw (-0.6,0)--(-0.1,0)
     (0.7,0) ++(0:0.8cm and 0.5cm) arc (0:45:0.8cm and 0.5cm) node (n1)  {}
     (0.7,0) ++(45:0.8cm and 0.5cm) arc (45:95:0.8cm and 0.5cm) node(n2) {}
              (0.7,0) ++(95:0.8cm and 0.5cm) arc (95:115:0.8cm and 0.5cm)  node(n3) {} 
          (0.7,0) ++(115:0.8cm and 0.5cm) arc (115:155:0.8cm and 0.5cm) node(n4) {}
            (0.7,0) ++(155:0.8cm and 0.5cm) arc (155:265:0.8cm and 0.5cm)  node(n5) {}
             (0.7,0) ++(265:0.8cm and 0.5cm) arc (265:315:0.8cm and 0.5cm) node(n6) {}
              (0.7,0) ++(315:0.8cm and 0.5cm) arc (315:360:0.8cm and 0.5cm)
            (n1.base) to (1.03,0.05)
            (0.95,-0.05) to (n5.base)
           (n2.base) to (n6.base) 
     (n3.base) to  [out= 320,in= 300] (n4.base) 
         (1.5,0)--(2,0);
                   \filldraw [gray] (n1) circle [radius=1.5pt]; 
          \filldraw [gray]  (-0.1,0) circle [radius=1.5pt];
         \filldraw [gray] (n3) circle [radius=1.5pt]; 
         \filldraw [gray] (n6) circle [radius=1.5pt]; 
          }
          + \epsilon^{A}_{43}
 \tikz[baseline=(vert_cent.base)]{
  \node (vert_cent) {\hspace{-13pt}$\phantom{-}$};
  \draw (-0.6,0)--(-0.1,0)
     (0.7,0) ++(0:0.8cm and 0.5cm) arc (0:25:0.8cm and 0.5cm) node (n1)
            {}
     (0.7,0) ++(25:0.8cm and 0.5cm) arc (25:65:0.8cm and 0.5cm) node(n2)
       {}
        (0.7,0) ++(65:0.8cm and 0.5cm) arc (65:85:0.8cm and 0.5cm) node(n3){}
        (0.7,0) ++(85:0.8cm and 0.5cm) arc (85:135:0.8cm and 0.5cm) node(n4){}
        (0.7,0) ++(135:0.8cm and 0.5cm) arc (135:225:0.8cm and 0.5cm)  node(n5) {} 
        (0.7,0) ++(225:0.8cm and 0.5cm) arc (225:275:0.8cm and 0.5cm) node(n6) {}
            (0.7,0) ++(275:0.8cm and 0.5cm) arc (275:360:0.8cm and 0.5cm)
             (n1.base) to  [out=240,in=240] (n2.base) 
             (n3.base) to  (0.43,0.05)
             (0.37,-0.05) to (n5.base) 
     (n4.base) to  (n6.base) 
         (1.5,0)--(2,0);
                   \filldraw [gray] (n1) circle [radius=1.5pt]; 
          \filldraw [gray]  (-0.1,0) circle [radius=1.5pt];
          \filldraw [gray] (n3) circle [radius=1.5pt]; 
           \filldraw [gray] (n6) circle [radius=1.5pt]; 
  }       
  \nn 
 &{}+ \epsilon^{A}_{44} \, 
  \tikz[baseline=(vert_cent.base)]{
  \node (vert_cent) {\hspace{-13pt}$\phantom{-}$};
  \draw (-0.6,0)--(-0.1,0)
     (0.7,0) ++(0:0.8cm and 0.5cm) arc (0:45:0.8cm and 0.5cm) node (n1)  {}
     (0.7,0) ++(45:0.8cm and 0.5cm) arc (45:73:0.8cm and 0.5cm) node(n5)  {}
     (0.7,0) ++(73:0.8cm and 0.5cm) arc (73:107:0.8cm and 0.5cm) node(n6)  {}
     (0.7,0) ++(107:0.8cm and 0.5cm) arc (107:135:0.8cm and 0.5cm) node(n2)  {}
        (0.7,0) ++(135:0.8cm and 0.5cm) arc (135:225:0.8cm and 0.5cm) node(n3){}
          (0.7,0) ++(225:0.8cm and 0.5cm) arc (225:315:0.8cm and 0.5cm)  node(n4) {} 
          (0.7,0) ++(315:0.8cm and 0.5cm) arc (315:360:0.8cm and 0.5cm) 
           (n1.base) to (0.76,0.05)
         (n2.base) to  (n4.base) 
         (0.65,-0.05) to (n3.base)
         (n5.base) to [out = 280,in = 260] (n6.base)
         (1.5,0)--(2,0);
                   \filldraw [gray] (n1) circle [radius=1.5pt]; 
          \filldraw [gray]  (-0.1,0) circle [radius=1.5pt];
          \filldraw [gray] (n4) circle [radius=1.5pt]; 
           \filldraw [gray] (n6) circle [radius=1.5pt];
          }
+ \epsilon^{A}_{45} \, 
\tikz[baseline=(vert_cent.base)]{
  \node (vert_cent) {\hspace{-13pt}$\phantom{-}$};
  \draw (-0.6,0)--(-0.1,0)
     (0.7,0) ++(0:0.8cm and 0.5cm) arc (0:35:0.8cm and 0.5cm) node (n1)
            {}
     (0.7,0) ++(35:0.8cm and 0.5cm) arc (35:75:0.8cm and 0.5cm) node(n2)
       {}
        (0.7,0) ++(75:0.8cm and 0.5cm) arc (75:105:0.8cm and 0.5cm) node(n3){}
      (n1.base) to  [out=270,in=240] (n2.base) 
          (0.7,0) ++(105:0.8cm and 0.5cm) arc (105:145:0.8cm and 0.5cm)  node(n4) {} 
          (0.7,0) ++(145:0.8cm and 0.5cm) arc (145:240:0.8cm and 0.5cm) node(n5){}
          (0.7,0) ++(240:0.8cm and 0.5cm) arc (240:300:0.8cm and 0.5cm) node(n6){}
          (0.7,0) ++(300:0.8cm and 0.5cm) arc (300:360:0.8cm and 0.5cm) 
       (n3.base) to  [out= 280,in= 320] (n4.base) 
       (n5.base) to  [out= 80,in= 100] (n6.base) 
         (1.5,0)--(2,0);
                   \filldraw [gray] (n1) circle [radius=1.5pt]; 
          \filldraw [gray]  (-0.1,0) circle [radius=1.5pt];
          \filldraw [gray] (n3) circle [radius=1.5pt]; 
           \filldraw [gray] (n6) circle [radius=1.5pt];
           }
+ \epsilon^{A}_{47} \, 
\tikz[baseline=(vert_cent.base)]{
  \node (vert_cent) {\hspace{-13pt}$\phantom{-}$};
  \draw (-0.6,0)--(-0.1,0)
     (0.7,0) ++(0:0.8cm and 0.5cm) arc (0:45:0.8cm and 0.5cm) node (n1)  {}
            (0.7,0) ++(135:0.8cm and 0.5cm) arc (135:360:0.8cm and 0.5cm) 
                (0.7,0.36) ++(0:0.55cm and 0.35cm) arc (0:50:0.55cm and 0.35cm) node (n2) {}
          (0.7,0.36) ++(50:0.55cm and 0.35cm) arc (50:130:0.55cm and 0.35cm)  node (n3)  {}
           (0.7,0.36) ++(130:0.55cm and 0.35cm) arc (130:230:0.55cm and 0.35cm) node (n4)  {}
            (0.7,0.36) ++(230:0.55cm and 0.35cm) arc (230:310:0.55cm and 0.35cm) node (n5)  {}
             (0.7,0.36) ++(310:0.55cm and 0.35cm) arc (310:360:0.55cm and 0.35cm) node (n6)  {}
             (n2.base) to (0.74,0.4)
             (0.62,0.3) to (n4.base)
             (n3.base) to (n5.base)
         (1.5,0)--(2,0);
                   \filldraw [gray] (n1) circle [radius=1.5pt]; 
         \filldraw [gray]  (-0.1,0) circle [radius=1.5pt];
         \filldraw [gray] (n3) circle [radius=1.5pt]; 
          \filldraw [gray] (n4) circle [radius=1.5pt];
          }\nn
&+\epsilon^{A}_{411} \, 
\tikz[baseline=(vert_cent.base)]{
  \node (vert_cent) {\hspace{-13pt}$\phantom{-}$};
  \draw (-0.6,0)--(-0.1,0)
     (0.7,0) ++(0:0.8cm and 0.5cm) arc (0:25:0.8cm and 0.5cm) node (n1)
            {}
     (0.7,0) ++(25:0.8cm and 0.5cm) arc (25:60:0.8cm and 0.5cm) node(n2)
       {}
        (0.7,0) ++(60:0.8cm and 0.5cm) arc (60:72.5:0.8cm and 0.5cm) node(n3){}
      (n1.base) to  [out=270,in=250] (n2.base) 
        (0.7,0) ++(72.5:0.8cm and 0.5cm) arc (72.5:107.5:0.8cm and 0.5cm) node(n4){}
          (0.7,0) ++(107.5:0.8cm and 0.5cm) arc (107.5:120:0.8cm and 0.5cm)  node(n5) {} 
          (0.7,0) ++(120:0.8cm and 0.5cm) arc (120:155:0.8cm and 0.5cm) node(n6) {}
            (0.7,0) ++(155:0.8cm and 0.5cm) arc (155:360:0.8cm and 0.5cm)
             (n3.base) to  [out= 280,in= 260] (n4.base) 
     (n5.base) to  [out= 320,in= 300] (n6.base) 
         (1.5,0)--(2,0);
                   \filldraw [gray] (n1) circle [radius=1.5pt]; 
          \filldraw [gray]  (-0.1,0) circle [radius=1.5pt];
          \filldraw [gray] (n3) circle [radius=1.5pt]; 
           \filldraw [gray] (n5) circle [radius=1.5pt]; 
          }
+ \epsilon^{A}_{412} \,
          \tikz[baseline=(vert_cent.base)]{
  \node (vert_cent) {\hspace{-13pt}$\phantom{-}$};
  \draw (-0.6,0)--(-0.1,0)
   (0.7,0) ++(0:0.8cm and 0.5cm) arc (0:25:0.8cm and 0.5cm) node (n1) {}
     (0.7,0) ++(25:0.8cm and 0.5cm) arc (25:42.5:0.8cm and 0.5cm) node (n2)
            {}
     (0.7,0) ++(42.5:0.8cm and 0.5cm) arc (42.5:80:0.8cm and 0.5cm) node(n3)
       {}
        (0.7,0) ++(80:0.8cm and 0.5cm) arc (80:100:0.8cm and 0.5cm) node(n4){}
      (n2.base) to  [out=240,in=240] (n3.base) 
          (0.7,0) ++(100:0.8cm and 0.5cm) arc (100:137.5:0.8cm and 0.5cm)  node(n5){}
           (0.7,0) ++(137.5:0.8cm and 0.5cm) arc (137.5:155:0.8cm and 0.5cm)  node(n6) {} 
            (0.7,0) ++(155:0.8cm and 0.5cm) arc (155:360:0.8cm and 0.5cm) 
         (n4.base) to  [out= 290,in= 320] (n5.base) 
         (n1.base) to  [out= 220,in= 320] (n6.base) 
         (1.5,0)--(2,0);
                   \filldraw [gray] (n1) circle [radius=1.5pt]; 
          \filldraw [gray]  (-0.1,0) circle [radius=1.5pt];
          \filldraw [gray] (n3) circle [radius=1.5pt]; 
           \filldraw [gray] (n5) circle [radius=1.5pt];
          }
\label{dg4susy}
\end{align}
with the non-zero values which may be read off using Eq.~\eqref{epsvals} from the $\zeta_5$, $\zeta_3$ coefficients in Eq.~\eqref{c4WZ}:
\begin{align}
\epsilon^{A}_{41}=-10Z_6,\quad \epsilon^{A}_{42} =& \epsilon^{A}_{43}=\tfrac12\epsilon^{A}_{44} = - \tfrac32Z_4 ,\nn
\epsilon^{A}_{45}=\epsilon^{A}_{411} = \tfrac14Z_4, \quad  \epsilon^{A}_{47} =& - \tfrac32Z_4 ,\quad \epsilon^{A}_{412} =-\tfrac{1}{2}Z_4, 
\label{epsmom}
\end{align}
The resulting variations in the five-loop anomalous dimension coefficients may be obtained from Eq.~\eqref{susyvars}. Here $X_{1,41}$ is defined in Eq.~\eqref{nlla1}, and the $\Xhat$ quantities are defined in Eqs.~\eqref{nlla2}, \eqref{Xa}, \eqref{Xb} (the latter two representing the effect of Eq.~\eqref{dg3susy} at this next order). We take $c_1=\frac12$ as before, $\epsilon_1=0$ since this is the $\MOp$ scheme, and we use the values in Eqs.~\eqref{zeroeps}, \eqref{epsmom} for $\epsilon^A_{41}-\epsilon^A_{413}$. We then obtain the values for $X^{A}_{1,41}$, $\Xhat^{A}_{1,42}-\Xhat^{A}_{1,413}$, $\Xhat^A_{2,31}-\Xhat^A_{2,34}$ given in Eq.~\eqref{MOMp}, and it is easily verified that on substitution in Eq.~\eqref{susyvars} we obtain variations in the five-loop coefficients which precisely remove the even $\zeta$s from the five-loop $\msbar$ results given in Ref.~\cite{gracey} (and have no other effect). Therefore the general Wess-Zumino model is free of even $\zeta$s up to five loops in the $\MOp$ scheme.


The above is certainly the easiest way to obtain the numbers required to construct the transformation to $\MOp$. However, as is implicit in the first expression on the right-hand side of Eq.~\eqref{delg4def}, we may also obtain the $\epsilon$ coefficients from the finite $\zeta_4$, $\zeta_6$ contributions evaluated in the $\MOp$ scheme at four loops, and this perhaps is more in line with the physical meaning of $\MOp$. These contributions are, in fact, mostly the same as the $\msbar$ contributions in Eq.~\eqref{zms4}, except in the case of $\epsilon_{47}$ which must be read off from Eq.~\eqref{zmom4}. As previously, these finite contributions need to be multiplied by a factor $\left(\frac12\right)$ to obtain the $\epsilon$ values. It is easy to check that we reproduce the values in Eq.~\eqref{epsmom}. Finally, as we see in Eq.~\eqref{delg4def}, these finite contributions computed within the $\MOp$ scheme must be written as functions of $y^{\msbar}$; 
in other words, the vertices of the diagrams in Eq.~\eqref{dg4susy} are understood as representing the $\msbar$ couplings. We also remark that the $\beta$-function contributions in the second expression on the right-hand side of Eq.~\eqref{delg4def} may, of course, also be read off from the $\msbar$ results in Eq.~\eqref{zms4}, this time from the simple poles in odd $\zeta$s; this, again of course, is a consequence of the relation between the simple poles for odd $\zeta$s and the finite parts for even $\zeta$s, as expressed in Eq.~\eqref{Grels:main}.

We now turn our attention again to MOM. As at four loops, the transformation to MOM involves introducing $\epsilon^B_1$ and now indeed additional higher-order $\epsilon$ quantities. But we must now also take account of the fact that (as explained in Appendix \ref{mstrans}) in making the transition from $\msbar$ to $\MOp$, the quantities like $\epsilon^A_{42}$ are derived from finite parts of two-point functions like $f^{\zeta_4\MOp}_{4,0}$ evaluated in $\MOp$, while in making the transition from $\msbar$ to $\MO$, the $\epsilon$ quantities like $\epsilon^B_{42}$ are derived from $f^{\zeta_4\msbar}_{4,0}$, i.e. evaluated in $\msbar$. This only starts to make a difference at this order, and in fact even here it is only $\epsilon_{47}$ which is affected, as we see comparing Eqs.~\eqref{zms4}, \eqref{zmom4}. So we can certainly write
\be
\epsilon_{41}^{\zeta_6B}=\epsilon_{41}^{\zeta_6A},\quad \epsilon_{4J}^{\zeta_4B}=\epsilon_{4J}^{\zeta_4A},\quad J\ne7.
\label{epsdiff}
\ee
For $X_{1,41}$ and $\Xhat_{1,4J}$ where $J=5,6,8-13$, the same argument as at four loops may then be applied. 
 In $X_{1,41}$, $\Xhat_{1,4J}$, we see from Eqs.~\eqref{nlla1}, \eqref{nlla2} that the new non-zero $\epsilon_1$ multiplies $c_{41}$, $\chat_{4J}$ respectively, which have no dependence on even $\zeta$s for $J=5,6,8-13$, as may be seen in Eqs.~\eqref{nllb}, \eqref{c4WZ}. Likewise in $\Xhat_{2,3K}$ for $K=1\ldots4$, defined in Eqs.~\eqref{Xa}, \eqref{Xb}, the new non-zero $\ehat_{21}$  multiplies $c_{31}$ or $\chat_{3K}$, $K=2,3,4$, which have no $\zeta_4$-dependence. Recalling also Eq.~\eqref{epsdiff}, we therefore have
\begin{align}
X^{\zeta_6B}_{1,41}=X^{\zeta_6A}_{1,41},\quad \Xhat^{\zeta_4B}_{1,4J}=&\Xhat^{\zeta_4A}_{1,4J}, \quad J=5,6,8-13, \nn
\Xhat^{\zeta_4B}_{2,3K}=&\Xhat^{\zeta_4A}_{2,3K},\quad K=1\ldots4.
\end{align}
 However for the remaining $\Xhat_{1,42}$,
$\Xhat_{1,43}$, $\Xhat_{1,44}$, $\Xhat_{1,47}$, defined as in Eq.~\eqref{nlla2}, more care is required. First of all, in Eq.~\eqref{nllb}, $c^{\msbar}_{42}$, and similarly $c^{\msbar}_{43}$, $c^{\msbar}_{44}$, $c^{\msbar}_{47}$ do have $\zeta_4$ dependence.  Furthermore, for related reasons, we also have for the first time the possibility of $\zeta_4$ dependent terms beyond leading order in the same $\Xhat$ quantities, in both $\ehat_{42}$ etc in Eq.~\eqref{nlla2}, and in $\delta'c_{42}$ etc in Eq.~\eqref{nllb}. To be precise, these second-order $\zeta_4$ terms arise from the presence of $\epsilon_{31}$ (as given in Eq.~\eqref{e3defb}).
The results for these $\Xhat$ quantities may be written using Eqs.~\eqref{nlla2}-\eqref{ctil} as
\begin{align}
\Xhat_{1,42}=&c_1\epsilon_{42}-\epsilon_1c_{42}-4c_1\epsilon_1\epsilon_{31},\nn
\Xhat_{1,43}=&c_1\epsilon_{43}-\epsilon_1c_{43}-4c_1\epsilon_1\epsilon_{31},\nn
\Xhat_{1,44}=&c_1\epsilon_{44}-\epsilon_1c_{44}-8c_1\epsilon_1\epsilon_{31},\nn
\Xhat_{1,47}=&c_1\epsilon_{47}-\epsilon_1c_{47}.
\label{x1}
\end{align}
 As already explained, we take $\epsilon^B_{31}$, $\epsilon^B_{42}$ etc to be given by the finite parts of the $\msbar$ renormalisation constants in Eqs.~\eqref{zms12},  \eqref{zms3}, \eqref{zms4}, multiplied by the usual factor of $(\frac12)$. The full values are given in Eqs.~\eqref{e3defb}, \eqref{eps4vals}; but here we are only interested in the $\zeta_4$ contributions, namely
\begin{align}
\epsilon_{31}^{\zeta_4B}=-\frac38,\quad \epsilon_{47}^{\zeta_4B}=&\frac{33}{32},\nn
\epsilon^{\zeta_4B}_{42}=\epsilon^{\zeta_4B}_{43}=\frac12\epsilon^{\zeta_4B}_{44}=\epsilon^{\zeta_4A}_{42}=&\frac{9}{32};
\label{x2}
\end{align}
together with $\epsilon^B_1=-\frac12$ as before.
We then finally find, combining Eqs.~\eqref{x1}, \eqref{x2} and \eqref{c4WZ} 
\begin{align}
\Xhat_{1,42}^{\zeta_4B}=\Xhat_{1,43}^{\zeta_4B}=\frac12\Xhat_{1,44}^{\zeta_4B}&=\frac{9}{64},\nn
\Xhat_{1,47}^{\zeta_4B}=&\frac{9}{64},
\label{MOMXa}
\end{align}
which are indeed the same $\zeta_4$ contributions as in the $\MOp$ case, as may be seen in Eq.~\eqref{MOMp}. As we have already argued, the even-$\zeta$ contributions to the remaining $\Xhat$ are also unchanged; so that the even $\zeta$s must cancel in $\MO$ as well as in $\MOp$. 
We can in fact go further and show that the full set of variations given in Eq.~\eqref{susyvars} can be applied to the five-loop $\msbar$ results given in Ref.~\cite{gracey} to yield the MOM results also given there. This is somewhat off the line of our main purpose here, but for completeness is explained in some detail in Appendix \ref{susyvar}; partly as a confirmation of our results and partly because we believe our methods give some insight into the processes involved.

\section{Towards a general proof}\label{genproof}
So far we have been using results derived using $\msbar$ for the single coupling case and for the multi-coupling supersymmetric Wess-Zumino model  in order to try to construct a coupling redefinition to a ``$\pi$-free'' scheme with no even $\zeta$s. However in order to prove that there exists a scheme which is free of even $\zeta$s to all orders and for a general multi-coupling AD theory, it seems like a good idea to define the scheme {\it ab initio} rather than starting from $\msbar$. In this section we shall report on progress in this direction. We do not yet have an all-orders proof, since working in non-minimal schemes has its own complications; however there are indications of how a systematic approach may be pursued to higher orders than we have reached so far. We shall show that the $\MOp$ $\beta$-function for a general multi-coupling theory is free of $\zeta_4$ contributions up to five loops, and free of $\zeta_6$ contributions up to six loops. In both cases this is one loop beyond where the even-$\zeta$ first appears, and in the $\zeta_6$ case presents the extra complication of the higher-order pole term in Eq.~\eqref{Grels:b}. It seems likely that extensions to higher loops and higher $\zeta$s will only introduce further algebraic complexity rather than issues of principle. We also address the issue of whether $\MO$ is equally free of even $\zeta$s, but here due to the additional algebraic complexity we have confined ourselves to the $\zeta_4$ case, which we have again pursued as far as five loops. 

We consider first the $\MOp$ scheme where we subtract the finite parts depending on even $\zeta$s. At lowest order, we may write
\begin{align}
g_B=&\mu^{\epsilon}\Bigl(g^{\MOp}+F_{3,0}^{\zeta_4\mop}\zeta_4+F_{4,0}^{\zeta_4\MOp}\zeta_4\nn
&+\frac{1}{\epsilon}\left\{F^{\mop}_{1,1}+F^{\zeta_3\mop}_{3,1}\zeta_3+F^{\zeta_4\MOp}_{4,1}\zeta_4\right\}\nn
&+\frac{1}{\epsilon^2}\left\{F^{\mop}_{2,2}+F^{\zeta_3\MOp}_{4,2}\zeta_3\right\}\Bigr)+\ldots
\end{align}
Our convention in this section is that scheme-dependent RG functions like $F_{4,0}^{\zeta_4\MOp}$ evaluated in the $\MOp$ scheme are functions of the $\MOp$ coupling $g^{\MOp}$. We also omit reference to quantities which do not concern us, in particular non-$\zeta$ dependent quantities such as $F^{\MOp}_{3,1}$. Similarly to Eq.~\eqref{betdefa}, we define the $\beta$-function to be 
\be
\bhat=\mu\frac{d}{d\mu} g^{\MOp}=-\epsilon g^{\MOp}+\beta^{\MOp}+\beta^{\zeta_3\MOp}\zeta_3+\epsilon\tbeta^{\zeta_4\MOp}\zeta_4+\ldots
\ee
and using Eq.~\eqref{dgzero} we find that
\be
\beta^{\MOp}_1=F^{\mop}_{1,1},\quad \tbeta^{\zeta_n\MOp}_L=LF^{\zeta_n\MOp}_{L,0}, \quad \beta^{\zeta_{n}\MOp}_L=LF^{\zeta_{n}\MOp}_{L,1}.
\label{genbets}
\ee
At four loops we have 
\begin{align}
\beta^{\zeta_4\MOp}_4=&4F^{\zeta_4\MOp}_{4,1}-\beta_1\cdot F^{\zeta_4\mop}_{3,0}-\tbeta^{\zeta_4\MOp}_3\cdot F^{\mop}_{1,1}\nn
=&4F^{\zeta_4\MOp}_{4,1}-F^{\mop}_{1,1}\cdot F^{\zeta_4\mop}_{3,0}-3F^{\zeta_4\mop}_{3,0}\cdot F^{\mop}_{1,1},
\label{genbet4}
\end{align}
and similarly
\be
F^{\zeta_3\MOp}_{4,2}=\frac14\left[3F^{\zeta_3\mop}_{3,1}\cdot F^{\mop}_{1,1}+F^{\mop}_{1,1}\cdot F^{\zeta_3\mop}_{3,1}\right].
\label{F3rel}
\ee
Now let $G^{\zeta_3\MOp}_{4,2}$ and $G^{\zeta_4\MOp}_{4,1}$ denote the contributions to $F^{\zeta_3\MOp}_{4,2}$ and $F^{\zeta_4\MOp}_{4,1}$ obtained by omitting $\zeta$-dependent counterterm contributions. We have (recalling Eq.~\eqref{Lmin} which, in view of the associated remarks about scheme-independence, applies in the $\MOp$ scheme as well)
\begin{subequations}\label{ftilrel:main}
\begin{align}
G^{\zeta_3\MOp}_{4,2}=&F^{\zeta_3\MOp}_{4,2}-F^{\zeta_3\mop}_{3,1}\cdot F^{\mop}_{1,1},\label{ftilrel:a}\\
G^{\zeta_4\MOp}_{4,1}=&F^{\zeta_4\MOp}_{4,1}-F^{\zeta_4\mop}_{3,0}\cdot F^{\mop}_{1,1},\label{ftilrel:b}\\
G^{\zeta_3\mop}_{3,1}=&F^{\zeta_3\mop}_{3,1},\label{ftilrel:c}\\
G^{\zeta_4\mop}_{3,0}=&F^{\zeta_4\mop}_{3,0}.\label{ftilrel:d}
\end{align}
\end{subequations}
The motivation for doing this is that the quantities such as $G^{\zeta_3\MOp}_{4,2}$ obey the relations such as Eq.~\eqref{Grels:a}, whereas the corresponding  $F^{\zeta_3\MOp}_{4,2}$ do not.
We then find, using Eqs.~\eqref{genbet4}, \eqref{F3rel}
\begin{align}
\beta^{\zeta_4\MOp}_4=&4G^{\zeta_4\MOp}_{4,1}-[F^{\mop}_{1,1},G^{\zeta_4\mop}_{3,0}],\nn
0=&4G^{\zeta_3\MOp}_{4,2}-[F^{\mop}_{1,1},G^{\zeta_3\mop}_{3,1}].
\end{align}
But then using Eq.~\eqref{Grels:a} for $m=0$ and also for $m=1$,  we immediately find
\be
\beta^{\zeta_4\MOp}_4=0;
\label{betz44}
\ee
in other words we have shown the absence of even $\zeta$s in the $\MOp$ scheme at four loops.
At five-loop order we have
\begin{align}
\beta^{\zeta_4\MOp}_5=&5F^{\zeta_4\MOp}_{5,1}-4F^{\zeta_4\MOp}_{4,0}\cdot F^{\mop}_{1,1}-3F^{\zeta_4\mop}_{3,0}\cdot F^{\MOp}_{2,1}-F^{\mop}_{1,1}\cdot F^{\zeta_4\MOp}_{4,0}\nn
&-2F^{\MOp}_{2,1}\cdot F^{\zeta_4\mop}_{3,0},\nn
0=&5F^{\zeta_3\MOp}_{5,2}-4F^{\zeta_3\MOp}_{4,1}\cdot F^{\mop}_{1,1}-3F^{\zeta_3\mop}_{3,1}\cdot F^{\MOp}_{2,1}-F^{\mop}_{1,1}\cdot F^{\zeta_3\MOp}_{4,1}\nn
&-2F^{\MOp}_{2,1}\cdot F^{\zeta_3\mop}_{3,1}
\label{bet5def}
\end{align}
and
\begin{subequations}\label{grela:main}
\begin{align}
G^{\zeta_4\MOp}_{5,1}=&F^{\zeta_4\MOp}_{5,1}-F^{\zeta_4\MOp}_{4,0}\cdot F^{\mop}_{1,1}-F^{\zeta_4\mop}_{3,0}\cdot F^{\MOp}_{2,1}-G^{\zeta_4\MOp}_{4,1}\cdot F^{\mop}_{1,0},\label{grela:a}\\
G^{\zeta_3\MOp}_{5,2}=&F^{\zeta_3\MOp}_{5,2}-F^{\zeta_3\MOp}_{4,1}\cdot F^{\mop}_{1,1}-F^{\zeta_3\mop}_{3,1}\cdot F^{\MOp}_{2,1}-G^{\zeta_3\MOp}_{4,2}\cdot F^{\mop}_{1,0},\label{grela:b}\\
G^{\zeta_3\MOp}_{4,1}=&F^{\zeta_3\MOp}_{4,1}-F^{\zeta_3\mop}_{3,1}\cdot F^{\mop}_{1,0},\label{grela:c}\\
G^{\zeta_4\MOp}_{4,0}=&F^{\zeta_4\MOp}_{4,0}-F^{\zeta_4\mop}_{3,0}\cdot F^{\mop}_{1,0},\label{grela:d}
\end{align}
\end{subequations}
where $G^{\zeta_4\MOp}_{4,1}$, $G^{\zeta_3\MOp}_{4,2}$ are defined in Eqs.~\eqref{ftilrel:a}, \eqref{ftilrel:b}. The presence of these terms (instead of $F^{\zeta_4\MOp}_{4,1}$, $F^{\zeta_3\MOp}_{4,2}$) requires some explanation. Recall that for instance, $G^{\zeta_4\MOp}_{5,1}$ is defined by omitting $\zeta_4$-dependent counterterms from $F^{\zeta_4\MOp}_{5,1}$, and so we need to check that this is indeed the effect of Eq.~\eqref{grela:a}. $F^{\MOp}_{2,1}$ in Eqs.~\eqref{grela:a}, \eqref{grela:b} is derived from the simple pole in $\Gamma_{21}-Z^{\MOp}_{11}\Gamma_{11}$; the second term of course being the counterterm diagram for $\Gamma_{21}$. When we insert this into Eq.~\eqref{grela:a}, bearing in mind the definition of $G^{\zeta_4\MOp}_{4,1}$ in Eq.~\eqref{ftilrel:a}, we find a cancellation of the terms involving $(F^{\zeta_4\mop}_{3,0}\cdot F^{\mop}_{1,1})\cdot F^{\mop}_{1,0}$, between the third and fourth terms of Eq.~\eqref{grela:a}. We are left with 
\begin{align}
G^{\zeta_4\MOp}_{5,1}=&F^{\zeta_4\MOp}_{5,1}-F^{\zeta_4\MOp}_{4,0}\cdot F^{\mop}_{1,1}-F^{\zeta_4\MOp}_{4,1}\cdot F^{\mop}_{1,0}-F^{\zeta_4\mop}_{3,0}\cdot F^{\MOp}_{2,1}\nn
&+F^{\zeta_4\mop}_{3,0}\cdot(|F^{\mop}_{1,0}|\cdot F^{\mop}_{1,1}),
\label{Gexam}
\end{align}
where the vertical lines introduced in Section~\ref{basic} again mean that $F^{\mop}_{1,0}$ is not acted upon by the derivative. 
The five-loop graphs with $\zeta_4$-dependent counterterms are $\Gamma_{540}$, $\Gamma_{541}$, $\Gamma_{542}$, $\Gamma_{531}$, $\Gamma_{545}$, $\Gamma_{546}$ and $\Gamma_{557}$. Here for instance $\Gamma_{540}$ is the graph corresponding to $c_{540}$ in Ref.~\cite{gracey}, as defined in Appendix \ref{feynres}. The first three graphs have no three-loop $\zeta_4$-dependent counterterms, and their 
four-loop $\zeta_4$-dependent counterterms (from subdiagrams $\Gamma_{42}$, $\Gamma_{43}$ and $\Gamma_{44}$) are removed by the second and third terms in Eq.~\eqref{Gexam}. The following three graphs all have both a three-loop $\zeta_4$-dependent counterterm, and also a four-loop $\zeta_4$ dependent subdivergence supplied by the combination of the three-loop subdiagram $\Gamma_{31}$ and the one-loop subdiagram $\Gamma_{11}$. The first  $\zeta_4$-dependent counterterm is subtracted by the fourth term in Eq.~\eqref{Gexam}, and the second by the fifth term. Finally, $\Gamma_{557}$ has a four-loop $\zeta_4$-dependent counterterm corresponding to the subdiagram $\Gamma_{47}$, and a three-loop $\zeta_4$-dependent counterterms corresponding as usual to $\Gamma_{31}$. These are removed by the second and third, and fourth, terms respectively in Eq.~\eqref{Gexam}. With a little thought one can convince oneself that the subtractions in Eq.~\eqref{grela:a} have no other effect than removing these $\zeta_4$-dependent counterterms.

The somewhat asymmetric form of Eq.~\eqref{grela:a}, \eqref{grela:b} is clearly due to the fact that in $\MOp$, $\Gamma_{21}$ has a simple pole subtraction but not a finite subtraction. We shall see later that in the $\MO$ case at this order,  we find similar but more uniform expressions with both  $F^{\zeta_4\MOp}_{4,0}$, $F^{\zeta_3\MOp}_{4,1}$ replaced by $G^{\zeta_4\MOp}_{4,0}$, $G^{\zeta_3\MOp}_{4,1}$, and $F^{\MOp}_{2,1}$ (the $\msbar$ quantity) replaced by $F^{\MO}_{2,1}$.
We then find using Eqs.~\eqref{ftilrel:main}, \eqref{bet5def} and \eqref{grela:main}
\begin{align}
\beta^{\zeta_4\MOp}_5=&5\left(G^{\zeta_4\MOp}_{5,1}-\frac34G^{\zeta_3\MOp}_{5,2}\right)+\left[G^{\zeta_4\MOp}_{4,0}-\frac34G^{\zeta_3\MOp}_{4,1},F^{\mop}_{1,1}\right]\nn
&
+\left[\left(G^{\zeta_4\MOp}_{3,0}-\frac34G^{\zeta_3\mop}_{3,1}\right)\cdot F^{\mop}_{1,0},F^{\mop}_{1,1}\right]\nn
&+2\left[G^{\zeta_4\mop}_{3,0}-\frac34G^{\zeta_3\mop}_{3,1},F^{\MOp}_{2,1}\right]\nn
&+5\left(G^{\zeta_4\MOp}_{4,1}-\frac34G^{\zeta_3\MOp}_{4,2}\right)\cdot F^{\mop}_{1,0}.
\end{align}
We immediately find from Eq.~\eqref{Grels:a}
\be
\beta^{\zeta_4\MOp}_5=0,
\ee
so that there are no $\zeta_4$ at five loops in the $\MOp$ scheme. We could continue to six and higher loops, but we already see a developing pattern in which $\beta^{\zeta_4\MOp}$ may be expressed in terms of pairs of $G$ functions which vanish according to Eq.~\eqref{Grels:a}, and there appears to be no reason why this should not continue to higher loops. However at this current loop order $\zeta_6$ contributions are also possible and certainly present in the $\msbar$ scheme; and furthermore Eq.~\eqref{Grels:b} contains an extra term (compared to Eq.~\eqref{Grels:a}) whose impact on the calculation needs to be examined.
So we now turn to the case of $\zeta_6$. In fact, at five loops we can apply a similar argument to that for $\zeta_4$.  We have, using Eq.~\eqref{genbets}, 
\begin{align}
\beta^{\zeta_6\MOp}_5=&5F^{\zeta_6\MOp}_{5,1}-4F_{4,0}^{\zeta_6\MOp}\cdot F^{\mop}_{1,1}-F^{\mop}_{1,1}\cdot F^{\zeta_6\MOp}_{4,0}\nn
5F^{\zeta_5\MOp}_{5,2}=&4F^{\zeta_5\MOp}_{4,1}\cdot F^{\mop}_{1,1}+F^{\mop}_{1,1}\cdot F^{\zeta_5\MOp}_{4,1}.
\label{bet6rels}
\end{align}
We also have
\begin{align}
G^{\zeta_6\MOp}_{4,0}=&F^{\zeta_6\MOp}_{4,0},\nn
G^{\zeta_5\MOp}_{4,1}=&F^{\zeta_5\MOp}_{4,1},\nn
G^{\zeta_6\MOp}_{5,1}=&F^{\zeta_6\MOp}_{5,1}-F^{\zeta_6\MOp}_{4,0}\cdot F^{\mop}_{1,1},\nn
G^{\zeta_5\MOp}_{5,2}=&F^{\zeta_5\MOp}_{5,2}-F^{\zeta_5\MOp}_{4,1}\cdot F^{\mop}_{1,1}.
\label{G6rels}
\end{align}
Using Eqs.~\eqref{bet6rels}, \eqref{G6rels}, and \eqref{Grels:b} for $m=1$ and then for $m=0$, and since $G^{\zeta_4}_{4,2}=G^{\zeta_4}_{5,3}=0$, we find
\be
\beta^{\zeta_6\MOp}_5=5\left(G^{\zeta_6\MOp}_{5,1}-\frac54G^{\zeta_5\MOp}_{5,2}\right)+[G^{\zeta_6\MOp}_{4,0}-\frac54G^{\zeta_5\MOp}_{4,1}, F^{\mop}_{1.1}]=0.
\ee
We have now shown the absence of even $\zeta$s through five loops. However, at six loops the extra term in Eq.~\eqref{Grels:b} starts to make its presence felt in the $\zeta_6$ calculations. Again using Eq.~\eqref{genbets}, together with Eq.~\eqref{betz44}, we find
\begin{align}
\beta^{\zeta_6\MOp}_6=&6F^{\zeta_6\MOp}_{6,1}-5F^{\zeta_6\MOp}_{5,0}\cdot F^{\mop}_{1,1}-4F^{\zeta_6\MOp}_{4,0}\cdot F^{\MOp}_{2,1}-F^{\mop}_{1,1}\cdot F^{\zeta_6\MOp}_{5,0}\nn
&-2F^{\MOp}_{2,1}\cdot F^{\zeta_6\MOp}_{4,0},\nn
0=&6F^{\zeta_5\MOp}_{6,2}-5F^{\zeta_5\MOp}_{5,1}\cdot F^{\mop}_{1,1}-4F^{\zeta_5\MOp}_{4,1}\cdot F^{\MOp}_{2,1}-F^{\mop}_{1,1}\cdot F^{\zeta_5\MOp}_{5,1}\nn
&-2F^{\MOp}_{2,1}\cdot F^{\zeta_5\MOp}_{4,1}\nn
0=&6F^{\zeta_4\MOp}_{6,3}\phantom{-5F^{\zeta_5\MOp}_{5,2}\cdot F^{\mop}_{1,1}}-3F^{\zeta_4\mop}_{3,0}\cdot F_{3,3}-F^{\mop}_{1,1}\cdot F^{\zeta_4\MOp}_{5,2}.
\end{align}
We also have
\begin{subequations}\label{zsixa:main}
\begin{align}
G^{\zeta_6\MOp}_{6,1}=&F^{\zeta_6\MOp}_{6,1}-F^{\zeta_6\MOp}_{5,0}\cdot F^{\mop}_{1,1}-F^{\zeta_6\MOp}_{4,0}\cdot F^{\MOp}_{2,1}-G^{\zeta_6\MOp}_{5,1}\cdot F^{\mop}_{1,0},\label{zsixa:a}\\
G^{\zeta_5\MOp}_{6,2}=&F^{\zeta_5\MOp}_{6,2}-F^{\zeta_5\MOp}_{5,1}\cdot F^{\mop}_{1,1}-F^{\zeta_5\MOp}_{4,1}\cdot F^{\MOp}_{2,1}-G^{\zeta_5\MOp}_{5,2}\cdot F^{\mop}_{1,0},\label{zsixa:b}\\
G^{\zeta_4\MOp}_{6,3}=&F^{\zeta_4\MOp}_{6,3}-G^{\zeta_4\MOp}_{5,2}\cdot F^{\mop}_{1,1}-G^{\zeta_4\MOp}_{4,1}\cdot F^{\mop}_{2,2}-F^{\zeta_4\mop}_{3,0}\cdot F^{\mop}_{3,3},\label{zsixa:c}
\end{align}
\end{subequations}
together with
\begin{subequations}\label{zsixb:main}
\begin{align}
G^{\zeta_6\MOp}_{5,0}=&F^{\zeta_6\MOp}_{5,0}-F^{\zeta_6\MOp}_{4,0}\cdot F^{\mop}_{1,0},\label{zsixb:a}\\
G^{\zeta_5\MOp}_{5,1}=&F^{\zeta_5\MOp}_{5,1}-F^{\zeta_5\MOp}_{4,1}\cdot F^{\mop}_{1,0},\label{zsixb:b}\\
G^{\zeta_4\MOp}_{5,2}=&F^{\zeta_4\MOp}_{5,2}-G^{\zeta_4\MOp}_{4,1}\cdot F^{\mop}_{1,1}-F^{\zeta_4\mop}_{3,0}\cdot F^{\mop}_{2,2}.
\label{zsixb:c}
\end{align}
\end{subequations}
As before, the appearance of $G^{\zeta_6\MOp}_{5,1}$, $G^{\zeta_5\MOp}_{5,2}$ rather than $F^{\zeta_6\MOp}_{5,1}$, $F^{\zeta_5\MOp}_{5,2}$ in Eqs.~\eqref{zsixa:a}, \eqref{zsixa:b} reflects the lower-order counterterm structure in $F^{\MOp}_{2,1}$ and $F^{\MOp}_{2,2}$. We have not checked in detail, but if these were evaluated in $\MO$ rather than $\msbar$, then this replacement of $F$ by $G$ in the first of each pair on the RHS might be universal, as we shall shortly see at five loops for the $\zeta_4$ case. The appearance of $G^{\zeta_4\MOp}_{5,2}$ and $G^{\zeta_4\MOp}_{4,1}$ in Eq.~\eqref{zsixa:c} is not easy to justify {\it a priori} but one can check that it does indeed remove the $\zeta_4$ dependent counterterms and is expressed in a convenient fashion for combination with the other equations. Combining Eqs.~\eqref{zsixa:main}, \eqref{zsixb:main}, we find
\begin{align}
\beta^{\zeta_6\MOp}_6=&6G^{\zeta_6\MOp}_{6,1}+[G^{\zeta_6\MOp}_{5,0},F^{\mop}_{1,1}]+[G^{\zeta_6\MOp}_{4,0}\cdot F^{\mop}_{1,0},F^{\mop}_{1,1}]\nn
&+2[G^{\zeta_6\MOp}_{4,0},F^{\MOp}_{2,1}]+6G^{\zeta_5\MOp}_{5,1}\cdot F^{\mop}_{1,0},\nn
0=&6G^{\zeta_5\MOp}_{6,2}+[G^{\zeta_5\MOp}_{5,1},F^{\mop}_{1,1}]+[G^{\zeta_5\MOp}_{4,1}\cdot F^{\mop}_{1,0},F^{\mop}_{1,1}]\nn
&+2[G^{\zeta_5\MOp}_{4,1},F^{\MOp}_{2,1}]+6G^{\zeta_5\MOp}_{5,2}\cdot F^{\mop}_{1,0},\nn
0=&6G^{\zeta_4\MOp}_{6,3}+6F^{\zeta_4\MOp}_{5,2}\cdot F^{\mop}_{1,1}-F^{\mop}_{1,1}\cdot F^{\zeta_4\MOp}_{5,2}+F^{\zeta_4\MOp}_{4,1}\cdot F^{\MOp}_{2,2}\nn
&-6(F^{\zeta_4\MOp}_{4,1}\cdot F^{\mop}_{1,1})\cdot F^{\mop}_{1,1}
+6F^{\zeta_4\mop}_{3,0}\cdot F^{\MOp}_{3,3}\nn
&-3F^{\mop}_{1,1}\cdot [(|F^{\zeta_4\mop}_{3,0}|\cdot F^{\mop}_{1,1})\cdot F^{\mop}_{1,1}].
\end{align}

 Once again using Eq.~\eqref{Grels:b} for $m=1$ and then for $m=0$, together with
\be
5F^{\zeta_4\MOp}_{5,2}=F^{\mop}_{1,1}\cdot F^{\zeta_4\MOp}_{4,1}+3F^{\zeta_4\mop}_{3,1}\cdot F^{\MOp}_{2,2}
\ee
we find after considerable algebra that
\be
\beta^{\zeta_6\MOp}_6=0.
\ee
Unfortunately it does not seem very clear how to construct a general proof of the absence of $\zeta_6$ beyond this six-loop order. 

Comparing the full set of even $\zeta$s considered in Eqs.~\eqref{Grels:main}, {\eqref{Grelsa:main}: $\zeta_4\zeta_3$
and $(\zeta_3)^2\zeta_4$ appear in a similar way to $\zeta_4$ and in these cases we can readily imagine sketching a proof of the absence of these to all orders. $\zeta_8$, $\zeta_{10}$ and $\zeta_6\zeta_3$ appear in a similar way to  $\zeta_6$, and we anticipate that we can similarly prove the absence of these even $\zeta$s up to one loop beyond where they first appear; however a general proof will require further work.
$(\zeta_4)^2$ and $\zeta_4\zeta_5$ will probably require separate consideration, the former due to being the square of an even $\zeta$ and the latter due to mixing with $\zbar_{5,3}$.

In Sections~\ref{single} and \ref{susysec}, we have presented evidence that the $\MO$ scheme is also free of even $\zeta$s, at least up to five loops. Here we shall consider this from the current point of view, working now entirely within the $\MO$ scheme. This is essentially defined by subtracting all the finite parts of graphs as well as their pole terms. This can be accommodated by writing
\begin{align}
g_B=&\mu^{\epsilon}\Biggl(g+F^{\mo}_{1,0}+F_{3,0}^{\zeta_3\mo}\zeta_3+F_{3,0}^{\zeta_4\mo}\zeta_4\nn
&+\frac{1}{\epsilon}\left\{F^{\mo}_{1,1}+F^{\zeta_3\mo}_{3,1}\zeta_3+F^{\zeta_4\MO}_{4,1}\zeta_4\right\}+\frac{1}{\epsilon^2}\left\{F^{\MO}_{2,2}+F^{\zeta_3\MO}_{4,2}\zeta_3\right\}\Biggr)+\ldots
\end{align}
where again we only include the extra terms which will be relevant for our purposes, namely $F^{\mo}_{1,0}$ and $F_{3,0}^{\zeta_3\mo}$.

The four-loop calculation is exactly the same as the $\MOp$ case. However, at 5 loops we now find
\begin{subequations}\label{fivemom:main}
\begin{align}
\beta_5^{\zeta_4\MO}=&5F^{\zeta_4\MO}_{5,1}-\tbeta^{\zeta_4\MO}_4\cdot F^{\mo}_{1,1}-\tbeta^{\zeta_4\MO}_3\cdot F^{\MO}_{2,1}-\beta^{\mo}_1\cdot F^{\zeta_4\MO}_{4,0}\nn
&-\beta^{\MO}_2\cdot F^{\zeta_4\mo}_{3,0}-\tbeta^{\MO}_1\cdot F^{\zeta_4\MO}_{4,1},\label{fivemom:a}\\
5F^{\zeta_3\MO}_{5,2}=&\beta^{\zeta_3\MO}_4\cdot F^{\mo}_{1,1}+\beta^{\zeta_3\MO}_3\cdot F^{\MO}_{2,1}+\beta^{\mo}_1\cdot F^{\zeta_3\MO}_{4,1}\nn
&+\beta^{\MO}_2\cdot F^{\zeta_3\mo}_{3,1}+\tbeta^{\MO}_1\cdot F^{\zeta_3\MO}_{4,2}+\tbeta^{\zeta_3\MO}_3\cdot F^{\MO}_{2,2},\label{fivemom:b}\\
G^{\zeta_4\MO}_{5,1}=&F^{\zeta_4\MO}_{5,1}-G^{\zeta_4\MO}_{4,0}\cdot F^{\mo}_{1,1}-F^{\zeta_4\mo}_{3,0}\cdot F^{\MO}_{2,1}-G^{\zeta_4\MO}_{4,1}\cdot F^{\mo}_{1,0},\label{fivemom:c}\\
G^{\zeta_3\MO}_{5,2}=&F^{\zeta_3\MO}_{5,2}-G^{\zeta_3\MO}_{4,1}\cdot F^{\mo}_{1,1}-F^{\zeta_3\mo}_{3,1}\cdot F^{\MO}_{2,1}\nn
&-F^{\zeta_3\mo}_{3,0}\cdot F^{\MO}_{2,2}
-G^{\zeta_3\MO}_{4,2}\cdot F^{\mo}_{1,0}.\label{fivemom:d}
\end{align}
\end{subequations}
Here, $G^{\zeta_3\MO}_{4,2}$, $G^{\zeta_4\MO}_{4,1}$ and $G^{\zeta_4\MO}_{4,0}$ are defined similarly to Eqs.~\eqref{ftilrel:a}, \eqref{ftilrel:b}, \eqref{grela:d}, respectively, but instead of Eq.~\eqref{grela:c} we now have
\be
G^{\zeta_3\MO}_{4,1}=F^{\zeta_3\MO}_{4,1}-F^{\zeta_3\mo}_{3,1}\cdot F^{\mo}_{1,0}-F^{\zeta_3\mo}_{3,0}\cdot F^{\mo}_{1,1}.
\label{g41}
\ee
Note that as we speculated earlier, in Eq.~\eqref{fivemom:c} we now have two $G$ terms, $G^{\zeta_4\MO}_{4,1}$ $G^{\zeta_4\MO}_{4,0}$, replacing $F^{\zeta_4\MO}_{4,1}$ and $F^{\zeta_4\MO}_{4,0}$, and this appears related to the fact that (in contrast to $\msbar$ or $\MOp$) $F^{\MO}_{2,1}$ has both an $F^{\mo}_{1,0}$ and a $F^{\mo}_{1,1}$ counterterm respectively; and similarly for Eq.~\eqref{fivemom:d}. 
Combining Eqs.~\eqref{fivemom:main} now leads to
\begin{align}
\beta^{\zeta_4\MO}_5=&5\left(G^{\zeta_4\MO}_{5,1}-\frac34G^{\zeta_3\MO}_{5,2}\right)
+5\left(G^{\zeta_4\MO}_{4,0}-\frac34G^{\zeta_3\MO}_{4,1}\right)\cdot F^{\mo}_{1,1}\nn
&+5\left(F^{\zeta_4\mo}_{3,0}-\frac34F^{\zeta_3\mo}_{3,1}\right)\cdot F^{\MO}_{2,1}
+5\left(G^{\zeta_4\MO}_{4,1}-\frac34G^{\zeta_3\MO}_{4,2}\right)\cdot F^{\mo}_{1,0}\nn
&-\left(\tbeta^{\zeta_4\MO}_4-\frac34\beta^{\zeta_3\MO}_4\right)\cdot F^{\mo}_{1,1}
-\left(\tbeta^{\zeta_4\MO}_3-\frac34\beta^{\zeta_3\MO}_3\right)\cdot F^{\MO}
_{2,1}\nn
&-\beta^{\mo}_1\cdot \left(F^{\zeta_4\MO}_{4,0}-\frac34F^{\zeta_3\MO}_{4,1}\right)
-\beta^{\MO}_2\cdot \left(F^{\zeta_4\mo}_{3,0}-\frac34F^{\zeta_3\mo}_{3,1}\right)\nn
&-\tbeta^{\MO}_1\cdot \left(F^{\zeta_4\MO}_{4,1}-\frac34F^{\zeta_3\MO}_{4,2}\right)
+\frac34(\tbeta_3^{\zeta_3\MO}-5F^{\zeta_3\mo}_{3,0})\cdot F^{\MO}_{2,2}.
\label{MOMfin}
\end{align}
Now using
\begin{align}
\beta_4^{\zeta_3\MO}=&4F^{\zeta_3\MO}_{4,1}-3F^{\zeta_3\mo}_{3,0}\cdot F^{\mo}_{1,1}-F^{\mo}_{1,1}\cdot F^{\zeta_3\mo}_{3,0}-3F^{\zeta_3\mo}_{3,1}\cdot F^{\mo}_{1,0}\nn
&-F^{\mo}_{1,0}\cdot F^{\zeta_3\mo}_{3,1},\nn
\tbeta_4^{\zeta_4\MO}=&4F^{\zeta_4\MO}_{4,0}-3F^{\zeta_4\mo}_{3,0}\cdot F^{\mo}_{1,0}-F^{\mo}_{1,0}\cdot F^{\zeta_4\mo}_{3,0},
\end{align}
and Eqs~\eqref{F3rel}, \eqref{ftilrel:a}, \eqref{ftilrel:b}, \eqref{grela:d}, \eqref{g41}, \eqref{Grels:a}, together with
\be
(X\cdot Y)\cdot Y+Y\cdot (X\cdot Y)=X\cdot (Y\cdot Y)+(Y\cdot X)\cdot Y
\ee
we find all the terms on the right-hand side of Eq.~\eqref{MOMfin} cancel and we are left with
\be
\beta^{\zeta_4\MO}_5=0.
\ee
As we have seen, the discussion of the $\MO$ case is attended by considerably more technical complexity compared to $\MOp$, and we have not so far proceeded to higher orders in this scheme; but we would expect to find that the even $\zeta$s continue to cancel at higher orders, in a similar fashion to the $\MOp$ case.

\section{Conclusions}
In this section we offer some concluding remarks and suggestions for future directions. Firstly to sum up: in Section \ref{basic} we set the scene, explaining the developments culminating in Ref.~\cite{baikov3} which showed how even $\zeta$ contributions to Feyman diagrams only appear in certain combinations with odd $\zeta$ functions. We then showed in detail (up to seven loop order) in Section \ref{single} how the expressions for even-$\zeta$ contributions to $\beta$-functions derived in Ref.~\cite{baikov3} and summarised in our Appendix \ref{betres} implied a renormalisation scheme in which the even $\zeta$s are absent. This is implicit in Ref.~\cite{baikov3}, but here we were concerned to show that this scheme (the $\MOp$ scheme) had a natural physical interpretation in term of removing finite even $\zeta$s from the bare coupling, at least for theories in which the $\beta$ function is determined by anomalous dimensions (``AD'' theories). As an example of such a theory, in Section \ref{susysec} we considered a general supersymmetric Wess-Zumino model where we were able to demonstrate the required scheme redefinition with explicit numerical values up to five loops; both for $\MOp$ and for the better-known $\MO$ scheme in which {\it all} finite contributions to the bare coupling are subtracted (but in which we believe the even $\zeta$s also cancel). Thus far, we had adopted the viewpoint of starting in $\msbar$ and making a scheme redefinition to the new, even-$\zeta$-free scheme. In contrast, in Section \ref{genproof} we started within either $\MOp$ or $\MO$ and showed the absence of even $\zeta$s order by order in a general theory through six loops for $\MOp$ and five loops for $\MO$.
There seems to be no obstacle to pursuing this approach to higher orders; however the technical complexity increases. It is not completely clear how the relation between the $F^{\zeta_n}$ quantities (the basic $\zeta_n$-dependent contributions to the bare coupling) and the $G^{\zeta_n}$ quantities (in which the $\zeta_n$-dependent contributions to the counterterms are omitted) may be generalised to higher orders. In the case of $\zeta_4$, where the relation with odd $\zeta$s in Eq.~\eqref{Grels:a} is very simple, the procedure may not depend too much on knowing the detailed definition of $G^{\zeta_n}$ in terms of $F^{\zeta_n}$; but in the case of $\zeta_6$, which is related to several other $\zeta$s via Eq.~\eqref{Grels:b}, and presumably other higher $\zeta$s, the structure of the calculation is less transparent, and the calculation was performed in a somewhat {\it ad hoc} way which does not give much insight into how to proceed to higher orders. A different definition of $G^{\zeta_n}$ (such as omitting {\it all} counterterms) which still preserved its basic property of obeying Eqs.~\eqref{Grels:main} might give a more natural, but unfortunately also more unwieldy, relation between $G^{\zeta_n}$ and $F^{\zeta_n}$. However it is also possible that we have become too focussed on the details, and one needs to step back and take a more general approach. Certainly in Ref.~\cite{baikov1}, the $\Ghat$ scheme is introduced and shown to be ``$\pi$-free'' to all orders. It seems likely that the $\Ghat$ scheme is equivalent to our $\MOp$; it would be satisfying to prove this. However it is not immediately obvious how to adapt the methods used here to higher orders.

The approach adopted in Section \ref{genproof} of working within $\MOp$ or $\MO$ seemed likely to be more efficient than starting in $\msbar$ and performing a scheme redefinition. However, the definitions of $\beta$-functions within $\MOp$ and $\MO$ are significantly more involved than in $\msbar$, so this turns out to be something of a moot point. Nevertheless, the necessity of including higher-order effects in scheme redefinitions would present a serious obstacle at higher loop orders, as we have seen here to some extent. Furthermore, although in the case of $\zeta_4$ we found the nice relation Eq.~\eqref{MS18} which was crucial in demonstrating the relation with $\MOp$, in the case of higher $\zeta$s the corresponding Eq.~\eqref{MS20a:main} was much more unwieldy. We should also remark that even Eq.~\eqref{MS18} is based on Eq.~\eqref{MS17a:main} which is something of a conjecture.

Finally, we have only been able to give a physical definition of the $\MOp$ and $\MO$ schemes within ``AD'' theories, since two-point diagrams have only a single external momentum, and hence setting $p^2=\mu^2$ yields a well-defined result for the integrals (the ``$p$-integrals'') and in particular their finite parts. Subtracting the whole finite part, or only the part depending on even $\zeta$s, gives the $\MO$ or $\MOp$ scheme respectively. In the case of more general theories, one can still reduce the computation to $p$-integrals by judiciously ``nullifying" external momenta, and thereby obtain a unique well-defined result for the simple pole (the process called ``infra-red rearrangement''). The scheme redefinitions may be derived from simple pole residues for odd $\zeta$s, as we see for instance with $F^{\zeta_3}_{3,1}$ in Eqs.~\eqref{bet3}, \eqref{bet4a}. Since Eqs.~\eqref{Grels:main}, \eqref{Grelsa:main} hold true down to finite parts of $p$ integrals, one can still define related quantities like $F_{3,0}^{\zeta_4}$ 
which are derived from the finite parts for even $\zeta$s; but there is no longer any natural physical meaning for these, the choice of which momenta to nullify being somewhat technical (based on the requirement of not introducing spurious infra-red divergences). This is the process we have effectively followed in Appendix \ref{genscal} where we have considered a general scalar field theory (i.e. not an ``AD theory'') and shown that we may remove the 
even $\zeta$s by a scheme redefinition, as shown at four and five loops in Eqs.~\eqref{gendel3}, \eqref{gendel4} respectively.

\section*{\it Note added} While this work was in progress J.A. Gracey kindly informed us of his own related work, which recently appeared in Ref.~\cite{gracey3}. Here the absence of even $\zeta$'s is demonstrated for the five-loop QCD $\beta$-function for a variety of momentum subtraction schemes.

\section*{Acknowledgments}
 We are very grateful for many useful conversations and correspondence with John Gracey, Tim Jones, Hugh Osborn and Andreas Vogt. We also thank Hugh Osborn and Colin Poole for providing the tikz code for many of the diagrams.

\appendix

\section{General scalar theory}\label{genscal}
In this section we consider a general scalar field theory. The $\beta$ functions for this theory have been computed up to five loops in Ref.~\cite{klein}, though a certain amount of work is required to extract the $\beta$-function coefficients corresponding to each tensor structure. In this Appendix we shall show that the even $\zeta$ terms up to this order may be removed by a scheme redefinition, though since this is not an AD theory, there is no natural physical interpretation of the scheme. The action is given by
\be
S=\int d^dx\left(\frac12\pa_{\mu}\phi^i \pa^{\mu}\phi^i-\frac12m^2\phi^i\phi^i-\frac{1}{4!}g_{ijkl}\phi^i\phi^j\phi^k\phi^l\right).
\ee
The $\beta$-function $\beta_{ijkl}$ may then be decomposed into 1PI pieces together with one-particle reducible  pieces determined by the anomalous dimension, in the form:
\be
 \beta=\btil+\Scal_4 \tikz[baseline=(vert_cent.base)]{
  \node (vert_cent) {\hspace{-13pt}$\phantom{-}$};
 \draw (-1.4,0.3)--(-1,0);
\draw (-1.4,-0.3)--(-1,0);
\draw (-1.4,0)--(-0.6,0);
 \draw (-0.3,0) circle [radius=0.3cm];
\draw (0,0)--(0.3,0);
\draw (-0.3,0) node {$\gamma$}
        }
\ee
with $\btil$ denoting the 1PI contributions and $\S_4$  the sum over the four terms where the anomalous dimension $\gamma$ is attached to each external line. Here and elsewhere we suppress indices as far as possible. Up to three loops the contributions to $\btil$ are given by
\begin{align}
          \btil^{(1)}=&c_1\Scal_3\tikz[baseline=(vert_cent.base)]{
  \node (vert_cent) {\hspace{-13pt}$\phantom{-}$};
 \draw (-0.4,0.3)--(0.1,0);
\draw (-0.4,-0.3)--(0.1,0);
    \draw (0.7,0) ++(0:0.6cm and 0.4cm) arc (0:180:0.6cm and 0.4cm) node(n1) {}
             (0.7,0) ++(180:0.6cm and 0.4cm) arc (180:360:0.6cm and 0.4cm) node(n2){};
             \draw (1.3,0)--(1.8,0.3); 
 \draw (1.3,0)--(1.8,-0.3); 
        },\nn
         \btil^{(2)}=&c_2\Scal_6\tikz[baseline=(vert_cent.base)]{
  \node (vert_cent) {\hspace{-13pt}$\phantom{-}$};
 \draw (-0.5,0.3)--(0,0);
\draw (-0.5,-0.3)--(0,0);
 \draw (0,0)--(1.4,0.7);
\draw (0,0)--(1.4,-0.7);
    \draw (1,0) ++(0:0.2cm and 0.5cm) arc (0:180:0.2cm and 0.5cm) node(n1) {}
             (1,0) ++(180:0.2cm and 0.5cm) arc (180:360:0.2cm and 0.5cm) node(n2){};
        },
\nn
     \btil^{(3)}=&
 \Scal_3\left(c_{3a}\tikz[baseline=(vert_cent.base)]{
  \node (vert_cent) {\hspace{-13pt}$\phantom{-}$};
\draw (-0.2,0.3)--(0.1,0);
\draw (-0.2,-0.3)--(0.1,0); 
    \draw (0.7,0) ++(0:0.6cm and 0.4cm) arc (0:180:0.6cm and 0.4cm) node(n1) {}
             (0.7,0) ++(180:0.6cm and 0.4cm) arc (180:360:0.6cm and 0.4cm) node(n2){};
 \draw (0.7,0) ++(0:0.1cm and 0.4cm) arc (0:180:0.1cm and 0.4cm) node(n1) {}
             (0.7,0) ++(180:0.1cm and 0.4cm) arc (180:360:0.1cm and 0.4cm) node(n2){};
    \draw (1.3,0)--(1.5,0.3); 
 \draw (1.3,0)--(1.5,-0.3);          
        }
+c_{3b}\tikz[baseline=(vert_cent.base)]{
  \node (vert_cent) {\hspace{-13pt}$\phantom{-}$};
 \draw (-0.4,0.3)--(0.1,0);
\draw (-0.4,-0.3)--(0.1,0);
    \draw (0.7,0) ++(0:0.6cm and 0.4cm) arc (0:180:0.6cm and 0.4cm) node(n1) {}
         (0.7,0) ++(180:0.6cm) arc (180:360:0.6cm and 0.4cm) node(n2){};
       \draw (1.3,0)--(1.8,0.3); 
 \draw (1.3,0)--(1.8,-0.3);      
\draw (0.7,0.4) circle [radius=0.3cm];
        }\right)
+\Scal_6\left(c_{3c} \tikz[baseline=(vert_cent.base)]{
  \node (vert_cent) {\hspace{-13pt}$\phantom{-}$};
 \draw (-0.5,0.3)--(0,0);
\draw (-0.5,-0.3)--(0,0);
 \draw (0,0)--(1.4,0.7);
\draw (0,0)--(1.4,-0.7);
    \draw (1,0.25) ++(0:0.1cm and 0.25cm) arc (0:180:0.1cm and 0.25cm) node(n1) {}
             (1,0.25) ++(180:0.1cm and 0.25cm) arc (180:360:0.1cm and 0.25cm) node(n2){};
 \draw (1,-0.25) ++(0:0.1cm and 0.5cm) arc (0:180:0.1cm and 0.25cm) node(n1) {}
             (1,-0.25) ++(180:0.1cm and 0.5cm) arc (180:360:0.1cm and 0.25cm) node(n2){};
        }
+c_{3d}\tikz[baseline=(vert_cent.base)]{
  \node (vert_cent) {\hspace{-13pt}$\phantom{-}$};
 \draw (-0.2,0.5)--(0.1,0.5);
\draw (-0.2,-0.5)--(0.1,-0.5);
    \draw (0.7,0.5) ++(0:0.6cm and 0.2cm) arc (0:180:0.6cm and 0.2cm) node(n1) {}
             (0.7,0.5) ++(180:0.6cm and 0.2cm) arc (180:360:0.6cm and 0.2cm) node(n2){};
 \draw (0.7,-0.5) ++(0:0.6cm and 0.2cm) arc (0:180:0.6cm and 0.2cm) node(n1) {}
             (0.7,-0.5) ++(180:0.6cm and 0.2cm) arc (180:360:0.6cm and 0.2cm) node(n2){};
\draw (0.1,0.5)--(0.1,-0.5);
\draw (1.3,0.5)--(1.3,-0.5);
 \draw (1.3,0.5)--(1.5,0.5);
\draw (1.3,-0.5)--(1.5,-0.5);
        }\,\right)\nn
&+c_{3e}\Scal_{12}\tikz[baseline=(vert_cent.base)]{
  \node (vert_cent) {\hspace{-13pt}$\phantom{-}$};
\draw (-0.4,0.3)--(0.1,0);
\draw (-0.4,-0.3)--(0.1,0); 
    \draw (0.7,0) ++(0:0.6cm and 0.4cm) arc (0:50:0.6cm and 0.4cm) node(n1) {}
     (0.7,0) ++(50:0.6cm and 0.4cm) arc (50:130:0.6cm and 0.4cm)  node(n2){}
             (0.7,0) ++(130:0.6cm and 0.4cm) arc (130:250:0.6cm and 0.4cm) 
     (0.7,0) ++(310:0.6cm and 0.4cm) arc (310:250:0.6cm and 0.4cm)  node(n3){}
 (0.7,0) ++(360:0.6cm and 0.4cm) arc (360:310:0.6cm and 0.4cm) node(n4) {}
(n1.base) to [out= 215, in = 325] (n2.base);
\draw (n2.base)--(n4.base);
\draw (n1.base)--(1.6,0.3);
\draw (n4.base)--(1.6,-0.3);
}
+c_{3f}\tikz[baseline=(vert_cent.base)]{
\node (vert_cent) {\hspace{-13pt}$\phantom{-}$};
\draw (-0.7,-0.7) -- (-0.05,-0.05);
\draw (0.05,0.05) -- (0.7,0.7);
\draw (-0.7,0.7) -- (0.7,-0.7);
\draw (-0.5,-0.5) -- (0.5,-0.5);
\draw (-0.5,-0.5) -- (-0.5,0.5);
\draw (-0.5,0.5) -- (0.5,0.5);
\draw (0.5,-0.5) -- (0.5,0.5);
}.
\label{twobet}
\end{align}
The values of the coefficients may be extracted from Ref.~\cite{kaz} and are given at one and two loops by
\be
c_1=1,\quad c_2=-1,
\label{cotwo}
\ee
and at three loops by
\be
 c^{\msbar}_{3a}=\tfrac12,\quad c^{\msbar}_{3b}=-\tfrac38,\quad c^{\msbar}_{3c}=c^{\msbar}_{3d}=-\tfrac12, \quad c^{\msbar}_{3e}=2,\quad c^{\msbar}_{3f}=12\zeta_3.
\label{cothree}
\ee
We do not need to concern ourselves with the structure or the coefficient values of the anomalous dimension $\gamma$ up to the order we are working, since there are no even-$\zeta$ contributions to $\gamma$ through five loops (and consequently no odd-$\zeta$ contributions through four loops).

The four-loop contributions to $\btil$ are given by
\begin{align}
\btil^{(4)}=&\Scal_3\left(c_{4a}\tikz[scale=0.6, baseline=(vert_cent.base)]{
	\node (vert_cent) {\hspace{-13pt}$\phantom{-}$};
	\draw circle[radius=1cm](0,0);
	\node[fill,circle,inner sep=0pt,minimum size=0pt] at \coord{-50} (A) {};
	\vertex at \coord{50} (B) {};
	\vertex at (0,-1) (C) {};
	\draw[bend right] (A) to (B);
	\draw (A)--(C)--(B);
	\draw (-1.5,-1)--(-1,0)--(-1.5,1);
	\draw (1.5,-1)--(1,0)--(1.5,1);
	}
+c_{4b}\tikz[baseline=(vert_cent.base)]{
  \node (vert_cent) {\hspace{-13pt}$\phantom{-}$};
\draw (-0.4,0.3)--(0.1,0);
\draw (-0.4,-0.3)--(0.1,0); 
    \draw (0.7,0) ++(0:0.6cm and 0.4cm) arc (0:180:0.6cm and 0.4cm) node(n1) {}
             (0.7,0) ++(180:0.6cm and 0.4cm) arc (180:360:0.6cm and 0.4cm) node(n2){};
 \draw (0.7,.2) ++(0:0.1cm and 0.2cm) arc (0:180:0.1cm and 0.2cm) node(n1) {}
             (0.7,.2) ++(180:0.1cm and 0.2cm) arc (180:360:0.1cm and 0.2cm) node(n2){};
 \draw (0.7,-.2) ++(0:0.1cm and 0.2cm) arc (0:180:0.1cm and 0.2cm) node(n1) {}
             (0.7,-.2) ++(180:0.1cm and 0.2cm) arc (180:360:0.1cm and 0.2cm) node(n2){};
    \draw (1.3,0)--(1.8,0.3); 
 \draw (1.3,0)--(1.8,-0.3);          
        }
+c_{4c}\tikz[scale=0.6, baseline=(vert_cent.base)]{
	\node (vert_cent) {\hspace{-13pt}$\phantom{-}$};
	\draw circle[radius=1cm](0,0);
	\node[fill,circle,inner sep=0pt,minimum size=0pt] at \coord{-70} (A) {};
	\vertex at \coord{70} (B) {};
	\vertex at (0,1) (C) {};
	\draw [bend right] (A) to (B) [bend left=50] (B) to (C) [bend left=50] (C) to (A);
	\draw (-1.5,-1)--(-1,0)--(-1.5,1);
	\draw (1.5,-1)--(1,0)--(1.5,1);
}\,\,\right)\nn
&+\Scal_6\Biggl(c_{4d}\tikz[scale=0.6, baseline=(vert_cent.base)]{
	\node (vert_cent) {\hspace{-13pt}$\phantom{-}$};
	\draw circle[radius=1cm](0,0);
	\vertex at \coord{40} (A) {};
	\vertex at \coord{140} (B) {};
	\draw [bend right=30] (1.25,1) to (A) [bend right] (A) to (B) [bend right=30] (B) to (1.25,-1);
	\vertex at \coord{-10} (C) {};
	\vertex at \coord{-170} (D) {};
	\draw [bend left] (C) to (D) [bend right] (C) to (D);
	\draw (-1.5,1)--(-1,0)--(-1.5,-1);
}
+c_{4e}\tikz[scale=0.6, baseline=(vert_cent.base)]{
	\node (vert_cent) {\hspace{-13pt}$\phantom{-}$};
	\draw circle[radius=1cm](0,0);
	\draw (-0.1,0) circle[radius=0.35cm];
	\vertex at \coord{20} (A) {};
	\vertex at \coord{160} (B) {};
	\draw [bend right=20] (1.25,1) to (A) [bend right=60] (A) to (B) [bend right=20] (B) to (1.25,-1);
	\draw (-1.5,1)--(-1,0)--(-1.5,-1);
}
+c_{4f} \tikz[baseline=(vert_cent.base)]{
  \node (vert_cent) {\hspace{-13pt}$\phantom{-}$};
 \draw (-0.5,0.3)--(0,0);
\draw (-0.5,-0.3)--(0,0);
 \draw (0,0)--(1.4,0.7);
\draw (0,0)--(1.4,-0.7);
    \draw (1,0) ++(0:0.2cm and 0.5cm) arc (0:180:0.2cm and 0.5cm) node(n1) {}
             (1,0) ++(180:0.2cm and 0.5cm) arc (180:360:0.2cm and 0.5cm) node(n2){};
\draw (1,0) ++(0:0.2cm and 0.1cm) arc (0:180:0.2cm and 0.1cm) node(n1) {}
             (1,0) ++(180:0.2cm and 0.1cm) arc (180:360:0.2cm and 0.1cm) node(n2){};
        }
+c_{4g} \tikz[baseline=(vert_cent.base)]{
  \node (vert_cent) {\hspace{-13pt}$\phantom{-}$};
 \draw (-0.5,0.3)--(0,0);
\draw (-0.5,-0.3)--(0,0);
 \draw (0,0)--(1.4,0.7);
\draw (0,0)--(1.4,-0.7);
    \draw (1.2,0.4) ++(0:0.1cm and 0.2cm) arc (0:180:0.1cm and 0.2cm) node(n1) {}
             (1.2,0.4) ++(180:0.1cm and 0.2cm) arc (180:360:0.1cm and 0.2cm) node(n2){};
 \draw (1.2,0) ++(0:0.1cm and 0.5cm) arc (0:180:0.1cm and 0.2cm) node(n1) {}
             (1.2,0) ++(180:0.1cm and 0.5cm) arc (180:360:0.1cm and 0.2cm) node(n2){};
 \draw (1.2,-0.4) ++(0:0.1cm and 0.5cm) arc (0:180:0.1cm and 0.2cm) node(n1) {}
             (1.2,-0.4) ++(180:0.1cm and 0.5cm) arc (180:360:0.1cm and 0.2cm) node(n2){};
        }\nn
&+c_{4h}\tikz[baseline=(vert_cent.base)]{
\node (vert_cent) {\hspace{-13pt}$\phantom{-}$};
 \draw (-0.5,0.3)--(0,0);
\draw (-0.5,-0.3)--(0,0);
\draw[name path=path1] [rotate=30] (.8,0) ellipse [x radius=.3, y radius=.07];
 \draw [name path=path5][rotate=30] (0,0)--(0.51,0);
\draw[name path=path2][rotate=30] (1.09,0)--(1.4,0);
\draw[name path=path3] [rotate=-30] (.8,0) ellipse [x radius=.3, y radius=.07];
 \draw [name path=path6][rotate=-30] (0,0)--(0.51,0);
\draw[name path=path4][rotate=-30] (1.09,0)--(1.4,0);
\draw [name intersections={of=path1 and path2, by=x}]
[name intersections={of=path3 and path4, by=y}] (x)-- (y);
\draw [name intersections={of=path1 and path5, by=x}]
[name intersections={of=path3 and path6, by=y}] (x)-- (y);}
+c_{4i}\tikz[baseline=(vert_cent.base)]{
\node (vert_cent) {\hspace{-13pt}$\phantom{-}$};
 \draw (-0.5,0.3)--(0,0);
\draw (-0.5,-0.3)--(0,0);
\draw[name path=path1] [rotate=30] (.8,0) ellipse [x radius=.3, y radius=.07];
 \draw [name path=path5][rotate=30] (0,0)--(0.51,0);
\draw[name path=path2][rotate=30] (1.09,0)--(1.4,0);
\draw[name path=path3] [rotate=-30] (.8,0) ellipse [x radius=.3, y radius=.07];
 \draw [name path=path6][rotate=-30] (0,0)--(0.51,0);
\draw[name path=path4][rotate=-30] (1.09,0)--(1.4,0);
\path [name path=path8][name intersections={of=path1 and path2, by=x1}]
[name intersections={of=path3 and path6, by=y1}] (x1)-- (y1);
\draw [name path=path7] [name intersections={of=path1 and path5, by=x2}]
[name intersections={of=path3 and path4, by=y2}] (x2)-- (y2);
\draw [name intersections={of=path7 and path8, by=y0}] let  \p1=(y0) in (x1)--({\x1+1.5},{\y1+1.5});
\draw [name intersections={of=path7 and path8, by=y0}] let  \p1=(y0) in (y1)--({\x1-1.5},{\y1-1.5});
}\,\Biggr)+\Scal_{12}\Biggl(c_{4j} \tikz[baseline=(vert_cent.base)]{
  \node (vert_cent) {\hspace{-13pt}$\phantom{-}$};
 \draw (-0.5,0.3)--(0,0);
\draw (-0.5,-0.3)--(0,0);
 \draw (0,0)--(1.4,0.7);
\draw (0,0)--(1.4,-0.7);
    \draw (1,0) ++(0:0.1cm and 0.5cm) arc (0:180:0.1cm and 0.5cm) node(n1) {}
             (1,0) ++(180:0.1cm and 0.5cm) arc (180:360:0.1cm and 0.5cm) node(n2){};
        \draw (0.5,-0.25) circle [radius=0.25cm];
}
+c_{4k}\tikz[baseline=(vert_cent.base)]{
\node (vert_cent) {\hspace{-13pt}$\phantom{-}$};
 \draw (-0.5,0.3)--(0,0);
\draw (-0.5,-0.3)--(0,0);
\draw[name path=path1] [rotate=30] (.8,0) ellipse [x radius=.3, y radius=.05];
 \draw [name path=path5][rotate=30] (0,0)--(0.51,0);
\draw[name path=path2][rotate=30] (1.09,0)--(1.4,0);
\draw (0,0)--(-30:1.4);
\draw ({0.8*cos(30)},-0.2) ellipse [x radius=.05, y radius=.2];
\draw [name intersections={of=path1 and path5, by=x}] (x)--({0.8*cos(30)},0);
\draw [name intersections={of=path2 and path1, by=y}] (y)--({0.8*cos(30)},0);
}\nn&
+c_{4l}\tikz[baseline=(vert_cent.base)]{
\node (vert_cent) {\hspace{-13pt}$\phantom{-}$};
 \draw (-0.5,0.3)--(0,0);
\draw (-0.5,-0.3)--(0,0);
\draw[name path=path1] [rotate=30] (.5,0) ellipse [x radius=.2, y radius=.05];
\draw[name path=path3] [rotate=30] (.9,0) ellipse [x radius=.2, y radius=.05];
 \draw [name path=path5][rotate=30] (0,0)--(0.31,0);
\draw[name path=path2][rotate=30] (1.09,0)--(1.4,0);
\draw (0,0)--(-30:1.4);
\draw [name intersections={of=path1 and path5, by=x}] (x)--({0.8*cos(30)},-0.4);
\draw [name intersections={of=path2 and path3, by=y}] (y)--({0.8*cos(30)},-0.4);
}+c_{4m}\tikz[baseline=(vert_cent.base)]{
\node (vert_cent) {\hspace{-13pt}$\phantom{-}$};
 \draw (-0.5,0.3)--(0,0);
\draw (-0.5,-0.3)--(0,0);
\draw[name path=path1] [rotate around={-30:({0.6*cos(30)},0)}] ({0.6*cos(30)-0.15},0) ellipse [x radius=.15, y radius=.05];
\draw (0,0) -- ({1.2*cos(30)},{-1.2*sin(30)});
\draw (0,0) -- ({1.2*cos(30)},{1.2*sin(30)});
\draw ({0.9*cos(30)},{-0.9*sin(30)}) -- ({0.6*cos(30)},0);
\draw ({0.9*cos(30)},{0.9*sin(30)}) -- ({0.6*cos(30)},0);
\draw  ({0.9*cos(30)},{-0.9*sin(30)}) --  ({0.9*cos(30)},{0.9*sin(30)})
}+c_{4n}\tikz[baseline=(vert_cent.base)]{
\node (vert_cent) {\hspace{-13pt}$\phantom{-}$};
 \draw (-0.5,0.3)--(0,0);
\draw (-0.5,-0.3)--(0,0);
\draw[name path=path2][rotate=30] (0,0)--(1.4,0);
\draw[name path=path2][rotate=-30] (0,0)--(1.4,0);
\draw ({cos(30)},{sin(30)})--({cos(30)},0);
\draw ({0.5*cos(30)},{-0.5*sin(30)})--({cos(30)},0);
\draw ({0.5*cos(30)},{-0.5*sin(30)})--({cos(30)},sin(30);
\draw ({cos(30)},-0.25) ellipse [x radius=.05, y radius=.25];
}
+c_{4o}\tikz[scale=0.6, baseline=(vert_cent.base)]{
	\node (vert_cent) {\hspace{-13pt}$\phantom{-}$};
	\draw circle[radius=1cm](0,0);
	\vertex at \coord{-40} (A) {};
	\vertex at \coord{-140} (B) {};
	\draw (-0.9,1.15)--(A) [bend left] (A) to (B) (B)--(-0.9,-1.15);
	\vertex at \coord{30} (C) {};
	\vertex at \coord{90} (D) {};
	\vertex at \coord{150} (E) {};
	\draw [bend right=10] (1,1.25) to (C) [bend right=60] (C) to (D) [bend right=60] (D) to (E) [bend right=10] (E) to (1,-1.25);
}
+c_{4p}\tikz[baseline=(vert_cent.base)]{
\node (vert_cent) {\hspace{-13pt}$\phantom{-}$};
\draw (-0.7,-0.7) -- (-0.5,-0.5);
\path [name path=path1] (-0.5,-0.5) -- (0.5,0);
\draw[name path=path2] (-0.7,0.7) -- (0.7,-0.7);
\draw [name intersections={of=path1 and path2, by=y0}] let  \p1=(y0) in (-0.5,-0.5)--({\x1-1.5},{\y1-1});
\draw [name intersections={of=path1 and path2, by=y0}] let  \p1=(y0) in (0.5,0)--({\x1+1.5},{\y1+1});
\draw (0.5,0.5)--(0.7,0.7);
\draw (-0.5,-0.5) -- (0.5,-0.5);
\draw (-0.5,-0.5) -- (-0.5,0.5);
\draw (-0.5,0.5) -- (0.5,0.5);
\draw (0.5,-0.5) -- (0.5,0);
 \draw (0.5,0.25) ++(0:0.1cm and 0.25cm) arc (0:180:0.1cm and 0.25cm) node(n1) {}
             (0.5,0.25) ++(180:0.1cm and 0.25cm) arc (180:360:0.1cm and 0.25cm) node(n2){};
}\,\Biggr)\nn
&
+c_{4q}\Scal_6\tikz[baseline=(vert_cent.base)]{
\node (vert_cent) {\hspace{-13pt}$\phantom{-}$};
\path [name path=path1] (-0.7,-0.7) -- (0.5,0.5);
\draw [name path=path2] (-0.7,0.7) -- (0.5,-0.5);
\draw [name intersections={of=path1 and path2, by=y0}] let  \p1=(y0) in (-0.7,-0.7)--({\x1-1.5},{\y1-1});
\draw [name intersections={of=path1 and path2, by=y0}] let  \p1=(y0) in (0.5,0.5)--({\x1+1.5},{\y1+1});
 \draw (0.5,0) ++(0:0.1cm and 0.5cm) arc (0:180:0.1cm and 0.5cm) node(n1) {}
             (0.5,0) ++(180:0.1cm and 0.5cm) arc (180:360:0.1cm and 0.5cm) node(n2){};
\draw (-0.5,-0.5) -- (0.5,-0.5);
\draw (-0.5,-0.5) -- (-0.5,0.5);
\draw (-0.5,0.5) -- (0.5,0.5);
\draw (0.6,0) -- (0.8,0.2);
\draw (0.6,0) -- (0.8,-0.2);
}
+c_{4r}\Scal_{24}\tikz[scale=0.6, baseline=(vert_cent.base)]{
	\node (vert_cent) {\hspace{-13pt}$\phantom{-}$};
	\draw circle[radius=1cm](0,0);
	\vertex at \coord{-40} (A) {};
	\vertex at \coord{-140} (B) {};
	\draw (-0.9,1.15)--(A) [bend left] (A) to (B) (B)--(-0.9,-1.15);
	\vertex at \coord{-20} (C) {};
	\vertex at \coord{60} (D) {};
	\draw [bend left] (1.25,1) to (D) [bend left] (D) to (C) (C)--(1,-1.25);
	}+c_{4s}\tikz[baseline=(vert_cent.base)]{
\node (vert_cent) {\hspace{-13pt}$\phantom{-}$};
\draw (-0.7,-0.7) -- (0.7,0.7);
\draw (-0.7,0.7) -- (0.7,-0.7);
\draw (-0.5,-0.5) -- (0.5,-0.5);
\draw (-0.5,-0.5) -- (-0.5,0.5);
\draw (-0.5,0.5) -- (0.5,0.5);
\draw (0.5,-0.5) -- (0.5,0.5);
}.
\label{fourbet}
\end{align}

At four loops (again extracted from Ref.~\cite{kaz}) the coefficients are 
\begin{align}
c^{\msbar}_{4a}=\tfrac13(6\zeta_3-11),\quad c^{\msbar}_{4b}=1-\zeta_3,\quad c^{\msbar}_{4c}&=\tfrac{7}{12},\quad\quad c^{\msbar}_{4d}=\tfrac12,\quad  c^{\msbar}_{4e}=\tfrac{121}{144},\nn c^{\msbar}_{4f}=1-2\zeta_3, \quad  c^{\msbar}_{4g}=c^{\msbar}_{4o}=\tfrac14(2\zeta_3-1),&\quad c^{\msbar}_{4h}=c^{\msbar}_{4l}=\tfrac{1}{6}(5-6\zeta_3),\nn
 c^{\msbar}_{4i}=\tfrac{5}{6},\quad 
  c^{\msbar}_{4j}=-\tfrac{37}{288},\quad 
 c^{\msbar}_{4k}=c^{\msbar}_{4r}=\tfrac23,\quad& c^{\msbar}_{4m}=4\zeta_3-5,\quad c^{\msbar}_{4n}=-5,\nn
c^{\msbar}_{4p}=3(\zeta_4-2\zeta_3),\quad& c^{\msbar}_{4q}=-3(2\zeta_3+\zeta_4),\quad
c^{\msbar}_{4s}=-40\zeta_5.
\label{cofour}
\end{align} 

In accordance with Eqs.~\eqref{delone}, we expect that we can remove the four-loop $\zeta_4$ dependence by a redefinition
\be
(\delta g)^{(3)} =-\frac14\beta^{\zeta_3\msbar}_3\zeta_4=-3\zeta_4\tikz[baseline=(vert_cent.base)]{
\node (vert_cent) {\hspace{-13pt}$\phantom{-}$};
\draw (-0.7,-0.7) -- (-0.05,-0.05);
\draw (0.05,0.05) -- (0.7,0.7);
\draw (-0.7,0.7) -- (0.7,-0.7);
\draw (-0.5,-0.5) -- (0.5,-0.5);
\draw (-0.5,-0.5) -- (-0.5,0.5);
\draw (-0.5,0.5) -- (0.5,0.5);
\draw (0.5,-0.5) -- (0.5,0.5);
}.
\label{gendel3}
\ee
This is similar to the supersymmetric case in Eq.~\eqref{dg3susy}, but with the difference that in this case there seems little point in relating the coefficients to the finite $\zeta_4$ contributions which now have little physical meaning.
Indeed, using Eqs.~\eqref{blin}, \eqref{bdef}, where now 
\be
B\cdot  \equiv B^{ijkl}\frac{\pa}{\pa g^{ijkl}}
\ee
we find
\be
(\delta \beta)^{(3)}=3\zeta_4\left(-\Scal_{12}\tikz[baseline=(vert_cent.base)]{
\node (vert_cent) {\hspace{-13pt}$\phantom{-}$};
\draw (-0.7,-0.7) -- (-0.5,-0.5);
\path [name path=path1] (-0.5,-0.5) -- (0.5,0);
\draw[name path=path2] (-0.7,0.7) -- (0.7,-0.7);
\draw [name intersections={of=path1 and path2, by=y0}] let  \p1=(y0) in (-0.5,-0.5)--({\x1-1.5},{\y1-1});
\draw [name intersections={of=path1 and path2, by=y0}] let  \p1=(y0) in (0.5,0)--({\x1+1.5},{\y1+1});
\draw (0.5,0.5)--(0.7,0.7);
\draw (-0.5,-0.5) -- (0.5,-0.5);
\draw (-0.5,-0.5) -- (-0.5,0.5);
\draw (-0.5,0.5) -- (0.5,0.5);
\draw (0.5,-0.5) -- (0.5,0);
 \draw (0.5,0.25) ++(0:0.1cm and 0.25cm) arc (0:180:0.1cm and 0.25cm) node(n1) {}
             (0.5,0.25) ++(180:0.1cm and 0.25cm) arc (180:360:0.1cm and 0.25cm) node(n2){};
}+\Scal_6\tikz[baseline=(vert_cent.base)]{
\node (vert_cent) {\hspace{-13pt}$\phantom{-}$};
\path [name path=path1] (-0.7,-0.7) -- (0.5,0.5);
\draw [name path=path2] (-0.7,0.7) -- (0.5,-0.5);
\draw [name intersections={of=path1 and path2, by=y0}] let  \p1=(y0) in (-0.7,-0.7)--({\x1-1.5},{\y1-1});
\draw [name intersections={of=path1 and path2, by=y0}] let  \p1=(y0) in (0.5,0.5)--({\x1+1.5},{\y1+1});
 \draw (0.5,0) ++(0:0.1cm and 0.5cm) arc (0:180:0.1cm and 0.5cm) node(n1) {}
             (0.5,0) ++(180:0.1cm and 0.5cm) arc (180:360:0.1cm and 0.5cm) node(n2){};
\draw (-0.5,-0.5) -- (0.5,-0.5);
\draw (-0.5,-0.5) -- (-0.5,0.5);
\draw (-0.5,0.5) -- (0.5,0.5);
\draw (0.6,0) -- (0.8,0.2);
\draw (0.6,0) -- (0.8,-0.2);
}\right)
\ee
so that the two $\zeta_4$ terms are cancelled.

We now turn to the five-loop order. The $\beta$-function results may be extracted from Appendix B of Ref.~\cite{klein}. Here five-loop results are tabulated diagram by diagram, and all one has to do is to multiply the simple pole by the usual loop factor of 5, together with the symmetry factor given in the column labelled $W_4$. Unfortunately the symmetry factors $\Scal_3$ etc are not given, but luckily these cancel out in our calculations, as we shall show(?).
At this loop order, following Eq.~\eqref{delg4def}, but again suppressing the expression in terms of the finite even-$\zeta$ quantities in this case, we expect to have to introduce a variation
\be
(\delta g)^{(4)}=-\frac{3}{16}\beta^{\zeta_3\msbar}_4\zeta_4-\frac{5}{16}\beta^{\zeta_5\msbar}_4\zeta_6.
\label{gendel4}
\ee
More explicitly, after extracting the $\zeta_3$ and $\zeta_5$ contributions to the $\beta$-function coefficients in Eq.~\eqref{cofour}, this results in 
\begin{align}
(\delta g)^{(4)}=&\Scal_3\left(\epsilon^A_{4a}\tikz[scale=0.6, baseline=(vert_cent.base)]{
	\node (vert_cent) {\hspace{-13pt}$\phantom{-}$};
	\draw circle[radius=1cm](0,0);
	\node[fill,circle,inner sep=0pt,minimum size=0pt] at \coord{-50} (A) {};
	\vertex at \coord{50} (B) {};
	\vertex at (0,-1) (C) {};
	\draw[bend right] (A) to (B);
	\draw (A)--(C)--(B);
	\draw (-1.5,-1)--(-1,0)--(-1.5,1);
	\draw (1.5,-1)--(1,0)--(1.5,1);
	}
+\epsilon^A_{4b}\tikz[baseline=(vert_cent.base)]{
  \node (vert_cent) {\hspace{-13pt}$\phantom{-}$};
\draw (-0.4,0.3)--(0.1,0);
\draw (-0.4,-0.3)--(0.1,0); 
    \draw (0.7,0) ++(0:0.6cm and 0.4cm) arc (0:180:0.6cm and 0.4cm) node(n1) {}
             (0.7,0) ++(180:0.6cm and 0.4cm) arc (180:360:0.6cm and 0.4cm) node(n2){};
 \draw (0.7,.2) ++(0:0.1cm and 0.2cm) arc (0:180:0.1cm and 0.2cm) node(n1) {}
             (0.7,.2) ++(180:0.1cm and 0.2cm) arc (180:360:0.1cm and 0.2cm) node(n2){};
 \draw (0.7,-.2) ++(0:0.1cm and 0.2cm) arc (0:180:0.1cm and 0.2cm) node(n1) {}
             (0.7,-.2) ++(180:0.1cm and 0.2cm) arc (180:360:0.1cm and 0.2cm) node(n2){};
    \draw (1.3,0)--(1.8,0.3); 
 \draw (1.3,0)--(1.8,-0.3);          
        }\right)\nn
&+\Scal_6\Biggl(\epsilon^A_{4f} \tikz[baseline=(vert_cent.base)]{
  \node (vert_cent) {\hspace{-13pt}$\phantom{-}$};
 \draw (-0.5,0.3)--(0,0);
\draw (-0.5,-0.3)--(0,0);
 \draw (0,0)--(1.4,0.7);
\draw (0,0)--(1.4,-0.7);
    \draw (1,0) ++(0:0.2cm and 0.5cm) arc (0:180:0.2cm and 0.5cm) node(n1) {}
             (1,0) ++(180:0.2cm and 0.5cm) arc (180:360:0.2cm and 0.5cm) node(n2){};
\draw (1,0) ++(0:0.2cm and 0.1cm) arc (0:180:0.2cm and 0.1cm) node(n1) {}
             (1,0) ++(180:0.2cm and 0.1cm) arc (180:360:0.2cm and 0.1cm) node(n2){};
        }
+\epsilon^A_{4g} \tikz[baseline=(vert_cent.base)]{
  \node (vert_cent) {\hspace{-13pt}$\phantom{-}$};
 \draw (-0.5,0.3)--(0,0);
\draw (-0.5,-0.3)--(0,0);
 \draw (0,0)--(1.4,0.7);
\draw (0,0)--(1.4,-0.7);
    \draw (1.2,0.4) ++(0:0.1cm and 0.2cm) arc (0:180:0.1cm and 0.2cm) node(n1) {}
             (1.2,0.4) ++(180:0.1cm and 0.2cm) arc (180:360:0.1cm and 0.2cm) node(n2){};
 \draw (1.2,0) ++(0:0.1cm and 0.5cm) arc (0:180:0.1cm and 0.2cm) node(n1) {}
             (1.2,0) ++(180:0.1cm and 0.5cm) arc (180:360:0.1cm and 0.2cm) node(n2){};
 \draw (1.2,-0.4) ++(0:0.1cm and 0.5cm) arc (0:180:0.1cm and 0.2cm) node(n1) {}
             (1.2,-0.4) ++(180:0.1cm and 0.5cm) arc (180:360:0.1cm and 0.2cm) node(n2){};
        }
+\epsilon^A_{4h}\tikz[baseline=(vert_cent.base)]{
\node (vert_cent) {\hspace{-13pt}$\phantom{-}$};
 \draw (-0.5,0.3)--(0,0);
\draw (-0.5,-0.3)--(0,0);
\draw[name path=path1] [rotate=30] (.8,0) ellipse [x radius=.3, y radius=.07];
 \draw [name path=path5][rotate=30] (0,0)--(0.51,0);
\draw[name path=path2][rotate=30] (1.09,0)--(1.4,0);
\draw[name path=path3] [rotate=-30] (.8,0) ellipse [x radius=.3, y radius=.07];
 \draw [name path=path6][rotate=-30] (0,0)--(0.51,0);
\draw[name path=path4][rotate=-30] (1.09,0)--(1.4,0);
\draw [name intersections={of=path1 and path2, by=x}]
[name intersections={of=path3 and path4, by=y}] (x)-- (y);
\draw [name intersections={of=path1 and path5, by=x}]
[name intersections={of=path3 and path6, by=y}] (x)-- (y);}\Biggr)\nn
&+\Scal_{12}\Biggl(\epsilon^A_{4l}\tikz[baseline=(vert_cent.base)]{
\node (vert_cent) {\hspace{-13pt}$\phantom{-}$};
 \draw (-0.5,0.3)--(0,0);
\draw (-0.5,-0.3)--(0,0);
\draw[name path=path1] [rotate=30] (.5,0) ellipse [x radius=.2, y radius=.05];
\draw[name path=path3] [rotate=30] (.9,0) ellipse [x radius=.2, y radius=.05];
 \draw [name path=path5][rotate=30] (0,0)--(0.31,0);
\draw[name path=path2][rotate=30] (1.09,0)--(1.4,0);
\draw (0,0)--(-30:1.4);
\draw [name intersections={of=path1 and path5, by=x}] (x)--({0.8*cos(30)},-0.4);
\draw [name intersections={of=path2 and path3, by=y}] (y)--({0.8*cos(30)},-0.4);
}
+\epsilon^A_{4m}\tikz[baseline=(vert_cent.base)]{
\node (vert_cent) {\hspace{-13pt}$\phantom{-}$};
 \draw (-0.5,0.3)--(0,0);
\draw (-0.5,-0.3)--(0,0);
\draw[name path=path1] [rotate around={-30:({0.6*cos(30)},0)}] ({0.6*cos(30)-0.15},0) ellipse [x radius=.15, y radius=.05];
\draw (0,0) -- ({1.2*cos(30)},{-1.2*sin(30)});
\draw (0,0) -- ({1.2*cos(30)},{1.2*sin(30)});
\draw ({0.9*cos(30)},{-0.9*sin(30)}) -- ({0.6*cos(30)},0);
\draw ({0.9*cos(30)},{0.9*sin(30)}) -- ({0.6*cos(30)},0);
\draw  ({0.9*cos(30)},{-0.9*sin(30)}) --  ({0.9*cos(30)},{0.9*sin(30)})
}
+\epsilon^A_{4o}\tikz[scale=0.6, baseline=(vert_cent.base)]{
	\node (vert_cent) {\hspace{-13pt}$\phantom{-}$};
	\draw circle[radius=1cm](0,0);
	\vertex at \coord{-40} (A) {};
	\vertex at \coord{-140} (B) {};
	\draw (-0.9,1.15)--(A) [bend left] (A) to (B) (B)--(-0.9,-1.15);
	\vertex at \coord{30} (C) {};
	\vertex at \coord{90} (D) {};
	\vertex at \coord{150} (E) {};
	\draw [bend right=10] (1,1.25) to (C) [bend right=60] (C) to (D) [bend right=60] (D) to (E) [bend right=10] (E) to (1,-1.25);
}
+\epsilon^A_{4p}\tikz[baseline=(vert_cent.base)]{
\node (vert_cent) {\hspace{-13pt}$\phantom{-}$};
\draw (-0.7,-0.7) -- (-0.5,-0.5);
\path [name path=path1] (-0.5,-0.5) -- (0.5,0);
\draw[name path=path2] (-0.7,0.7) -- (0.7,-0.7);
\draw [name intersections={of=path1 and path2, by=y0}] let  \p1=(y0) in (-0.5,-0.5)--({\x1-1.5},{\y1-1});
\draw [name intersections={of=path1 and path2, by=y0}] let  \p1=(y0) in (0.5,0)--({\x1+1.5},{\y1+1});
\draw (0.5,0.5)--(0.7,0.7);
\draw (-0.5,-0.5) -- (0.5,-0.5);
\draw (-0.5,-0.5) -- (-0.5,0.5);
\draw (-0.5,0.5) -- (0.5,0.5);
\draw (0.5,-0.5) -- (0.5,0);
 \draw (0.5,0.25) ++(0:0.1cm and 0.25cm) arc (0:180:0.1cm and 0.25cm) node(n1) {}
             (0.5,0.25) ++(180:0.1cm and 0.25cm) arc (180:360:0.1cm and 0.25cm) node(n2){};
}\,\Biggr)\nn
&+\epsilon^A_{4q}\Scal_6\tikz[baseline=(vert_cent.base)]{
\node (vert_cent) {\hspace{-13pt}$\phantom{-}$};
\path [name path=path1] (-0.7,-0.7) -- (0.5,0.5);
\draw [name path=path2] (-0.7,0.7) -- (0.5,-0.5);
\draw [name intersections={of=path1 and path2, by=y0}] let  \p1=(y0) in (-0.7,-0.7)--({\x1-1.5},{\y1-1});
\draw [name intersections={of=path1 and path2, by=y0}] let  \p1=(y0) in (0.5,0.5)--({\x1+1.5},{\y1+1});
 \draw (0.5,0) ++(0:0.1cm and 0.5cm) arc (0:180:0.1cm and 0.5cm) node(n1) {}
             (0.5,0) ++(180:0.1cm and 0.5cm) arc (180:360:0.1cm and 0.5cm) node(n2){};
\draw (-0.5,-0.5) -- (0.5,-0.5);
\draw (-0.5,-0.5) -- (-0.5,0.5);
\draw (-0.5,0.5) -- (0.5,0.5);
\draw (0.6,0) -- (0.8,0.2);
\draw (0.6,0) -- (0.8,-0.2);
}
+\epsilon^A_{4s}\tikz[baseline=(vert_cent.base)]{
\node (vert_cent) {\hspace{-13pt}$\phantom{-}$};
\draw (-0.7,-0.7) -- (0.7,0.7);
\draw (-0.7,0.7) -- (0.7,-0.7);
\draw (-0.5,-0.5) -- (0.5,-0.5);
\draw (-0.5,-0.5) -- (-0.5,0.5);
\draw (-0.5,0.5) -- (0.5,0.5);
\draw (0.5,-0.5) -- (0.5,0.5);
}.
\label{fourbeta}
\end{align}
with coefficients given by
\begin{align}
\epsilon^A_{4a}=2Z_4,\quad \epsilon^A_{4b}=-Z_4,\quad \epsilon^A_{4f}=-2Z_4, \quad&  \epsilon^A_{4g}=\epsilon^A_{4o}=\tfrac12Z_4,\quad \epsilon^A_{4h}=\epsilon^A_{4l}=-Z_4,\nn
 \quad \epsilon^A_{4m}=4Z_4,\quad 
\epsilon^A_{4p}=-6Z_4,\quad& \epsilon^A_{4q}=-6Z_4,\quad
\epsilon^A_{4s}=-40Z_6,\nn
\label{cotfour}
\end{align} 
where $Z_4$, $Z_6$ are defined in Eq.~\eqref{Zdefs} and we use the $A$ superscript to denote the minimal redefinition for removing the even $\zeta$s.
We must also take into account the effect of $(\delta g)^{(3)}$ in Eq.~\eqref{gendel3} at this order.
We list here the variations of the five-loop coefficients, derived from the first-order variations in Eq.~\eqref{blin}. The notation of the coefficients follows that of Ref.~\cite{klein}, so that for instance $c_{547}$ means the five-loop coefficient corresponding to diagram 47 in Ref.~\cite{klein}. Also, we define
\be
X^A_{1,4a}=c_1\epsilon^A_{4a}
\ee
where $c_1$ is given in Eq.~\eqref{cotwo} (with some redundancy in notation, in order to harmonise with Appendix~\ref{susyvar}). We then have
\begin{align}
\delta c_{524}=4\delta c_{525}=-2\delta c_{526}=&2\delta c_{527}\nn
=8\delta c_{528}=-4\delta c_{529}=&24X^A_{1,4s}=300\zeta_6,\nn
\delta c_{530}=\delta c_{540}=\delta c_{544}=&-4\delta c_{532}\nn
=\frac12\delta c_{539}=2\delta c_{541}=&24X^A_{1,4p}=27\zeta_4,\nn
\delta c_{533}=\delta c_{534}=-\delta c_{536}=&24X^A_{1,4p}+12X^A_{2,3f}=63\zeta_4,\nn
\delta c_{535}=&12X^A_{1,4q}-6X^A_{2,3f}=-\frac92\zeta_4,\nn
\delta c_{537}=\delta c_{538}=2\delta c_{542}=2\delta c_{543}=&-24X^A_{1,4p}+24X^A_{1,4q}=0,\nn
\delta c_{545}=&48X^A_{1,4o}=-\frac92\zeta_4,\nn
\delta c_{547}=&12X^A_{1,4a}-12X^A_{1,4m}=\frac92\zeta_4,\nn
\delta c_{549}=&24X^A_{1,4l}=\frac{3}{16}\zeta_4,\nn
\delta c_{550}=&-24X^A_{1,4l}=-\frac92\zeta_4,\nn
\delta c_{551}=&-24X^A_{1,4l}+24X^A_{1,4m}=-\frac{45}{2}\zeta_4,\nn
\delta c_{553}=&24X^A_{1,4m}=-\frac34\zeta_4,\nn
\delta c_{555}=&24X^A_{1,4m}=-18\zeta_4,\nn
\delta c_{557}=&24X^A_{1,4l}=\frac92\zeta_4,\nn
\delta c_{558}=&24X^A_{1,4l}+12X^A_{1,4m}=-\frac92\zeta_4,\nn
\delta c_{558a}=&24X^A_{1,4h}+12X^A_{1,4m}=-\frac92\zeta_4,\nn
\delta c_{559}=&12X^A_{1,4m}=-9\zeta_4,\nn
\delta c_{560}=&48X^A_{1,4o}=-\frac92\zeta_4,\nn
\delta c_{561}=&12X^A_{1,4a}=-\frac38\zeta_4,\nn
\delta c_{563}=&24X^A_{1,4h}=\frac92\zeta_4,\nn
\delta c_{564}=&-24X^A_{1,4h}+12X^A_{1,4m}=-\frac{27}{2}\zeta_4,\nn
\delta c_{566}=&12X^A_{1,4h}=\frac{3}{16}\zeta_4,\nn
\delta c_{569}=&12X^A_{1,4m}=-9\zeta_4,\nn
\delta c_{571}=&24X^A_{1,4h}=\frac92\zeta_4,\nn
\delta c_{573}=&3X^A_{1,4a}=-\frac98\zeta_4,\nn
\delta c_{574}=&-24X^A_{1,4m}=18\zeta_4,\nn
\delta c_{579}=&-6X^A_{1,4g}+9X^A_{1,4b}=\frac94\zeta_4,\nn
\delta c_{582}=&-24X^A_{1,4g}+36X^A_{1,4l}=9\zeta_4,\nn
\delta c_{588}=&3X^A_{1,4a}-6X^A_{1,4h}=-\frac94\zeta_4,\nn
\delta c_{589}=&12X^A_{1,4l}=\frac94\zeta_4,\nn
\delta c_{591}=&-12X^A_{1,4o}+12X^A_{1,4b}=\frac{27}{8}\zeta_4,\nn
\delta c_{592}=&-12X^A_{1,4o}=\frac98\zeta_4,\nn
\delta c_{593}=&3X^A_{1,4a}+12X^A_{1,4b}=\frac{3}{16}\zeta_4,\nn
\delta c_{595}=&6X^A_{1,4a}-12X^A_{1,4l}=-\frac92\zeta_4,\nn
\delta c_{596}=&24X^A_{1,4o}=-\frac94\zeta_4,\nn
\delta c_{5101}=&24X^A_{1,4f}=9\zeta_4,\nn
\delta c_{5104}=&24X^A_{1,4g}=-\frac94\zeta_4,\nn
\delta c_{5107}=&-12X^A_{1,4b}+12X^A_{1,4f}=\frac94\zeta_4,\nn
\delta c_{5109}=&6X^A_{1,4b}-6X^A_{1,4f}=-\frac98\zeta_4,\nn
\delta c_{5111}=&-12X^A_{1,4a}+24X^A_{1,4f}=\frac{27}{2}\zeta_4,\nn
\delta c_{5112}=&12X^A_{1,4f}+24X^A_{1,4o}=\frac94\zeta_4,\nn
\delta c_{5113}=&-24X^A_{1,4f}+24X^A_{1,4l}=-\frac92\zeta_4,\nn
\delta c_{5118}=&24X^A_{1,4g}=-\frac94\zeta_4,\nn
\delta c_{5123}=&12X^A_{1,4f}+24X^A_{1,4g}=\frac94\zeta_4.
\end{align}
This list is incomplete in two ways. Firstly we only list variations of those five-loop coefficients which contain an even-$\zeta$-function contribution, and secondly we only include those $X$ which contain an even-$\zeta$ contribution. Indeed, one finds that all the even-$\zeta$ terms are cancelled at this order.

\section{$\beta$-function results}\label{betres}
In this Appendix we recapitulate the results of Ref.~\cite{baikov1}, expressing (for a single-coupling theory) $\beta$-functions for even $\zeta$s in terms of those for odd $\zeta$s. We denote  the $L$-loop $\zeta_n$ dependent contribution to $\beta$ by $\beta^{\zeta_n}_L$. The results of Ref.~\cite{baikov1} for the $\beta$-functions up to seven loops are as follows (after adjusting with a factor of 2 for our conventions).

The $\zeta_4$ contributions are given by
\begin{subequations}\label{z4:main}
\begin{align}
\beta_4^{\zeta_4}=&\frac12\beta_1\beta^{\zeta_3}_3,\label{z4:a}\\
\beta_5^{\zeta_4}=&\frac{9}{16}\beta_1\beta^{\zeta_3}_4+\frac14\beta_2\beta^{\zeta_3}_3,\label{z4:b}\\
\beta_6^{\zeta_4}=&\frac35\beta_1\beta^{\zeta_3}_5+\frac38\beta_2\beta^{\zeta_3}_4,\label{z4:c}\\
\beta_7^{\zeta_4}=&\frac58\beta_1\beta^{\zeta_3}_6+\frac{9}{20}\beta_2\beta^{\zeta_3}_5
+\frac{3}{16}\beta_3\beta^{\zeta_3}_4-\frac14\beta_4\beta^{\zeta_3}_3\label{z4:d}.
\end{align}
\end{subequations}
The $\zeta_6$ contributions are given by
\begin{subequations}\label{z6:main}
\begin{align}
\beta^{\zeta_6}_5=&\frac{15}{16}\beta_1\beta^{\zeta_5}_4,\label{z6:a}\\
\beta^{\zeta_6}_6=&\beta_1\beta^{\zeta_5}_5-\frac18(\beta_1)^3\beta^{\zeta_3}_3+\frac58\beta_2\beta^{\zeta_5}_4,\label{z6:b}\\
\beta^{\zeta_6}_7=&\frac{25}{24}\beta_1\beta^{\zeta_5}_6+\frac{3}{4}\beta_2\beta^{\zeta_5}_5-\frac{5}{32}(\beta_1)^3\beta^{\zeta_3}_4+\frac{5}{16}\beta_3\beta^{\zeta_5}_4-\frac14\beta_2(\beta_1)^2\beta^{\zeta_3}_3.\label{z6:c}
\end{align}
\end{subequations}
The $\zeta_8$ contributions are given by
\begin{subequations}\label{z8:main}
\begin{align}
\beta^{\zeta_8}_{6}=&\frac75\beta_1\beta^{\zeta_7}_{5},\label{z8:a}\\
\beta^{\zeta_8}_{7}=&\frac{35}{24}\beta_1\beta^{\zeta_7}_{6}+\frac{7}{16}(\beta_1)^2\beta^{(\zeta_3)^2}_5-\frac{35}{64}(\beta_1)^3\beta^{\zeta_5}_4+\frac{21}{20}\beta_2\beta^{\zeta_7}_5-\frac{7}{96}\beta_1\left(\beta^{\zeta_3}_3\right)^2.\label{z8:b}
\end{align}
\end{subequations}
The $\zeta_{10}$ contribution is given by
\be
\beta^{\zeta_{10}}_{7}=\frac{15}{8}\beta_1\beta^{\zeta_9}_6.
\label{z10}
\ee
The mixed $\zeta_4\zeta_3$ contributions are given by
\begin{subequations}\label{z43:main}
\begin{align}
\beta_6^{\zeta_4\zeta_3}=&\frac65\beta_1\beta_5^{(\zeta_3)^2},\label{z43:a}\\
\beta_7^{\zeta_4\zeta_3}=&\frac54\beta_1\beta^{(\zeta_3)^2}_6+\frac{9}{10}\beta_2\beta^{(\zeta_3)^2}_5-\frac{1}{16}\beta^{\zeta_3}_3\beta^{\zeta_3}_4.\label{z43:b}
\end{align}
\end{subequations}
The remaining non-vanishing mixed-$\zeta$ $\beta$-functions up to this order are
\begin{subequations}\label{z45:main}
\begin{align}
\beta^{\zeta_4\zeta_5}_7=&-\frac{25}{8}\beta_1\beta^{\overline\zeta_{5,3}}_6+\frac{5}{8}\beta_1\beta^{\zeta_3\zeta_5}_6-\frac{1}{4}\beta^{\zeta_3}_3\beta^{\zeta_5}_4,\label{z45:a}\\
\beta^{\zeta_6\zeta_3}_7=&\frac{25}{24}\beta_1\beta^{\zeta_3\zeta_5}_6+\frac{5}{16}\beta^{\zeta_3}_3\beta^{\zeta_5}_4,\label{z45:b}
\end{align}
\end{subequations}
and
\be
\beta^{\zeta_4(\zeta_3)^2}_7=\frac{15}{8}\beta_1\beta^{(\zeta_3)^3}_6.
\label{z433}
\ee

\section{Feynman diagram results}\label{feynres}
In this section we record the results for the two-point graphs which contribute to the anomalous dimension for the fields in the supersymmetric Wess-Zumino model. These results may be used to check our detailed calculations in various ways. Most obviously, the anomalous dimensions quoted for the Wess-Zumino model in Section 4 may readily be extracted. Secondly, the various $F$ values in Eq.~\eqref{FMOMres} have been computed starting from Eq.~\eqref{zphimom} which in turn followed from specialising the results in the current Section to the single-coupling case. 
For convenience we write 
\be
f=\sum_{Ln_g}f_{Ln_g}+\ldots
\label{ffdef}
\ee
 Here $f$ is defined as for $F$ in Eq.~\eqref{rbar}, but now corresponds to a two-point function rather than a four-point function. Of course the values of these quantities depend on the particular theory under consideration. Here we give results for the example of the supersymmetric Wess-Zumino model, where we have explicit results in the general case up to five loops thanks to Ref.~\cite{gracey}. We write 
\be
f_{Ln_g}=\Rbar\left(\Gamma_{Ln_g}\right)
\ee
for the contribution to $f$ from $\Gamma_{Ln_g}$, the $L$-loop graph corresponding to the anomalous dimension coefficient $c_{Ln_g}$ in the notation of \cite{gracey}. Up to four loops these graphs are also displayed in Eqs.~\eqref{gam3}, \eqref{gam4}. The ellipsis in Eq.~\eqref{ffdef} indicates that we ignore $O(\epsilon^p)$ terms with $p>0$. Typically the $\Rbar$ operation is defined for $\msbar$ but here we allow for the possibility of other schemes such as $\MOp$, $\MO$. For each graph we have
\be
f^{\msbar}_{Ln_g}=\fbar^{\msbar}_{Ln_g}C^{\msbar}_{Ln_g}
\ee
(with similar expressions for the other schemes)
where $\fbar^{\msbar}_{Ln_g}$ is purely numerical and $C^{\msbar}_{Ln_g}$ contains all the dependence on the couplings. 
For instance, in the case of the simple one-loop graph we have $\left(C^{\msbar}_{11}\right)^i{}_j=g^{\msbar ikl}g^{\msbar}_{jkl}$.
The results in the standard $\msbar$ scheme are given at one and two loops by
\begin{align}
\fbar^{\msbar}_{11}=&-\frac12\left(2\frac{1}{\epsilon}+2\right),\nn
\fbar^{\msbar}_{21}=&\frac12\left(-2\frac{1}{\epsilon^2}+\frac{1}{\epsilon}+\frac{11}{2}\right),
\label{zms12}
\end{align}
at three loops
\begin{align}
\fbar^{\msbar}_{31}=&-\frac14\left\{4\zeta_3\frac{1}{\epsilon}+12\zeta_3+3\zeta_4\right\},\nn
\fbar^{\msbar}_{32}=&-\frac18\left\{\frac83\frac{1}{\epsilon^3}-\frac{4}{3}\frac{1}{\epsilon^2}-\frac{2}{3}\frac{1}{\epsilon}
+\frac{37}{3}-\frac43\zeta_3\right\},\nn
\fbar^{\msbar}_{33}=&2\fbar^{\msbar}_{32}-\frac12\zeta_3,\nn
\fbar^{\msbar}_{34}=&-\frac12\left\{\frac43\frac{1}{\epsilon^3}-2\frac{1}{\epsilon^2}+\frac{4}{3}\frac{1}{\epsilon}+17-\frac23\zeta_3\right\},
\label{zms3} 
\end{align}
and at four loops
\begin{align}
\fbar^{\msbar}_{41}=&2\left\{\frac52\frac{1}{\epsilon}+\frac{25}{8}\zeta_6+\frac{35}{4}\zeta_5-\frac14(\zeta_3)^2\right\},\nn
\fbar^{\msbar}_{42}=\fbar^{\msbar}_{43}=\frac12\fbar^{\msbar}_{44}=&\frac14\left\{-2\zeta_3\frac{1}{\epsilon^2}+\left(3\zeta_3-\frac32\zeta_4\right)\frac{1}{\epsilon}+\frac{59}{2}\zeta_3+\frac{9}{4}\zeta_4+\frac{13}{2}\zeta_5\right\},\nn
\fbar^{\msbar}_{45}=&\frac18\left\{-4\frac{1}{\epsilon^4}+2\frac{1}{\epsilon^3}+\frac{1}{\epsilon^2}+\left(\frac12-\zeta_3\right)\frac{1}{\epsilon}+\frac{169}{4}-\frac{23}{2}\zeta_3-\frac34\zeta_4\right\},\nn
\fbar^{\msbar}_{46}=&\frac14\left\{-2\frac{1}{\epsilon^4}+\frac{8}{3}\frac{1}{\epsilon^3}-\frac{5}{6}\frac{1}{\epsilon^2}-\frac{2}{3}\frac{1}{\epsilon}+\frac{971}{24}-\frac{22}{3}\zeta_3\right\},\nn
\fbar^{\msbar}_{47}=&\frac14\left\{-6\zeta_3\frac{1}{\epsilon^2}+\left(\frac32\zeta_4+3\zeta_3\right)\frac{1}{\epsilon}+\frac{67}{2}\zeta_3+\frac{33}{4}\zeta_4+\frac{7}{2}\zeta_5,\right\},\nn
\fbar^{\msbar}_{48}=&\frac18\left\{-\frac43\frac{1}{\epsilon^4}+2\frac{1}{\epsilon^3}-\frac{1}{3}\frac{1}{\epsilon^2}-\frac56\frac{1}{\epsilon}+\frac{532}{12}-\frac{16}{3}\zeta_3\right\},\nn
\fbar^{\msbar}_{49}=\fbar^{\msbar}_{410}=&\fbar^{\msbar}_{46}+2\zeta_3,\nn
\fbar^{\msbar}_{411}=&\fbar^{\rm{MOM}}_{45}+\frac32\zeta_3,\nn
\fbar^{\msbar}_{412}=&2\fbar^{\msbar}_{48}+\frac14\left\{\zeta_3\frac{1}{\epsilon}+\frac{13}{2}\zeta_3+\frac34\zeta_4\right\},\nn
\fbar^{\msbar}_{413}=&\frac12\left\{-\frac23\frac{1}{\epsilon^4}+2\frac{1}{\epsilon^3}-\frac{19}{6}\frac{1}{\epsilon^2}+\frac52\frac{1}{\epsilon}+\frac{447}{8}-\frac{10}{3}\zeta_3\right\},
\label{zms4}
\end{align}

  The expressions in curly brackets are the results of evaluating the pure momentum integral with its associated subtractions, while the fractions multiplying these expressions are symmetry factors, also incorporating the overall minus sign which implements the final subtraction of poles. We suppress throughout a factor of $(16\pi^2)^{-1}$ for each loop order. It is straightforward to determine the corresponding momentum integral for each Feynman diagram from the explanations given in Ref.~\cite{gracey}, then the results for the momentum integrals may be extracted from Eqs.~\eqref{zms12}, \eqref{zms3}, \eqref{zms4} and used for other theories such as the scalar theory of Appendix A. The coefficient of the $\epsilon^{-p}$ pole in $f^{\msbar}_{Ln_g}$ is equal to $2^{p+1}a^{\msbar}_{Lp|Ln_g}$, where $a^{\msbar}_{Lp|Ln_g}$ is defined in Appendix A of Ref.~\cite{gracey}. The factor of $2^{p+1}$ is due to a factor two difference in our respective definitions of $\epsilon$, and a difference in our definition of $Z_{\phi}$.
We have for the counterterm in $\msbar$ for the $L$-loop graph $n_g$
\be
Z^{\msbar}_{Ln_g}=\Khat(f^{\msbar}_{Ln_g})=\sum_{p=1}^{L}Z^{\msbar}_{Ln_g,p}\epsilon^{-p},
\ee
where $\Khat$ denotes the operation of extracting the pole terms, also defining here the pole coefficients $Z^{\msbar}_{Ln_g,p}$. Note that with our definitions, the $f^{\msbar}_{Ln_g}$ also contain finite terms. We have included these since they are required in defining the scheme redefinition which transforms from $\msbar$ to $\MO$, see Appendix \ref{mstrans}. However, as we said earlier, we drop $O(\epsilon)$ terms in the definition of $f^{\msbar}_{Ln_g}$. The contribution to the anomalous dimension from a given diagram is given by
\be
\gamma^{\msbar}_{Ln_g}=-\frac12LZ^{\msbar}_{Ln_g,1};
\label{gamdefb}
\ee
though the corresponding expression in other schemes is more involved, as explained in the context of $\MO$ at the beginning of Appendix \ref{alt} (the minus sign in Eq.~\eqref{gamdefb} derives from the ``$-\epsilon g$'' in $\beta(g)$ in Eq.~\eqref{betdefa} inserted into Eq.~\eqref{gamdefa}), and the factor of $\frac12$ is conventionally inserted in Eq.~\eqref{gamdefa}.  
The full two-point renormalisation constant is given in $\msbar$ by
\be
Z^{\msbar}_{\phi}=\sum_{L,n_g}Z^{\msbar}_{Ln_g},
\ee
with similar expressions for other schemes. 

For the case of $\MOp$ we have
\be
\fbar^{\MOp}_{Ln_g}=\fbar^{\msbar}_{Ln_g},\quad L=1,2,3,
\ee
where the $\fbar^{\msbar}_{Ln}$ may be read off from Eqs.~\eqref{zms12}, \eqref{zms3}. Then
\be
Z^{\MOp}_{Ln_g}=\Khat^{\MOp}(f^{\MOp}_{Ln_g})=\sum_{p=0}^{L}Z^{\MOp}_{Ln_g,p}\epsilon^{-p},
\label{zmomdef}
\ee
where $\Khat^{\MOp}$ denotes the operation of extracting the pole terms, together with the finite part proportional to $\zeta_4$ (and any other finite part containing even $\zeta$s, in general); so that the sum over coefficients now starts at $p=0$.
 At four loops we have
\be
\fbar^{\MOp}_{4n}=\fbar^{\msbar}_{4n},\quad n\ne7
\ee
where the $\msbar$ results are as in Eq.~\eqref{zms4}. For the case of $\fbar^{\MOp}_{47}$ we have
\be
\fbar^{\MOp}_{47}=\frac14\left\{-6\zeta_3\frac{1}{\epsilon^2}+\left(-\frac92\zeta_4+3\zeta_3\right)\frac{1}{\epsilon}+\frac{67}{2}\zeta_3+\frac{9}{4}\zeta_4+\frac{7}{2}\zeta_5,\right\}.
\label{zmom4}
\ee
We then again have $Z^{\MOp}_{47}$ defined as in Eq.~\eqref{zmomdef}.
The only differences here between the $\MOp$ and $\msbar$ results lie in the simple poles and finite parts proportional to $\zeta_4$. We note that as expected, the finite part proportional to $\zeta_4$ in $\MOp$ is $\frac34$ times the simple pole proportional to $\zeta_3$ in $\msbar$/$\MOp$.

We now give results for the corresponding quantities evaluated in MOM. 
The one and two-loop contributions are given by
\begin{align}
\fbar^{\rm{MOM}}_{11}=&-\frac12\left\{2\frac{1}{\epsilon}+2\right\},\nn
\fbar^{\rm{MOM}}_{21}=&\frac12\left\{-2\frac{1}{\epsilon^2}-3\frac{1}{\epsilon}+\frac32\right\},
\label{z12def}
\end{align}
the three-loop contributions by 
\begin{align}
\fbar^{\rm{MOM}}_{31}=&-\frac14\left\{4\zeta_3\frac{1}{\epsilon}+12\zeta_3+3\zeta_4\right\},\nn
\fbar^{\rm{MOM}}_{32}=&-\frac18\left\{\frac83\frac{1}{\epsilon^3}+\frac{20}{3}\frac{1}{\epsilon^2}+\frac{10}{3}\frac{1}{\epsilon}-\frac43\zeta_3-\frac53\right\},\nn
\fbar^{\rm{MOM}}_{33}=&2\fbar^{\rm{MOM}}_{32}-\frac12\zeta_3,\nn
\fbar^{\rm{MOM}}_{34}=&-\frac12\left\{\frac43\frac{1}{\epsilon^3}-2\frac{1}{\epsilon^2}-\frac{11}{3}\frac{1}{\epsilon}-\frac23\zeta_3+3\right\}.
\label{zdef}
\end{align}
and the four-loop contributions by 
\begin{align}
\fbar^{\rm{MOM}}_{41}=&2\left\{\frac52\frac{1}{\epsilon}+\frac{25}{8}\zeta_6+\frac{35}{4}\zeta_5-\frac14(\zeta_3)^2\right\},\nn
\fbar^{\rm{MOM}}_{42}=\fbar^{\rm{MOM}}_{43}=\frac12\fbar^{\rm{MOM}}_{44}=&\frac14\Bigl\{-2\zeta_3\frac{1}{\epsilon^2}+\left(-\frac32\zeta_4-5\zeta_3\right)\frac{1}{\epsilon}\nn
&+\frac{11}{2}\zeta_3-\frac{15}{4}\zeta_4+\frac{13}{2}\zeta_5\Bigr\},\nn
\fbar^{\rm{MOM}}_{45}=&\frac18\Bigl\{-4\frac{1}{\epsilon^4}-14\frac{1}{\epsilon^3}-15\frac{1}{\epsilon^2}+\left(\frac12-\zeta_3\right)\frac{1}{\epsilon}\nn
&+\frac{73}{4}-\frac{15}{2}\zeta_3-\frac34\zeta_4\Bigr\},\nn
\fbar^{\rm{MOM}}_{46}=&\frac14\left\{-2\frac{1}{\epsilon^4}-\frac{16}{3}\frac{1}{\epsilon^3}+\frac{5}{6}\frac{1}{\epsilon^2}+\frac{13}{2}\frac{1}{\epsilon}+\frac{37}{24}-\frac{10}{3}\zeta_3\right\},\nn
\fbar^{\rm{MOM}}_{47}=&\frac14\Bigl\{-6\zeta_3\frac{1}{\epsilon^2}+\left(-\frac92\zeta_4-21\zeta_3\right)\frac{1}{\epsilon}\nn
&+\frac{19}{2}\zeta_3+\frac{9}{4}\zeta_4+\frac{7}{2}\zeta_5\Bigr\},\nn
\fbar^{\rm{MOM}}_{48}=&\frac18\left\{-\frac43\frac{1}{\epsilon^4}-\frac{10}{3}\frac{1}{\epsilon^3}-\frac{1}{3}\frac{1}{\epsilon^2}+\left(\frac{7}{6}+\frac83\zeta_3\right)\frac{1}{\epsilon}+\frac{11}{12}\right\},\nn
\fbar^{\rm{MOM}}_{49}=\fbar^{\rm{MOM}}_{410}=&\fbar^{\rm{MOM}}_{46}+\zeta_3,\nn
\fbar^{\rm{MOM}}_{411}=&\fbar^{\rm{MOM}}_{45}+\frac12\zeta_3,\nn
\fbar^{\rm{MOM}}_{412}=&\frac14\Bigl\{-\frac43\frac{1}{\epsilon^4}-\frac{10}{3}\frac{1}{\epsilon^3}-\frac{1}{3}\frac{1}{\epsilon^2}+\left(\frac{7}{6}-\frac13\zeta_3\right)\frac{1}{\epsilon}\nn
&+\frac{11}{12}+\frac52\zeta_3+\frac34\zeta_4\Bigr\},\nn
\fbar^{\rm{MOM}}_{413}=&\frac12\Bigl\{-\frac23\frac{1}{\epsilon^4}-\frac23\frac{1}{\epsilon^3}+\frac{23}{6}\frac{1}{\epsilon^2}+\left(-\frac{23}{3}+\frac43\zeta_3\right)\frac{1}{\epsilon}\nn
&+\frac{61}{8}-\frac23\zeta_3\Bigr\}.\nn
\label{z4mom}
\end{align}
We have for the counterterm in $\MO$ for the $L$-loop graph $n_g$
\be
Z^{\MO}_{Ln_g}=\Khat^{\MO}(f^{\MO}_{Ln_g})=\sum_{p=0}^{L}Z^{\MO}_{Ln_g,p}\epsilon^{-p},
\ee
where $\Khat^{\MO}$ denotes the operation of extracting the pole terms together with all the finite parts, and the $\fbar^{\MO}_{Ln_g}$ are defined in Eqs.~\eqref{z12def}, \eqref{zdef} and \eqref{z4mom}.
The difference between $\msbar$ and MOM is that the finite parts of graphs are also subtracted. There is a cumulative effect, since the finite parts present in lower-order $Z$s affect the pole terms in higher order counterterms so that, as we see, the simple poles in $\fbar^{\msbar}_{21}$ and $\fbar^{\rm{MOM}}_{21}$ differ, and this can be traced back to the finite difference between $Z^{\msbar}_{11}$ and $Z^{\rm{MOM}}_{11}$.

\section{Single-coupling Wess-Zumino model}\label{wzsingle}
We now discuss the derivation of the  $\beta$-function and and anomalous dimension for a single coupling theory within $\msbar$, $\MO$ and $\MOp$ in the case of the supersymmetric Wess-Zumino model. This permits a simple and transparent comparison with results given in Ref.~\cite{gracey} and hence provides a useful check of all our results. Following Ref.~\cite{gracey}, we define 
\be
g=y^2,
\ee
where $y$ is now the single coupling corresponding to $y^{ijk}$ in Section \ref{susysec}, and we express all our results in terms of $g$. Note again the change in notation from Ref.~\cite{gracey}, where $a$ is used for the coupling which we denote as $g$.
The relation between the MOM and MS couplings is given by
\be
g_B=\mu^{\epsilon}Z_gg=\mu^{\epsilon} Z_{g}^{\rm{MOM}}g^{\rm{MOM}}
\label{azdef}
\ee
where
\be
Z_gZ_{\phi}^{3}=Z^{\rm{MOM}}_g\left(Z_{\phi}^{\rm {MOM}}\right)^{3}=1.
\label{aZMOM}
\ee
In this Appendix, quantities such as $g$ without a superscript (as in Eq.~\eqref{azdef}) are presumed to be evaluated in $\msbar$.
$Z_{\phi}$ is the two-point (or field) renormalisation which was defined in Section \ref{feynres}. As explained there, it may be obtained by adding the contributions from each diagram which are given there with the appropriate symmetry factors. The relation given in Eq.~\eqref{aZMOM} between $Z_{\phi}$ and $Z_g$, the renormalisation constant for the coupling $g$, is a consequence of the non-renormalisation theorem for supersymmetry.
Focussing on the finite terms, we find
\be
g=\left(C_{\Phi}^{\rm{MOM}}\right)^{-3}g^{\rm{MOM}}
\label{adefa}
\ee
where
\be
C_{\Phi}^{\rm{MOM}}=1+c^{\MO}_1g^{\MO}+c^{\MO}_2(g^{\MO})^2+c^{\MO}_3(g^{\MO})^3+\ldots,
\label{cphi}
\ee
with
\begin{align}
c^{\MO}_1=Z_{[10]}^{\MO}=&-1,\quad c^{\MO}_2=Z_{[20]}^{\MO}=\frac34,\nn
c^{\MO}_3=&Z_{[30]}^{\MO}=-\frac83\zeta_3-\frac34\zeta_4-\frac78,\nn
c^{\MO}_4=&Z_{[40]}^{\MO}=\frac{79}{8}+\frac{151}{24}\zeta_3-\frac{51}{16}\zeta_4+\frac{199}{8}\zeta_5+\frac{25}{4}\zeta_6-\frac12(\zeta_3)^2,
\label{cdef}
\end{align}
where we introduce the notation $Z^{\MO}_{[L,m]}$ to denote the $O(\epsilon^{-m})$ contribution to $Z^{\MO}_{\phi}$ at $L$ loops. 
This easily leads to 
\begin{align}
g=&g^{\MO}\Biggl(1+3g^{\MO}+\frac{15}{4}\left(g^{\MO}\right)^2+[29+18\zeta_4+64\zeta_3]\frac{\left(g^{\MO}\right)^3}{8}\nn
&+\left[-372+297\zeta_4-300\zeta_6-504\zeta_5+24(\zeta_3)^2+210\zeta_3\right]\frac{\left(g^{\MO}\right)^4}{16}\Biggr)+\ldots.
\label{aaMOM}
\end{align}
Solving for $g^{\MO}$ in terms of $g$, we find alternatively
\begin{align}
g^{\MO}=&g\Bigl(1-3g+\frac{57}{4}g^2-[64\zeta_3+18\zeta_4+659]\frac{g^3}{8}\nn
&+[2094\zeta_3-24(\zeta_3)^2+351\zeta_4+504\zeta_5+300\zeta_6+8895]\frac{g^4}{16}\Bigr)+\ldots,
\label{aMOM}
\end{align}
which agrees with Ref.~\cite{gracey}. 
We also remark that rewriting $g^{\MO}$ in terms of $g$ in $C_{\Phi}^{\rm{MOM}}$ as given by Eq.~\eqref{cphi}, we obtain
\begin{align}
C_{\Phi}(g)=&1-g+\frac{15}{4}g^2-[64\zeta_3+18\zeta_4+471]\frac{g^3}{24}\nn
&+[1838\zeta_3-24(\zeta_3)^2+279\zeta_4+504\zeta_5+300\zeta_6+6156]\frac{g^4}{48}+\ldots
\end{align}
which again agrees with Ref.~\cite{gracey}.

The anomalous dimension in $\msbar$ is given by\cite{gracey}
\begin{align}
\gamma(g)=&\frac16\beta(g)\nn
=&\frac12g-\frac12g^2+(12\zeta_3+5)\frac{g^3}{8}+[18\zeta_4-60\zeta_3-80\zeta_5-9]\frac{g^4}{8}\nn
&+[504(\zeta_3)^2+858\zeta_3-441\zeta_4+1828\zeta_5-900\zeta_6+2646\zeta_7+79]\frac{g^5}{32}+\ldots
\label{gamres}
\end{align}
(note our definition
\be
\beta(g)=\mu\frac{d}{d\mu} g =2y\mu\frac{d}{d\mu} y=2y.3\gamma y=6g\gamma(g)
\ee
appears to differ by a factor of 2 from \cite{gracey}).
We may now obtain the $\beta$-function in MOM by inserting Eq.~\eqref{aMOM} into 
\be
\beta^{\rm {MOM}}=\mu\frac{d}{d\mu}g^{\MO}=\beta(g)\frac{d}{dg}g^{\MO}
\ee
and then rewriting $g$ in terms of $g^{\MO}$ via Eq.~\eqref{aaMOM}, we find 
\be
\beta^{\rm {MOM}}=6\gamma^{\rm {MOM}}
\label{betMOM}
\ee
where
\begin{align}
\gamma^{\rm {MOM}}=&\frac12g^{\MO}-\frac12\left(g^{\MO}\right)^2+(6\zeta_3+7)\frac{\left(g^{\MO}\right)^3}{4}-[13\zeta_3+20\zeta_5+20]\frac{\left(g^{\MO}\right)^4}{2}\nn
&+[216(\zeta_3)^2+772\zeta_3+230\zeta_5+1323\zeta_7+1222]\frac{\left(g^{\MO}\right)^5}{16}+\ldots.
\label{gamMOM}
\end{align}

We may also consider the simpler variant scheme $\rm MOM'$, where we only subtract the finite pieces which involve even $\zeta$s. We again use Eq.~\eqref{adefa} with Eq.~\eqref{cphi}, but now using coefficients derived using $Z^{\MOp}$ rather than $Z^{\MO}$ in Eq.~\eqref{cdef}. It may be helpful to give the detailed numbers in this case since this result has not appeared before. We have
\be
Z^{\MOp}_{[L0]}=\sum_{i=1}^{13}Z^{\MOp}_{Li,0},
\label{Zbrack}
\ee
where $Z^{\MOp}_{Li,0}$ was defined in Eq.~\eqref{zmomdef}. The non-zero contributions are
\begin{align}
Z^{\MOp}_{31,0}=&-\frac34\zeta_4,\nn
Z^{\MOp}_{41,0}=\frac{25}{4}\zeta_6,\quad Z^{\MOp}_{42,0}=&Z^{\MOp}_{43,0}=\frac12Z^{\MOp}_{44,0}=Z^{\MOp}_{47,0}=\frac{9}{16}\zeta_4,\nn
Z^{\MOp}_{45,0}=Z^{\MOp}_{411,0}=&-\frac{3}{32}\zeta_4,\quad Z^{\MOp}_{412,0}=\frac{3}{16}\zeta_4.
\end{align}
One might naively expect that all we need to do to obtain $Z^{\MOp}_{41,0}$ etc is take only the even-$\zeta$ contributions to $Z^{\MO}_{41,0}$ etc , but in fact $Z^{\MOp}_{42,0}$, $Z^{\MOp}_{43,0}$ and $Z^{\MOp}_{44,0}$ differ from the even-$\zeta$ contributions to the corresponding $\MO$ quantities. This is due to a similar cumulative effect to that mentioned at the end of Section \ref{feynres}.
This leads using Eq.~\eqref{Zbrack} to replacing $g^{\MO}$, $c_{\Phi}^{\MO}$ in Eqs.~\eqref{cphi}, \eqref{cdef} by $g^{\MOp}$, $c_{\Phi}^{\MOp}$ with coefficients given by
\begin{align}
c^{\MOp}_1=c^{\MOp}_2=&0\nn
c^{\MOp}_3=Z^{\MOp}_{[30]}=&-\frac34\zeta_4,\nn
c^{\MOp}_4=Z^{\MOp}_{[40]}=&\frac{25}{4}\zeta_6+\frac{45}{16}\zeta_4.
\end{align}
We now obtain
\be
g=g^{\MOp}\Biggl(1+9\zeta_4\frac{\left(g^{\MOp}\right)^3}{4}-[135\zeta_4+300\zeta_6]\frac{\left(g^{\MOp}\right)^4}{16}\Biggr)+\ldots,
\ee
with the trivial inverse up to this order
\be
g^{\MOp}=g\Biggl(1-9\zeta_4\frac{g^3}{4}+[135\zeta_4+300\zeta_6]\frac{g^4}{16}\Biggr)+\ldots.
\label{aMOp}
\ee
The same argument which led to Eq.~\eqref{gamMOM} now results in
\begin{align}
\gamma^{\MOp}=&\frac12g^{\MOp}-\frac12\left(g^{\MOp}\right)^2+(12\zeta_3+5)\frac{\left(g^{\MOp}\right)^3}{8}-[9+60\zeta_3+80\zeta_5]\frac{\left(g^{\MOp}\right)^4}{8}\nn
&+[504(\zeta_3)^2+858\zeta_3+1828\zeta_5+2646\zeta_7+79]\frac{\left(g^{\MOp}\right)^5}{32}+\ldots,
\end{align}
i.e. the same functional form as Eq.~\eqref{gamres}, but simply with the $\zeta_4$ and $\zeta_6$ terms removed.

\section{Alternative computation of WZ $\beta$-function in MOM}\label{alt}
In this Appendix we record the details of a computation of the $\beta$-function for the single-coupling Wess-Zumino model using the $\MO$ scheme {\it ab initio} instead of making a transformation from the $\msbar$ scheme. There are two ways to achieve this. One could exploit the non-renormalisation theorem and compute the anomalous dimension according to the standard definition
\be
\gamma= \frac12\mu\frac{d}{d\mu}\ln Z_{\phi}=\frac12Z_{\phi}^{-1}\beta(g)\frac{d}{dg}Z_{\phi};
\label{gamdefa}
\ee
or one can simply compute the $\beta$-function directly, computing $Z_g$ from Eq.~\eqref{aZMOM}. We have performed the computation using both methods; the first is somewhat more efficient, though one has to be careful to include cross-terms which would not be present in $\msbar$. However, we present details only for the second method, which is more in line with our general procedure in Sections 2, 3 and 5. We start with the result for $Z_{\phi}$, which is given by
\begin{align}
Z_{\phi}^{\MO}=&1-\frac{1}{\epsilon}g^{\MO}(1+\epsilon)
+\frac{3}{\epsilon}\left(g^{\MO}\right)^2\left(-\frac12+\frac14\epsilon\right)\nn
&+\frac{1}{\epsilon}\left(g^{\MO}\right)^3\left(\frac{7}{12}-\zeta_3-\left[\frac78
+\frac83\zeta_3+\frac34\zeta_4\right]\epsilon\right)\nn
&+\frac{1}{\epsilon}\left(g^{\MO}\right)^4\left(\frac{77}{48}-\frac{115}{12}\zeta_3-\frac{21}{8}\zeta_4+\left[\frac{79}{8}+\frac{151}{24}\zeta_3-\frac{51}{16}\zeta_4\right]\epsilon\right)\nn
&+\frac{1}{\epsilon}\left(g^{\MO}\right)^{5}\left(\frac{359}{40}+\frac{151}{40}\zeta_3-\frac{1269}{80}\zeta_4\right)+\ldots
\label{zphimom}
\end{align}
where we have only displayed the simple pole and finite term up to four loops, and only the simple pole at five loops. This result may be obtained by adding the diagram-by-diagram multi-coupling $\MO$ results for $\fbar$ in Appendix~\ref{feynres} and specialising to the single-coupling case.
This then leads using Eq.~\eqref{aZMOM} to 
\begin{align}
Z_{a}^{\MO}=&1+\frac{3}{\epsilon}g^{\MO}(1+\epsilon)
+\frac{3}{\epsilon}\left(g^{\MO}\right)^2\left(\frac{11}{2}+\frac54\epsilon\right)\nn
&+\frac{1}{\epsilon}\left(g^{\MO}\right)^3\left(\frac{149}{4}+3\zeta_3+\left[\frac{29}{8}
+8\zeta_3+\frac94\zeta_4\right]\epsilon\right)\nn
&+\frac{3}{\epsilon}\left(g^{\MO}\right)^4\left(\frac{241}{16}+\frac{97}{4}\zeta_3+\frac{45}{8}\zeta_4+\left[-\frac{31}{4}+\frac{35}{8}\zeta_3+\frac{99}{16}\zeta_4\right]\epsilon\right)\nn
&+\frac{3}{\epsilon}\left(g^{\MO}\right)^{5}\left(-\frac{1683}{40}+\frac{3429}{40}\zeta_3+\frac{4689}{80}\zeta_4\right)+\ldots,
\end{align}
from which we can trivially extract
\begin{align}
F^{\mo}_{1,1}=3\left(g^{\MO}\right)^2,\quad& F^{\mo}_{1,0}=3\left(g^{\MO}\right)^2,\nn
F^{\MO}_{2,1}=\frac{33}{2}\left(g^{\MO}\right)^3,\quad& F^{\MO}_{2,0}=\frac{15}{4}\left(g^{\MO}\right)^3,\nn
F^{\zeta_4\MO}_{3,1}=0,\quad F^{\zeta_3\mo}_{3,1}=&3\left(g^{\MO}\right)^4,\quad F^{\MO}_{3,1}=\frac{149}{4}\left(g^{\MO}\right)^4,\nn
F^{\zeta_4\mo}_{3,0}=\frac{9}{4}\left(g^{\MO}\right)^4,\quad F^{\zeta_3\mo}_{3,0}=&8\left(g^{\MO}\right)^4,\quad F^{\MO}_{3,0}=\frac{29}{8}\left(g^{\MO}\right)^4,\nn
F^{\zeta_4\MO}_{4,1}=\frac{135}{8}\left(g^{\MO}\right)^5,\quad F^{\zeta_3\MO}_{4,1}=&\frac{291}{4}\left(g^{\MO}\right)^5,\quad F^{\MO}_{4,1}=\frac{723}{16}\left(g^{\MO}\right)^5,\nn
F^{\zeta_4\MO}_{4,0}=\frac{297}{16}\left(g^{\MO}\right)^5,\quad F^{\zeta_3\MO}_{4,0}=&\frac{105}{8}\left(g^{\MO}\right)^5,\quad F^{\MO}_{4,0}=-\frac{93}{4}\left(g^{\MO}\right)^5,\nn
F^{\zeta_4\MO}_{5,1}=\frac{14067}{80}\left(g^{\MO}\right)^6,\quad& F^{\zeta_3\MO}_{5,1}=\frac{10287}{40}\left(g^{\MO}\right)^6,\nn
F^{\MO}_{5,1}=&-\frac{5049}{40}\left(g^{\MO}\right)^6,
\label{FMOMres}
\end{align}
We then find
\begin{align}
\tbeta^{\MO}_1=&F^{\mo}_{1,0}=3\left(g^{\MO}\right)^2,\nn
\tbeta^{\MO}_2=&2F^{\MO}_{2,0}-2\tbeta_1^{\MO} F^{\mo}_{1,0}=-\frac{21}{2}\left(g^{\MO}\right)^3,\nn
\tbeta^{\MO}_3=&3F^{\MO}_{3,0}-2\tbeta_2^{\MO} F^{\mo}_{1,0}-3\tbeta_1 F^{\MO}_{2,0}=\frac{321}{8}\left(g^{\MO}\right)^4\nn
\tbeta^{\zeta_3\MO}_3=&3F^{\zeta_3\mo}_{3,0}=24\left(g^{\MO}\right)^4,\quad \tbeta^{\zeta_4\MO}_3=3F^{\zeta_4\mo}_{3,0}=\frac{27}{4}\left(g^{\MO}\right)^4,\nn
\tbeta^{\MO}_4=&4F^{\MO}_{4,0}-4\tbeta^{\MO}_1 F^{\MO}_{3,0}-3\tbeta^{\MO}_2 F^{\MO}_{2,0}
-2\tbeta^{\MO}_3 F^{\mo}_{1,0}\nn
&=-\frac{2073}{8}\left(g^{\MO}\right)^5,\nn
\tbeta^{\zeta_3\MO}_4=&4F^{\zeta_3\MO}_{4,0}-4\tbeta^{\MO}_1 F^{\zeta_3\mo}_{3,0}
-2\tbeta^{\zeta_3\MO}_3 F^{\mo}_{1,0}=-\frac{375}{2}\left(g^{\MO}\right)^5,\nn
\tbeta^{\zeta_4\MO}_4=&4F^{\zeta_4\MO}_{4,0}-4\tbeta^{\MO}_1 F^{\zeta_4\mo}_{3,0}
-2\tbeta^{\zeta_4\MO}_3 F^{\mo}_{1,0}=\frac{27}{4}\left(g^{\MO}\right)^5,
\label{tbetMOM}
\end{align}
where we have inserted values from Eq.~\eqref{FMOMres} to obtain explicit values. 
We also have, bearing in mind that within $\MO$ we are subtracting all finite parts,
\begin{align}
\beta^{\MO}_1=&F^{\mo}_{1,1},\nn
\beta^{\MO}_2=&2F^{\MO}_{2,1}-2\beta_1^{\mo} F^{\mo}_{1,0}-2\tbeta^{\MO}_1 F^{\mo}_{1,1},\nn
\beta^{\MO}_3=&3F^{\MO}_{3,1}-2\beta_2^{\MO} F^{\mo}_{1,0}-3\tbeta^{\MO}_1 F^{\MO}_{2,1}\nn
&-2\tbeta_2^{\MO} F^{\mo}_{1,0}-3\beta^{\mo}_1 F^{\MO}_{2,0},\nn
\beta^{\zeta_3\MO}_3=&3F^{\zeta_3\mo}_{3,1},\quad \beta^{\zeta_4\MO}_3=0,\nn
\beta^{\MO}_4=&4F^{\MO}_{4,1}-2\beta^{\MO}_3 F^{\mo}_{1,0}-4\tbeta^{\MO}_1 F^{\MO}_{3,1}
\nn
&-3\beta^{\MO}_2 F^{\MO}_{2,0}-3\tbeta^{\MO}_2 F^{\MO}_{2,1}
-2\tbeta^{\MO}_3 F^{\mo}_{1,1}-4\beta^{\mo}_1 F^{\MO}_{3,0},\nn
\beta^{\zeta_3\MO}_4=&4F^{\zeta_3\MO}_{4,1}-2\beta^{\zeta_3\MO}_3 F^{\mo}_{1,0}-4\tbeta^{\MO}_1 F^{\zeta_3\mo}_{3,1}
\nn
&-2\tbeta^{\zeta_3\MO}_3 F^{\mo}_{1,1}-4\beta^{\mo}_1 F^{\zeta_3\mo}_{3,0},\nn
\beta^{\zeta_4\MO}_4=&4F^{\zeta_4\MO}_{4,1}
-2\tbeta^{\zeta_4\MO}_3 F^{\mo}_{1,1}-4\beta^{\mo}_1 F^{\zeta_4\mo}_{3,0},\nn
\beta^{\MO}_5=&5F^{\MO}_{5,1}
-2\tbeta^{\MO}_4 F^{\mo}_{1,1}-5\beta^{\MO}_1 F^{\MO}_{4,0}\nn
&-5\tbeta_1^{\MO} F^{\MO}_{4,1}-3\tbeta^{\MO}_3 F^{\MO}_{2,1}-4\beta^{\MO}_2 F^{\MO}_{3,0},\nn
\beta^{\zeta_3\MO}_5=&5F^{\zeta_3\MO}_{5,1}
-2\tbeta^{\zeta_3\MO}_4 F^{\mo}_{1,1}-5\beta^{\mo}_1 F^{\zeta_3\MO}_{4,0}\nn
&-5\tbeta_1^{\MO} F^{\zeta_3\MO}_{4,1}-3\tbeta^{\zeta_3\MO}_3 F^{\MO}_{2,1}-4\beta^{\MO}_2 F^{\zeta_3\mo}_{3,0},\nn
&-2\beta^{\zeta_3\MO}_4 F^{\mo}_{1,0}-3\beta^{\zeta_3\MO}_3 F^{\MO}_{2,0}-4\tbeta^{\MO}_2 F^{\zeta_3\mo}_{3,1},\nn
\beta^{\zeta_4\MO}_5=&5F^{\zeta_4\MO}_{5,1}
-2\tbeta^{\zeta_4\MO}_4 F^{\mo}_{1,1}-5\beta^{\mo}_1 F^{\zeta_4\MO}_{4,0}\nn
&-5\tbeta_1^{\MO} F^{\zeta_4\MO}_{4,1}-3\tbeta^{\zeta_4\MO}_3 F^{\MO}_{2,1}-4\beta^{\MO}_2 F^{\zeta_4\mo}_{3,0},
\end{align}
Inserting the values in Eq.~\eqref{tbetMOM} together with those of Eq.~\eqref{FMOMres}, we indeed reproduce the results of Eqs.~\eqref{betMOM}, \eqref{gamMOM}.

 \section{Variations in the SUSY case}\label{susyvar}
In this appendix we describe the scheme variations of the anomalous dimension coefficients in the supersymmetric Wess-Zumino model. We give general results which in particular may be used for the transformation from the $\msbar$ scheme to the $\MO$ scheme. In the main body of the text we are only interested in the even-$\zeta$ contributions, but here for completeness we give the full results which confirm the transformations detailed in Ref.~\cite{gracey}. For those who want to skip the details, the four-loop transformations from $\msbar$ to $\MO$ may be obtained by combining Eqs.~\eqref{susyvars4} with Eqs.~\eqref{xthree}, and the five-loop results are given by combining Eqs~\eqref{susyvars} with Eq.~\eqref{MOMX}. Furthermore, by specialisation the results may also be used to derive the transformation from the $\msbar$ scheme to the $\MOp$ scheme, and this is given at five loops by combining Eqs~\eqref{susyvars} with Eq.~\eqref{MOMp}.

As described in the main text, a formalism for scheme redefinitions was introduced in Ref.~\cite{scheme} with particular reference to the Wess-Zumino model, and elaborated in Refs.~\cite{jot} and \cite{JP}. The transformations are expressed in terms of a small set of simpler units, which at higher loops are themselves recursively expressed in terms of lower-loop quantities.
In our case, at three loops we have for the variations of the coefficients defined in Eq.~\eqref{gam3}
\begin{align}
\delta c_{31}=&0,\nn
\delta c_{32}=&2\Xhat_{1,21},\nn
\delta c_{33}=&4\Xhat_{1,21},\nn
\delta c_{34}=&0,
\end{align}
where
\be
\Xhat_{1,21}=c_1\ehat_{21}-\epsilon_1c_{21},
\label{xhat}
\ee
and with
\be
\ehat_{21}=\epsilon_{21}-\frac124\epsilon_1^2.
\ee
Relations like these and similar later ones at higher loops are valid for transformations between any pair of schemes. The values $c_1$ etc are computed in the initial scheme, while the $\epsilon$ quantities depend on both the initial and final scheme. In this section we are always thinking of the initial scheme as $\msbar$, with the values of the coefficients $c_{1}^{\msbar}$ etc as given earlier in Eq.~\eqref{c3WZ}. We will later give results for the particular values of the $\epsilon$ (and derived quantities such as the $\Xhat$) appropriate to the special cases of final schemes $\MOp$ and $\MO$, denoted by superscripts $A$ and $B$ respectively.
The form of the quantity $\Xhat_{1,21}$ reflects the commutator structure in the scheme redefinition formula in Eqs.~\eqref{blin}, \eqref{comdef}, but also incorporates higher-order corrections in a somewhat natural way, in this case through $\ehat_{21}$ (as mentioned earlier, we use the ``hat'' to signal the presence of these higher-order corrections, as explained in more detail later). $\Xhat_{1,21}$ has a natural interpretation in terms of the insertion of one-loop subgraphs in two-loop graphs, and vice-versa.
At four loops we have for the variations of the coefficients defined in Eq.~\eqref{gam4}
\begin{align}
\delta c_{41}=&0,\nn
\delta c_{42}=&4X_{1,31},\nn
\delta c_{43}=&4X_{1,31},\nn
\delta c_{44}=&8X_{1,31},\nn
\delta c_{45}=&8\Xhat_{1,32}+2\Xhat_{1,33},\nn
\delta c_{46}=&8\Xhat_{1,32}+2\Xhat_{1,34},\nn
\delta c_{47}=&-4X_{1,31},\nn
\delta c_{48}=&-4\Xhat_{1,32}+2\Xhat_{1,34},\nn
\delta c_{49}=&4\Xhat_{1,33}+2\Xhat_{1,34},\nn
\delta c_{410}=&4\Xhat_{1,33}+2\Xhat_{1,34},\nn
\delta c_{411}=&6\Xhat_{1,33},\nn
\delta c_{412}=&-4\Xhat_{1,33}+4\Xhat_{1,34},\nn
\delta c_{413}=&0.
\label{susyvars4}
\end{align}
where
\be
X_{1,31}=c_1\epsilon_{31}-\epsilon_1c_{31},
\label{X31def}
\ee
and
\be
\Xhat_{1,32}=c_1\ehat_{32}-\epsilon_1\chat_{32},
\label{X32def}
\ee
with similar definitions for $\Xhat_{1,33}$ and $\Xhat_{1,34}$, and where
\begin{align}
\ehat_{32}=&\epsilon_{32}-\epsilon_1\ehat_{21}-\frac43\epsilon_1^3,\nn
\ehat_{33}=&\epsilon_{33}-2\epsilon_1\ehat_{21}-\frac{8}{3}\epsilon_1^3,\nn
\ehat_{34}=&\epsilon_{34}-4\epsilon_1\ehat_{21}-\frac{8}{3}\epsilon_1^3,
\label{e3defa}
\end{align}
and
\begin{align}
\chat_{32}=&c_{32}+\frac122\Xhat_{1,21},\nn
\chat_{33}=&c_{33}+\frac124\Xhat_{1,21},\nn
\chat_{34}=&c_{34},
\label{chat}
\end{align}
with  the $\msbar$ values of the coefficients $c_{31}^{\msbar}$ etc as given earlier in Eq.~\eqref{c3WZ}, and $\Xhat_{1,21}$ again defined by Eq.~\eqref{xhat}. Using the values up to three loops (which may be read off from the finite parts of the $\msbar$ results in Eq.~\eqref{zms3})
\begin{align}
\epsilon^B_1=-\frac12,\quad \epsilon^B_{21}=&\frac{11}{8}\nn
\epsilon^B_{31}=&-\frac32\zeta_3-\frac38\zeta_4,\nn
\epsilon^B_{32}=&-\frac{37}{48}+\frac{1}{12}\zeta_3,\nn
\epsilon^B_{33}=&2\epsilon^B_{32}-\frac14,\nn
\epsilon^B_{34}=&-\frac{17}{4}+\frac16\zeta_3,
\label{e3defb}
\end{align}
we find
\begin{align}
\Xhat^B_{1,21}=&\frac{3}{16},\nn
\Xhat^B_{1,31}=&-\frac{3}{16}\zeta_4,\nn
\Xhat^B_{1,32}=&-\frac{5}{96}+\frac{1}{24}\zeta_3,\nn
\Xhat^B_{1,33}=&-\frac{5}{48}-\frac{1}{24}\zeta_3,\nn
\Xhat^B_{1,34}=&-\frac{7}{12}+\frac{1}{12}\zeta_3.
\label{xthree}
\end{align}
At five loops (with the coefficients as defined in Ref.~\cite{gracey}) we have
\begin{align}
\delta c_{51}=&0,\nn
\delta c_{52}=\delta c_{53}=&2\Xhat_{1,42},\nn
\delta c_{54}=&4\Xhat_{1,45},\nn
\delta c_{55}=&8\Xhat_{1,42},\nn
\delta c_{56}=&4\Xhat_{1,45}+4\Xhat_{1,46}+2\Xhat_{2,33},\nn
\delta c_{57}=\delta c_{58}=&4\Xhat_{1,42}+2\Xhat_{1,44},\nn
\delta c_{59}=\delta c_{510}=&4X_{1,41},\nn
\delta c_{512}=&4X_{1,41},\nn
\delta c_{513}=&2\Xhat_{1,44},\nn
\delta c_{515}=&4\Xhat_{1,46}+2\Xhat_{2,34},\nn
\delta c_{516}=\delta c_{517}=&4X_{1,41},\nn
\delta c_{518}=&2X_{1,41},\nn
\delta c_{521}=\delta c_{522}=&2\Xhat_{1,44},\nn
\delta c_{523}=\delta c_{524}=&4\Xhat_{2,31}+4\Xhat_{1,42},\nn
\delta c_{525}=\delta c_{526}=&4\Xhat_{1,42}+2\Xhat_{1,44},\nn
\delta c_{527}=\delta c_{528}=&4\Xhat_{1,42},\nn
\delta c_{529}=&4\Xhat_{1,44},\nn
\delta c_{530}=&8\Xhat_{2,31}+4\Xhat_{1,44},\nn
\delta c_{531}=&-2\Xhat_{2,31}+2\Xhat_{1,47},\nn
\delta c_{532}=&2\Xhat_{1,46}+2\Xhat_{1,48}-2\Xhat_{2,32},\nn
\delta c_{533}=\delta c_{534}=&4\Xhat_{1,45}+2\Xhat_{1,46}+2\Xhat_{1,410}+4\Xhat_{2,32},\nn
\delta c_{535}=&6\Xhat_{1,45}+2\Xhat_{1,411},\nn
\delta c_{536}=&4\Xhat_{1,46}+2\Xhat_{1,412}-2\Xhat_{2,33},\nn
\delta c_{537}=&4\Xhat_{1,46}+2\Xhat_{1,413}+8\Xhat_{2,32}-2\Xhat_{2,34},\nn
\delta c_{538}=&8\Xhat_{1,49}+4\Xhat_{2,34},\nn
\delta c_{539}=&-4X_{1,41},\nn
\delta c_{540}=\delta c_{541}=&-4\Xhat_{1,42}+4\Xhat_{1,47},\nn
\delta c_{542}=&-4\Xhat_{1,44}+8\Xhat_{1,47},\nn
\delta c_{543}=&8\Xhat_{1,48}+2\Xhat_{1,412}-4\Xhat_{1,45},\nn
\delta c_{544}=&8\Xhat_{1,48}+2\Xhat_{1,413}+2\Xhat_{2,34}-4\Xhat_{1,46},\nn
\delta c_{545}=\delta c_{546}=&-2\Xhat_{2,31}+2\Xhat_{1,47},\nn
\delta c_{547}=\delta c_{548}=&2\Xhat_{1,48}+2\Xhat_{1,49}-2\Xhat_{2,32},\nn
\delta c_{549}=\delta c_{550}=&4\Xhat_{1,49}+4\Xhat_{1,411}+2\Xhat_{2,33},\nn
\delta c_{551}=\delta c_{552}=&2\Xhat_{1,412}+4\Xhat_{1,49}-2\Xhat_{2,33},\nn
\delta c_{553}=\delta c_{554}=&2\Xhat_{1,413}+4\Xhat_{1,49}+4\Xhat_{2,33}-2\Xhat_{2,34},\nn
\delta c_{555}=&8\Xhat_{1,411},\nn
\delta c_{556}=&4\Xhat_{1,49}+4\Xhat_{1,411}+2\Xhat_{2,33},\nn
\delta c_{557}=&-4\Xhat_{2,31}-4\Xhat_{1,47},\nn
\delta c_{558}=&2\Xhat_{1,413}-4\Xhat_{2,32}-4\Xhat_{1,48},\nn
\delta c_{559}=\delta c_{560}=&4\Xhat_{1,412}+2\Xhat_{1,413}+2\Xhat_{2,34}-4\Xhat_{1,49},\nn
\delta c_{561}=&6\Xhat_{1,412}-4\Xhat_{1,411},\nn
\delta c_{562}=&4\Xhat_{1,413}-4\Xhat_{2,33}-4\Xhat_{1,412},\nn
\delta c_{563}=&0,
\label{susyvars}
\end{align}
where $X_{1,41}$ is defined by
\be
X_{1,41}=c_1\epsilon_{41}-\epsilon_1c_{41},
\label{nlla1}
\ee
and quantities like $\Xhat_{1,42}$ are defined by
\be
\Xhat_{1,42}=c_1\ehat_{42}-\epsilon_1\chat_{42},
\label{nlla2}
\ee
with similar definitions for $\Xhat_{1,43}-\Xhat_{1,413}$ and with
\begin{align}
\ehat_{42}=&\epsilon_{42}-2\epsilon_1\epsilon_{31},\nn
\ehat_{43}=&\epsilon_{43}-2\epsilon_1\epsilon_{31},\nn
\ehat_{44}=&\epsilon_{44}-4\epsilon_1\epsilon_{31},\nn
\ehat_{45}=&\epsilon_{45}-4\epsilon_1\ehat_{32}-\epsilon_1\ehat_{33}-4\epsilon_1^2\ehat_{21}-4\epsilon_1^4,\nn
\ehat_{46}=&\epsilon_{46}-4\epsilon_1\ehat_{32}-\epsilon_1\ehat_{34}-\ehat_{21}^2-\frac{20}{3}\epsilon_1^2\ehat_{21}-4\epsilon_1^4,\nn
\ehat_{47}=&\epsilon_{47}-2\epsilon_1\epsilon_{31},\nn
\ehat_{48}=&\epsilon_{48}-2\epsilon_1\ehat_{32}-\epsilon_1\ehat_{34}-\frac83\epsilon_1^2\ehat_{21}-\frac43\epsilon_1^4,\nn
\ehat_{49}=&\epsilon_{49}-2\epsilon_1\ehat_{33}-\epsilon_1\ehat_{34}-\ehat_{21}^2-\frac{20}{3}\epsilon_1^2\ehat_{21}-4\epsilon_1^4,\nn
\ehat_{410}=&\epsilon_{410}-2\epsilon_1\ehat_{33}-\epsilon_1\ehat_{34}-\ehat_{21}^2-\frac{20}{3}\epsilon_1^2\ehat_{21}-4\epsilon_1^4,\nn
\ehat_{411}=&\epsilon_{411}-3\epsilon_1\ehat_{33}-4\epsilon_1^2\ehat_{21}-4\epsilon_1^4,\nn
\ehat_{412}=&\epsilon_{412}-2\epsilon_1\ehat_{33}-2\epsilon_1\ehat_{34}-\frac{16}{3}\epsilon_1^2\ehat_{21}-\frac83\epsilon_1^4,\nn
\ehat_{413}=&\epsilon_{413}-4\epsilon_1\ehat_{34}-2\ehat_{21}^2-8\epsilon_1^2\ehat_{21}-\frac83\epsilon_1^4.
\label{ehats}
\end{align}
and
\begin{align}
\chat_{42}=c^{\msbar}_{42}+\frac12\delta' c_{42},
\label{nllb}
\end{align}
with similar definitions for $\chat_{43}-\chat_{413}$.
Here, the $\msbar$ values of the coefficients $c_{41}^{\msbar}$ etc are as given earlier in Eq.~\eqref{c4WZ}, and $\delta' c_{42}$ is defined similarly to $\delta c_{42}$ in Eq.~\eqref{susyvars4}, but with hatted quantities such as $\Xhat_{1,32}$ replaced by tilded quantities such as $\Xtil_{1,32}$, defined by
\be
\Xtil_{1,32}=c_1\epsilon_{32}-\epsilon_1\ctil_{32},
\ee
with similar equations for $\Xtil_{1,32}$, $\Xtil_{1,33}$, $\Xtil_{1,34}$,
and with 
\begin{align}
\ctil_{31}=&c_{31},\nn
\ctil_{32}=&c_{32}+\frac23\Xhat_{1,21},\nn
\ctil_{33}=&c_{33}+\frac43\Xhat_{1,21},\nn
\ctil_{34}=&c_{34},
\label{ctil}
\end{align}
with $\Xhat_{1,21}$ as defined in Eq.~\eqref{xhat}. So the difference between the hatted and tilded quantities is simply the combinatoric factors in Eqs.~\eqref{chat}, \eqref{ctil}. Finally, in analogy with Eqs.~\eqref{X31def}, \eqref{X32def}, we have
\be
\Xhat_{2,31}=c_{21}\epsilon_{31}-\ehat_{21}c_{31},
\label{Xa}
\ee
and
\be
\Xhat_{2,32}=c_{21}\ehat_{32}-\ehat_{21}\chat_{32},
\label{Xb}
\ee
with similar expressions for $\Xhat_{2,33}$ and $\Xhat_{2,34}$.
The four-loop values of the $\epsilon$ quantities within $\msbar$ may be read off from the finite parts of the $\msbar$ results in Eq.~\eqref{zms4}, and we find
\begin{align}
\epsilon^B_{41}=&\frac{35}{4}\zeta_5+\frac{25}{8}\zeta_6-\frac14\zeta_3^2,\nn
\epsilon^B_{42}=&\frac{59}{16}\zeta_3+\frac{9}{32}\zeta_4,\nn
\epsilon^B_{45}=&\frac{169}{64}-\frac{23}{32}\zeta_3-\frac{3}{64}\zeta_4,\nn
\epsilon^B_{46}=&\frac{971}{192}-\frac{11}{12}\zeta_3,\nn
\epsilon^B_{47}=&\frac{67}{16}\zeta_3+\frac{33}{32}\zeta_4,\nn
\epsilon^B_{48}=&\frac{523}{192}-\frac13\zeta_3,\nn
\epsilon^B_{49}=\epsilon^B_{410}=&\epsilon^B_{46}+\zeta_3,\nn
\epsilon^B_{411}=&\epsilon^B_{45}+\frac34\zeta_3,\nn
\epsilon^B_{412}=&2\epsilon^B_{48}+\frac{13}{16}\zeta_3+\frac{3}{32}\zeta_4,\nn
\epsilon^B_{413}=&\frac{447}{32}-\frac{5}{6}\zeta_3.
\label{eps4vals}
\end{align}
Combining with the four-loop coefficients in Eq.~\eqref{c4WZ}, and with lower-order results used already, we find from Eqs.~\eqref{nlla1}-\eqref{ctil} for the values of the $\Xhat$ appropriate to the transformation from $\msbar$ to $\MO$
\begin{align}
X^{B}_{1,41}=&-\frac{5}{8}\zeta_5+\frac{25}{16}\zeta_6-\frac18\zeta_3^2,\nn
\Xhat^{B}_{1,42}=\Xhat^{B}_{1,43}=&\frac{11}{32}\zeta_3+\frac{9}{64}\zeta_4,\nn
\Xhat^{B}_{1,44}=&2\Xhat^{B}_{1,42},\nn
\Xhat^{B}_{1,45}=&\frac{25}{128}-\frac{7}{64}\zeta_3-\frac{3}{128}\zeta_4,\nn
\Xhat^{B}_{1,46}=&\frac{7}{24}-\frac{5}{24}\zeta_3,\nn
\Xhat^{B}_{1,47}=&\frac{19}{32}\zeta_3+\frac{9}{64}\zeta_4,\nn
\Xhat^{B}_{1,48}=&\frac{115}{384}-\frac{1}{12}\zeta_3,\nn
\Xhat^{B}_{1,49}=\Xhat^{B}_{1,410}=&\frac{7}{24}+\frac{1}{24}\zeta_3,\nn
\Xhat^{B}_{1,411}=&\frac{25}{128}+\frac{1}{64}\zeta_3-\frac{3}{128}\zeta_4,\nn
\Xhat^{B}_{1,412}=&\frac{115}{192}-\frac{1}{96}\zeta_3+\frac{3}{64}\zeta_4,\nn
\Xhat^{B}_{1,413}=&\frac{59}{32}-\frac14\zeta_3,\nn
\Xhat^{B}_{2,31}=&-\frac{9}{16}\zeta_3+\frac{3}{16}\zeta_4,\nn
\Xhat^{B}_{2,32}=&\frac{11}{384}-\frac{1}{24}\zeta_3,\nn
\Xhat^{B}_{2,33}=&\frac{11}{192}+\frac{1}{24}\zeta_3,\nn
\Xhat^{B}_{2,34}=&\frac{5}{24}-\frac{1}{12}\zeta_3,
\label{MOMX}
\end{align}
Substituting into Eq.~\eqref{susyvars}, we readily find that these variations transform the 5-loop $\msbar$ results in Ref.~\cite{gracey} into the $\MO$ results also given there.
 
In the main text we are also interested in the five-loop transformation from $\msbar$ to $\MOp$. The values of the $\Xhat$ appropriate to the transformation to $\MOp$ are simply given by
\begin{align}
X^{A}_{1,41}=&\frac{25}{16}\zeta_6,\nn
\Xhat^{A}_{1,42}=\Xhat^{A}_{1,43}=&\frac{9}{64}\zeta_4,\nn
\Xhat^{A}_{1,44}=&2\Xhat^{A}_{1,42},\nn
\Xhat^{A}_{1,45}=&-\frac{3}{128}\zeta_4,\nn
\Xhat^{A}_{1,46}=&0,\nn
\Xhat^{A}_{1,47}=&\frac{9}{64}\zeta_4,\nn
\Xhat^{A}_{1,48}=&0,\nn
\Xhat^{A}_{1,49}=\Xhat^{A}_{1,410}=&0,\nn
\Xhat^{A}_{1,411}=&-\frac{3}{128}\zeta_4,\nn
\Xhat^{A}_{1,412}=&\frac{3}{64}\zeta_4,\nn
\Xhat^{A}_{1,413}=&0,\nn
\Xhat^{A}_{2,31}=&\frac{3}{16}\zeta_4,\nn
\Xhat^{A}_{2,32}=\Xhat^{A}_{2,33}=\Xhat^{A}_{2,34}=&0.
\label{MOMp}
\end{align}
in other words simply the even-$\zeta$ terms from Eq.~\eqref{MOMX}.

\section{Transformation from $\msbar$}\label{mstrans}
In Sect. 2 we have taken the results of Ref.~\cite{baikov3}, relating even-$\zeta$ to odd-$\zeta$ $\beta$-functions, and demonstrated a transformation to a new renormalisation scheme in which there were no even-$\zeta$ terms in the $\beta$-functions; at least in the case of AD theories in which the $\beta$ functions could be expressed in terms of the anomalous dimensions. We also argued that this scheme was a variant of MOM which we called $\rm MOM'$, but that the same cancellation of even-$\zeta$ terms was expected in MOM. However, all this relied on relations such as that given in Eq.~\eqref{FMOM} between $F^{\zeta_3\msbar}_{L,1}$ and $F^{\zeta_4\MOp}_{L,0}$. Furthermore the results of Ref.~\cite{baikov3} have only been derived for single-coupling theories. In this Appendix we show how the results of Ref.~\cite{baikov3} and also the calculations of Sect. 3 may be generalised to the multi-coupling case, at least at low orders; and we sketch out an all-orders proof of Eq.~\eqref{FMOM} and other similar relations required in Section 3, again for the multi-coupling case. We emphasise, however, that once again we are thinking of an AD theory.

We start by showing in some detail how to derive results analogous to the four and five loop expressions for the $\zeta_4$ $\beta$-function contributions in Eqs.~\eqref{z4:a} and \eqref{z4:b} for a general theory. We then go on to show how these  $\zeta_4$ contributions may be removed by a transition to the $\MOp$ scheme, and also the $\MO$ scheme. This will provide a model for the extension to higher loops and other even $\zeta$s. Starting at four loops, then, we have
\be
F^{\zeta_3\msbar}_{4,2}=\frac14\left(3F^{\zeta_3\msba}_{3,1}\cdot F^{\msba}_{1,1}+F^{\msba}_{1,1}\cdot F^{\zeta_3\msba}_{3,1}\right)
\label{MS2}
\ee
and so  
\be
G^{\zeta_3\msbar}_{4,2}=F^{\zeta_3\msbar}_{4,2}-F^{\zeta_3\msba}_{3,1}\cdot F^{\msba}_{1,1}=\frac14[F^{\msba}_{1,1},F^{\zeta_3\msba}_{3,1}].
\label{MS2a}
\ee
We also have
\begin{subequations}\label{MS3:main}
\begin{align}
G^{\zeta_3\msba}_{3,1}=&F^{\zeta_3\msba}_{3,1},\label{MS3:a}\\
G^{\zeta_4\msbar}_{4,1}=&F^{\zeta_4\msbar}_{4,1}=\frac14\beta^{\zeta_4\msbar}_4.\label{MS3:b}
\end{align}
\end{subequations}
We then find combining Eqs.~\eqref{MS2a}, \eqref{MS3:b}, and using Eq.~\eqref{Grels:a} for $m=1$ to relate $G^{\zeta_3\msbar}_{4,2}$ and $G^{\zeta_4\msbar}_{4,1}$,
\be
\beta^{\zeta_4\msbar}_4=\frac34[\beta_1,F^{\zeta_3\msba}_{3,1}]=\frac14[\beta_1,\beta^{\zeta_3\msbar}_{3}].
\label{MS3a}
\ee
Eq.~\eqref{MS3a} is the generalisation to the many coupling case of Eqs.~\eqref{z4:a}, derived in Ref.~\cite{baikov3}. 
The same argument that led from Eqs.~\eqref{bet3} to \eqref{delone} then implies that the four-loop $\zeta_4$ term in the $\beta$-function may be removed by a coupling redefinition $\delta g^{(3)}=-F^{\zeta_4\msbar}_{3,0}$; the generalisation to the multi-coupling case is somewhat trivial. Also, at this level there is no distinction between the $\msbar$ and $\MOp$ values for $F^{\zeta_4\msbar}_{3,0}$. This coupling redefinition precisely effects the change to $\MOp$; we shall see this in more detail as
we proceed to five loops, where we shall find some additional subtleties. We have at five loops
\be
0=5F^{\zeta_3\msbar}_{5,2}-4F^{\zeta_3\msbar}_{4,1}\cdot F^{\msba}_{1,1}-3F^{\zeta_3\msba}_{3,1}\cdot F^{\msbar}_{2,1}-F^{\msba}_{1,1}\cdot F^{\zeta_3\msbar}_{4,1}-2F^{\msbar}_{2,1}\cdot F^{\zeta_3\msba}_{3,1}
\label{MS4}
\ee
and
\begin{subequations}\label{MS5:main}
\begin{align}
G^{\zeta_4\msbar}_{5,1}=&F^{\zeta_4\msbar}_{5,1}-F^{\zeta_4\msbar}_{4,1}\cdot F^{\msba}_{1,0},\label{MS5:a}\\
G^{\zeta_3\msbar}_{5,2}=&F^{\zeta_3\msbar}_{5,2}-F^{\zeta_3\msbar}_{4,1}\cdot F^{\msba}_{1,1}-F^{\zeta_3\msba}_{3,1}\cdot F^{\msbar}_{2,1}-G^{\zeta_3\msbar}_{4,2}\cdot F^{\msba}_{1,0},\label{MS5:b}\\
G^{\zeta_3\msbar}_{4,1}=&F^{\zeta_3\msbar}_{4,1}-F^{\zeta_3\msba}_{3,1}\cdot F^{\msba}_{1,0},\label{MS5:c}\\
G^{\zeta_4\msbar}_{4,0}=&F^{\zeta_4\msbar}_{4,0},\label{MS5:d}
\end{align}
\end{subequations}
where $G^{\zeta_3\msbar}_{4,2}$ is defined in Eq.~\eqref{MS2a} and appears for reasons similar to those discussed after Eq.~\eqref{grela:main}.
 Combining Eqs.~\eqref{MS4} and \eqref{MS5:b}, we find
\be
5G^{\zeta_3\msbar}_{5,2}=[F^{\msba}_{1,1},F^{\zeta_3\msbar}_{4,1}]+2[F^{\msbar}_{2,1},F^{\zeta_3\msba}_{3,1}]-5G^{\zeta_3\msbar}_{4,2}\cdot F^{\msba}_{1,0}.
\label{MS6}
\ee
We then find using Eqs.~\eqref{MS5:a}, \eqref{MS6}, \eqref{Grels:a} and \eqref{MS3:b}
\begin{align}
\beta^{\zeta_4\msbar}_5=&5F^{\zeta_4\msbar}_{5,1},\nn
=&\frac34[\beta_1,F^{\zeta_3\msbar}_{4,1}]+\frac34[\beta_2,F^{\zeta_3\msba}_{3,1}]
=\frac{3}{16}[\beta_1,\beta^{\zeta_3\msbar}_{4}]+\frac14[\beta_2,\beta^{\zeta_3\msbar}_{3}]
\label{MS7}
\end{align}
Eq.~\eqref{MS7} is the generalisation to the many coupling case of Eq.~\eqref{z4:b}, derived in Ref.~\cite{baikov3}. We would like to go on and show that $\beta^{\zeta_4}$ vanishes in general up to five loops in $\rm MOM'$. As for the previous loop order, we would now like to relate $F^{\zeta_3\msbar}_{4,1}$ to a finite quantity which may be interpreted in terms of a scheme change to $\MOp$.  As we have stressed, it is the $G$ quantities which are related at different  orders in $\epsilon$ via Eq.~\eqref{Grels:main}, so we shall need to use Eq.~\eqref{FGrel:main}. But notice the difference between Eqs.~\eqref{FGrel:a} and \eqref{FGrel:b}, due to the fact that there is no subtraction of finite terms in $\msbar$. This seems likely to create a mismatch which would spoil the simplicity of the scheme redefinitions. We are saved by the fact that in discussing the redefinition required to access $\MOp$, we need to start within the $\MOp$ scheme itself; and in the $\MOp$ scheme, the analogue of Eq.~\eqref{FGrel:b}, i.e. Eq.~\eqref{grela:d}, has the same form as Eq.~\eqref{FGrel:a}. Let us see how this works in detail. We start with deriving the required coupling redefinition to take us from $\msbar$ to $\MOp$. In the $\MOp$ scheme we include finite parts in the even-$\zeta$ contributions to the bare coupling, while in the $\msbar$ scheme there are no finite terms in the bare coupling, and so the relation between the $\rm MOM'$ and $\msbar$ couplings is given by
\be
g^{\msbar}=g^{\MOp}+F^{\zeta_4\mop}_{3,0}\zeta_4+F^{\zeta_4\MOp}_{4,0}\zeta_4+\ldots
\label{MS9}
\ee
which may be rewritten
\be
g^{\MOp}=g^{\msbar}-F^{\zeta_4\msba}_{3,0}\zeta_4-F^{\zeta_4\MOp}_{4,0}(g^{\msbar})\zeta_4+\ldots,
\label{MS10p}
\ee
i.e.
\begin{align}
g^{\MOp}=&g^{\msbar}+\delta g(g^{\msbar}),\nn
\delta g(g^{\msbar})=&-F^{\zeta_4\msba}_{3,0}\zeta_4-F^{\zeta_4\MOp}_{4,0}(g^{\msbar})\zeta_4+\ldots
\label{MS10}
\end{align}

In going from Eq.~\eqref{MS9} to \eqref{MS10p} we have changed the argument in $F^{\zeta_4\mop}_{3,0}$ and $F^{\zeta_4\MOp}_{4,0}$ from $g^{\MOp}$ to $g^{\msbar}$. Since, as we see from Eq.~\eqref{MS9}, $g^{\MOp}=g^{\msbar}+O(\zeta_4)$, this may be done at the cost of introducing $O(\zeta_4)^2$ terms which do not concern us at present and are subsumed into the ellipsis (they will however be considered in more detail at the end of this Appendix). Our derivation of Eq.~\eqref{MS10} may appear somewhat longwinded, and indeed one might feel that one could have guessed the result; but we shall see later that it difficult to see how one could correctly guess both the $\MOp$ and the $\MO$ transformations.

We now go on to relate $F^{\zeta_3\msbar}_{4,1}$ as in Eq.~\eqref{MS7} to the coupling redefinition in Eq.~\eqref{MS10}. We find using Eqs.~\eqref{Grels:a}, \eqref{MS3:a}, \eqref{MS5:c}
\begin{subequations}\label{MS10f:main}
\begin{align}
\frac34F^{\zeta_3\msba}_{3,1}=&\frac34G^{\zeta_3\msba}_{3,1}=G^{\zeta_4\msba}_{3,0},\label{MS10f:a}\\
\frac34F^{\zeta_3\msbar}_{4,1}=&\frac34\left(G^{\zeta_3\msbar}_{4,1}+G^{\zeta_3\msba}_{3,1}\cdot F^{\msba}_{1,0}
\right)\nn
=&G^{\zeta_4\msbar}_{4,0}+G^{\zeta_4\msba}_{3,0}\cdot F^{\msba}_{1,0}.\label{MS10f:b}
\end{align}
\end{subequations}
Quantities such as $G_{4,0}^{\zeta_4\msbar}$ are the result of omitting the even-$\zeta$ dependent counterterms (the sole source of difference between the $\msbar$ and $\MOp$ schemes) from $F_{4,0}^{\zeta_4\msbar}$ and therefore their form is the same in both $\msbar$ and $\MOp$. Hence $G_{4,0}^{\zeta_4\msbar}$, $G_{4,0}^{\zeta_4\MOp}$ will be the same functions, but of  $g^{\msbar}$, $g^{\MOp}$, respectively. We may therefore rewrite Eq.~\eqref{MS10f:b} in the form 
\be
\frac34F^{\zeta_3\msbar}_{4,1}=G^{\zeta_4\MOp}_{4,0}(g^{\msbar})+G^{\zeta_4\msba}_{3,0}\cdot F^{\msba}_{1,0}.
\label{MS10e}
\ee
If we work within $\rm MOM'$, we find
\begin{align}
F_{4,0}^{\zeta_4\MOp}=&G_{4,0}^{\zeta_4\MOp}+F^{\zeta_4\mop}_{3,0}\cdot F_{1,0}^{\mop}\nn
=&G_{4,0}^{\zeta_4\MOp}+G^{\zeta_4\mop}_{3,0}\cdot F_{1,0}^{\mop}
\label{MS8}
\end{align}
the second term on the right-hand side of Eq.~\eqref{MS8} of course due to the finite three-loop subtraction of $F^{\zeta_4\mop}_{3,0}$ in $\rm MOM'$.

Combining Eqs.~\eqref{MS8} and \eqref{MS10e}, we find
\begin{align}
\frac34F^{\zeta_3\msba}_{3,1}=& F^{\zeta_4\msba}_{3,0},\nn
\frac34F^{\zeta_3\msbar}_{4,1}=&F_{4,0}^{\zeta_4\MOp}(g^{\msbar}).
\label{msmom}
\end{align}
Finally we have a relation at four loops which is the obvious extension of that at three loops, and which will enable us to show how the even-$\zeta$ contributions are removed by the scheme change from $\msbar$ to $\MOp$ up to five loops.
Indeed we now have, combining Eqs.~\eqref{MS3a}, \eqref{MS7} and \eqref{msmom}
\begin{align}
\beta^{\zeta_4\msbar}_4=&[\beta_1^{\msba},F^{\zeta_4\MOp}_{3,0}(g^{\msbar})],\nn
\beta^{\zeta_4\msbar}_5=&[\beta_1^{\msba},F^{\zeta_4\MOp}_{4,0}(g^{\msbar})]+[\beta_2^{\msbar},F^{\zeta_4\MOp}_{3,0}(g^{\msbar})].
\label{MS10d}
\end{align}
The $\beta$-function in the $\rm MOM'$ scheme will be given by 
\be
\beta^{\zeta_4\MOp}=\beta^{\zeta_4\msbar}+\delta_1\beta^{\zeta_4\msbar},
\ee
where $\delta_1\beta^{\zeta_4\msbar}$ represents the terms in Eq.~\eqref{blin} linear in $\delta g$ (since the quadratic terms do not contribute at this order). Reading off $\delta g$ from Eq.~\eqref{MS10}, we have
\begin{align}
\delta_1\beta_{1,2,3}^{\zeta_4\msbar}=&0,\nn
\delta_1\beta_4^{\zeta_4\msbar}=&-[\beta_1^{\msba},F^{\zeta_4\MOp}_{3,0}(g^{\msbar})],\nn
\delta_1\beta_5^{\zeta_4\msbar}=&-[\beta_1^{\msba},F^{\zeta_4\MOp}_{4,0}(g^{\msbar})]-[\beta_2^{\msbar},F^{\zeta_4\MOp}_{3,0}(g^{\msbar})]
\label{MS10a}
\end{align}
It is now clear, comparing with Eqs.~\eqref{MS10d}, that the change from $\msbar$ to $\rm MOM'$ removes the $\zeta_4$ terms in the $\beta$-function through five loops.

As we stated in the main text, we can show at least up to the five-loop level that the $\zeta_4$ $\beta$-function terms are removed by the change to $\MO$ as well as by the change to $\MOp$. We already know from Ref.~\cite{gracey} that this is true for the supersymmetric Wess-Zumino model, so it is not surprising that this works in general. Let us proceed with the demonstration. For $\MO$, we include all finite parts in the subtractions, so that
\be
g^{\msbar}=g^{\MO}+F^{\mo}_{1,0} +F^{\zeta_4\mo}_{3,0}\zeta_4+F^{\zeta_4\MO}_{4,0}\zeta_4+\ldots
\label{MS12}
\ee
We have 
\begin{subequations}\label{MS11:main}
\begin{align}
F_{3,0}^{\zeta_4\mo}=&\Gtil_{3,0}^{\zeta_4\mo},\label{MS11:a}\\
F_{4,0}^{\zeta_4\MO}=&\Gtil_{4,0}^{\zeta_4\mo}+F_{3,0}^{\zeta_4\mo}\cdot F_{1,0}^{\mo}+F_{1,0}^{\MO}\cdot F_{3,0}^{\zeta_4\mo}\nn
&+F^{\mo}_{1,1}\cdot F_{3,-1}^{\zeta_4\mo}\label{MS11:b},
\end{align}
\end{subequations}
where $\Gtil_{3,0}^{\zeta_4\mo}$, $\Gtil_{4,0}^{\zeta_4\mo}$ are precisely the contributions from the diagrams, to be distinguished from $G_{3,0}^{\zeta_4\mo}$, $G_{4,0}^{\zeta_4\MO}$ which have contributions from non-$\zeta_4$ counterterms. The $\Gtil^{\zeta_4}$ are  scheme-independent, but as in previous examples we have written $\Gtil^{\zeta_4\msbar}$ to emphasise the functional dependence on $g^{\msbar}$. Substituting using Eq.~\eqref{MS11:main} and rewriting Eq.~\eqref{MS12} to make $g^{\MO}$ the subject, we obtain
\begin{align}
g^{\MO}=&g^{\msbar}-F^{\msba}_{1,0}-\Gtil^{\zeta_4\msba}_{3,0}\zeta_4-\Gtil_{4,0}^{\zeta_4\msba}\zeta_4-F^{\msba}_{1,1}\cdot F_{3,-1}^{\zeta_4\msba}\zeta_4\nn
=&g^{\msbar}-F^{\msba}_{1,0}-F^{\zeta_4\msba}_{3,0}\zeta_4-F_{4,0}^{\zeta_4\msbar}\zeta_4.
\label{MS13a}
\end{align}
We should emphasise here that there are additional terms in Eqs.~\eqref{MS12}, \eqref{MS13a} from two through four loops (including $\zeta_3$-dependent terms) which we have not displayed explicitly since they do not affect our calculation. 

It is not surprising that Eq.~\eqref{MS13a} differs in structure from Eq.~\eqref{MS10}; but the apparent puzzle is how the $\zeta_4$ terms can cancel in both $\rm MOM'$ and MOM (as we certainly know in the WZ case). However in the case of the transformation from $\msbar$ to MOM we also need to take account of higher-order terms at five loops; i.e. those quadratic in $\delta g$ in Eq.~\eqref{blin}. So in this case, reading off $\delta g$ from Eq.~\eqref{MS13a}, we write 
\be
\beta^{\zeta_4\MO}=\beta^{\zeta_4\msbar}+\delta_1\beta^{\zeta_4\msbar}+\delta_2\beta^{\zeta_4\msbar},
\ee
where now $\delta_2\beta^{\zeta_4\msbar}$ represents the terms in Eq.~\eqref{blin} quadratic in $\delta g$. 
There is no difference between the transformations from $\msbar$ to $\rm MOM'$ and to MOM through 4 loops, as far as $\zeta_4$ terms are concerned. However, at five loops, the linear terms in Eq.~\eqref{blin} now give (using Eq.~\eqref{MS5:d})
\be
\delta_1\beta_5^{\zeta_4\msbar}=-[\beta_4^{\zeta_4\msbar},F^{\msba}_{1,0}]-[\beta^{\msba}_1,G^{\zeta_4\msbar}_{4,0}]-[\beta^{\msbar}_2,G^{\zeta_4\msbar}_{3,0}]
\label{MS14}
\ee
and the additional second-order terms give
\be
\delta_2\beta_5^{\zeta_4\msbar}=-F^{\msba}_{1,0}\cdot[\beta^{\msba}_1,G^{\zeta_4\msbar}_{3,0}]-G^{\zeta_4\msbar}_{3,0}\cdot[\beta^{\msba}_1,F^{\msba}_{1,0}]-F^{\msba}_{1,0}\cdot|G^{\zeta_4\msbar}_{3,0}|\cdot\beta^{\msba}_1.
\label{MS15}
\ee
Adding Eqs.~\eqref{MS14}, \eqref{MS15}, we find
\begin{align}
\delta\beta_5^{\zeta_4\msbar}=&\delta_1\beta_5^{\zeta_4}+\delta_2\beta_5^{\zeta_4}\nn
=&F^{\msba}_{1,0}\cdot\left(\beta_4^{\zeta_4\msbar}-[\beta^{\msba}_1,G^{\zeta_4\msba}_{3,0}\right)-\beta_4^{\zeta_4\msbar}\cdot F^{\msba}_{1,0}\nn
&-G^{\zeta_4\msbar}_{3,0}\cdot[\beta^{\msba}_1,F^{\msba}_{1,0}]-F^{\msba}_{1,0}\cdot|G^{\zeta_4\msba}_{3,0}|\cdot\beta^{\msba}_1\nn
&-[\beta^{\msba}_1,G^{\zeta_4\msbar}_{4,0}]-[\beta^{\msbar}_2,G^{\zeta_4\msba}_{3,0}]
\end{align}
Using Eq.~\eqref{MS3a} we obtain after some algebra
\begin{align}
\delta\beta_5^{\zeta_4\msbar}=&-[\beta^{\msba}_1,G^{\zeta_4\msbar}_{4,0}+G^{\zeta_4\msba}_{3,0}\cdot F^{\msba}_{1,0}]-[\beta^{\msbar}_2,G^{\zeta_4\msba}_{3,0}]\nn
=&-[\beta^{\msba}_1,G^{\zeta_4\MOp}_{4,0}(g^{\msbar})+G^{\zeta_4\msba}_{3,0}\cdot F^{\msba}_{1,0}]-[\beta^{\msbar}_2,G^{\zeta_4\msba}_{3,0}]\nn
=&-[\beta^{\msba}_1,F^{\zeta_4\MOp}_{4,0}(g^{\msbar})]-[\beta^{\msbar}_2,F^{\zeta_4\msba}_{3,0}]
\end{align}
where we have used the fact that $G^{\zeta_4\msbar}$ and $G^{\zeta_4\MOp}$ have the same functional form, together with Eq.~\eqref{MS8}.
This is the same change as for the transformation to $\rm MOM'$ in Eq.~\eqref{MS10a}, and therefore once again cancels the $\zeta_4$ terms. We emphasise that it is only in the vanishing of the even-$\zeta$ terms that the MOM and $\rm MOM'$ results for the $\beta$-function coincide. Finally, comparing the $\MOp$ transformation Eq.~\eqref{MS10} and the $\MO$ transformation Eq.~\eqref{MS13a}; as we stated earlier, if one tried to use either one to guess the other, it seems unlikely that one would obtain the correct result, since one would probably evaluate $F^{\zeta_4}_{4,0}$ in the wrong scheme.


We would hope that this demonstration of scheme redefinitions, from $\msbar$ to the $\MOp$ and $\MO$ schemes where the even $\zeta$s cancel, may be extended for the multi-coupling case beyond five loops and to the full range of even $\zeta$s. We have also indicated in Section 5 how to work within the $\MOp$ scheme {\it ab initio}; but as we observed there, it is not clear if this is the easiest way to proceed to higher orders. It therefore seems worthwhile to see how far we can follow this alternative route, starting from $\msbar$. The process we have followed has two main aspects. The first is the derivation of relations like Eqs.~\eqref{MS3a} and \eqref{MS7}, and the second is the derivation of relations like Eqs.~\eqref{msmom}, which provides the link to the scheme redefinitions through Eq.~\eqref{MS10}. There does not appear to be any great obstacle to extending the first step to higher orders, though equally it is difficult to make any general statements. On the other hand, it does appear to us that one can anticipate how the second step works to higher orders, at least in broad outline, and propose a useful generalisation of Eq.~\eqref{msmom}. Furthermore Eq.~\eqref{MS10} may also be readily extended for single even $\zeta$s (though the case of products of even $\zeta$s is less clear, see the end of this Appendix). Accordingly, even in the absence of higher loop results extending 
Eqs.~\eqref{MS3a} and \eqref{MS7} for the multi-coupling case, one can use the results given in Ref.~\cite{baikov3} and repeated in Eq.~\eqref{z4:main} to extend the single-coupling results to show that the even-$\zeta$ terms vanish as far as seven loops in $\rm MOM'$; as we did in Section 3.

Let us start the generalisation of Eq.~\eqref{msmom} by extending it to higher loops for $\zeta_4$, before turning to other even $\zeta$s. At the next loop order, i.e at six loops, we have
\be
G^{\zeta_3\msbar}_{5,1}=F^{\zeta_3\msbar}_{5,1}-F^{\zeta_3\msbar}_{4,1}\cdot F^{\msba}_{1,0}-G^{\zeta_3\msbar}_{4,2}\cdot F^{\msba}_{1,-1}-F^{\zeta_3\msba}_{3,1}\cdot F^{\msbar}_{2,0},
\label{MS16}
\ee
with $G^{\zeta_3\msbar}_{4,2}$ defined as in Eq.~\eqref{MS2a}; which may be rewritten using Eqs.~\eqref{MS2a} and \eqref{Grels:a} as
\begin{align}
F^{\zeta_3\msbar}_{5,1}=&G^{\zeta_3\msbar}_{5,1}+[G^{\zeta_3\msbar}_{4,1}+G^{\zeta_3\msbar}_{3,1}\cdot F^{\msba}_{1,0}]\cdot F^{\msba}_{1,0}+G^{\zeta_3\msba}_{3,1}\cdot F^{\msbar}_{2,0}+G^{\zeta_3\msbar}_{4,2}\cdot F^{\msba}_{1,-1}\nn
=&\frac43\{G^{\zeta_4\msbar}_{5,0}+[G^{\zeta_4\msbar}_{4,0}+G^{\zeta_4\msbar}_{3,0}\cdot F^{\msba}_{1,0}]\cdot F^{\msba}_{1,0}+G^{\zeta_4\msba}_{3,0}\cdot F^{\msbar}_{2,0}+G^{\zeta_4\msbar}_{4,1}\cdot F^{\msba}_{1,-1}\}.
\label{MS17}
\end{align}
It is the counterterms which differ between different schemes, and since we have removed the $\zeta_4$-dependent counterterms to obtain the $G^{\zeta_4}$-functions, these will have the same structure in $\msbar$ and $\rm MOM'$ (which only differ in respect of even $\zeta$-dependent counterterms). Of course we must bear in mind that $G^{\zeta_4\msbar}$, $G^{\zeta_4\MOp}$ are expressed as functions of $g^{\msbar}$, $g^{\MOp}$ respectively. The outcome is that we may replace $G^{\zeta_4\msbar}$ by $G^{\zeta_4\MOp}(g^{\msbar})$ in Eq.~\eqref{MS17}. Finally, using Eqs.~\eqref{ftilrel:b}, \eqref{MS8}, 
we may rewrite Eq.~\eqref{MS17} as
\begin{align}
F^{\zeta_3\msbar}_{5,1}=&\frac43\{G^{\zeta_4\MOp}_{5,0}(g^{\msbar})+[G^{\zeta_4\MOp}_{4,0}(g^{\msbar})+G^{\zeta_4\msba}_{3,0}\cdot F^{\msba}_{1,0}]\cdot F^{\msba}_{1,0}\nn
&+G^{\zeta_4\msba}_{3,0}\cdot F^{\msbar}_{2,0}+G^{\zeta_4\MOp}_{4,1}(g^{\msbar})\}\cdot F^{\msba}_{1,-1}\}\nn
=&\frac43\{G^{\zeta_4\MOp}_{5,0}(g^{\msbar})+F^{\zeta_4\MOp}_{4,0}(g^{\msbar})\cdot F^{\msba}_{1,0}+F^{\zeta_4\msba}_{3,0}\cdot F^{\msbar}_{2,0}\nn
&+G^{\zeta_4\MOp}_{4,1}(g^{\msbar})\cdot F^{\msba}_{1,-1}\}\nn
=&\frac43F^{\zeta_4\MOp}_{5,0}(g^{\msbar}).
\end{align}
In general, it would appear that there are functions $\Ftil_i[L_1,p,n,g]$ for each loop order, depending on $G^{\zeta_n}_{l,p}(g)$ for $l=1,\ldots L_1$, such that
\begin{subequations}\label{MS17a:main}
\begin{align}
F^{\zeta_3\msbar}_{L,1}(g)=&\Ftil_{L-2}[L,1,3,g],\quad L\ge 3,\label{MS17a:a}\\
F^{\zeta_4\MOp}_{L,0}(g)=&\Ftil_{L-2}[L,0,4,g],\quad L\ge 3.\label{MS17a:b}
\end{align}
\end{subequations}
In Eq.~\eqref{MS17a:main}, we have for the three lowest-order $\Ftil$
\begin{subequations}\label{MS17b:main}
\begin{align}
\Ftil_1[L,p,n,g]=&G^{\zeta_n}_{L,p}(g),\label{MS17b:a}\\
\Ftil_2[L,p,n,g]=&G^{\zeta_n}_{L,4,p}(g)+G^{\zeta_n}_{L-1,p}(g)\cdot F_{1,0}(g),\label{MS17b:b}\\
\Ftil_3[L,p,n,g]=&G^{\zeta_n}_{L,p}(g)+[G^{\zeta_n}_{L-1,p}(g)+G^{\zeta_n}_{L-2,p}(g)\cdot F_{1,0}(g)]\cdot F_{1,0}(g)\nn
&+G^{\zeta_n}_{L-1,p}(g)\cdot F_{2,0}(g)+G^{\zeta_n}_{L-1,p+1}(g)\cdot F_{1,-1}(g),\label{MS17b:c}
\end{align}
\end{subequations}
Note that $F_{1,0}$, $F_{1,-1}$ and $F_{2,0}$ have the same form in both the schemes of current interest, $\msbar$ and $\MOp$. Moreover, there is no need to specify which $G^{\zeta}$ is meant here, since as mentioned above, they have the same functional form in both $\msbar$ and $\MOp$\footnote{In general we do need to specify a scheme for the $G^{\zeta}$; since they differ in $\MO$ (where it is not just the even-$\zeta$ counterterms which are different from the other schemes) from their form in $\msbar$/$\MOp$.}. So once we have expressed $F^{\zeta_3\msbar}_{L,1}$ and $F^{\zeta_4\MOp}_{L,0}$ via Eq.~\eqref{MS17a:main} in terms of the $G^{\zeta}$ quantities, the difference between $\msbar$ and $\MOp$ resides solely in the functional form of the $\Ftil$; which we have emphasised by writing in terms of a generic coupling $g$. Now since Eq.~\eqref{MS17a:a} depends on $G^{\zeta_3}_{L,1}(g)$, and Eq.~\eqref{MS17a:b} depends in an identical way on $G^{\zeta_4}_{L,0}(g)$, we have in view of Eqs.~\eqref{Grels:a},
\be
F^{\zeta_4\msbar}_{L,1}=\frac43F^{\zeta_3\MOp}_{L,0}(g^{\msbar})
\label{MS18}
\ee
where we have now specialised to $g\equiv g^{\msbar}$.
This is the required generalisation of Eq.~\eqref{msmom} for the $\zeta_4$ case.
We now turn to the $\zeta_6$, $\zeta_8$ and $\zeta_{10}$ contributions. It is straightforward just using the structure of $\Ftil_i[L_1,p,n,g]$ as exemplified in Eqs.~\eqref{MS17b:main} to obtain a modified version of Eq.~\eqref{Grels:b} analogous to \eqref{MS18}, namely
\begin{subequations}\label{MS20a:main}
\begin{align}
F_{L,0}^{\zeta_6\MOp}(g^{\msbar})=&\frac54\left(F_{L,1}^{\zeta_5\msbar}-\frac13\Ftil_{L-3}[L,2,4,g^{\msbar}]\right),\label{MS20a:a}\\
F_{L,0}^{\zeta_8\MOp}(g^{\msbar})=&\frac74\left(F_{L,1}^{\zeta_7\msbar}-\frac12\Ftil_{L-4}[L,2,6,g^{\msbar}]
+\frac{1}{24}\Ftil_{L-4}[L,4,4,g^{\msbar}]\right),\label{MS20a:b}\\
F_{L,0}^{\zeta_{10}\MOp}(g^{\msbar})=&\frac94\Bigl(F_{L,1}^{\zeta_9\msbar}-\frac23\Ftil_{L-5}[L,2,8,g^{\msbar}]+\frac{7}{60}\Ftil_{L-5}[L,4,6,g^{\msbar}]\nn
&-\frac{1}{48}\Ftil_{L-5}[L,6,4,g^{\msbar}]\Bigr).\label{MS20a:c}
\end{align}
\end{subequations}
Here
\begin{subequations}\label{MS22:main}
\begin{align}
F_{L,0}^{\zeta_6\MOp}=&\Ftil_{L-3}[L,0,6,g^{\MOp}],\quad L\ge 4,\label{MS22:a}\\
F_{L,0}^{\zeta_8\MOp}=&\Ftil_{L-4}[L,0,8,g^{\MOp}],\quad L\ge 5,\label{MS22:b}\\
F_{L,0}^{\zeta_{10}\MOp}=&\Ftil_{L-5}[L,0,{10},g^{\MOp}],\quad L\ge 6,\label{MS22:c}
\end{align}
\end{subequations}
and 
\begin{subequations}\label{MS23:main}
\begin{align}
F_{L,1}^{\zeta_5\msbar}=&\Ftil_{L-3}[L,1,5,g^{\msbar}],\quad L\ge 4,\label{MS23:a}\\
F_{L,1}^{\zeta_7\msbar}=&\Ftil_{L-4}[L,1,7,g^{\msbar}],\quad L\ge 5,\label{MS23:b}\\
F_{L,1}^{\zeta_9\msbar}=&\Ftil_{L-5}[L,1,9,g^{\msbar}],\quad L\ge 6.\label{MS23:c}
\end{align}
\end{subequations}
where $\Ftil_{1,2,3}$ are defined in Eq.~\eqref{MS17b:main}.
We emphasise that here, structures like $\Ftil_{L-3}[L,2,4,g^{\msbar}]$ in Eq.~\eqref{MS20a:a} unfortunately have no simple relation to $F_{L,2}^{\zeta_{4}\msbar}$.

We may expect that the relations in \eqref{Grelsa:main} also translate to similar relations between $F^{\MOp}$ and $F^{\msbar}$ quantities. However, the computation of relations is slightly more involved for mixed $\zeta$s. At  lowest order in each case, we immediately find respectively
\begin{align}
F^{\zeta_4\zeta_3\msba}_{5,0}=&\frac32F^{(\zeta_3)^2\msba}_{5,1}=\frac{3}{10}\beta^{(\zeta_3)^2\msba}_5,\nn
F^{\zeta_6\zeta_3\msba}_{6,0}=&\frac54F^{\zeta_5\zeta_3\msba}_{6,1}=\frac{1}{4}\beta^{\zeta_5\zeta_3\msba}_6,
\label{MS25}
\end{align}
but at higher loops more work is required. We have, using 
Eq.~\eqref{Grelsa:a},
\begin{align}
\frac32F^{(\zeta_3)^2\msbar}_{6,1}=&\frac32\left(G^{(\zeta_3)^2\msbar}_{6,1}+G^{(\zeta_3)^2\msba}_{5,1}\cdot F^{\msba}_{1,0}
+G^{\zeta_3\msba}_{3,1}\cdot G^{\zeta_3\msba}_{3,0}\right)\nn
=&G^{\zeta_4\zeta_3\msbar}_{6,0}+G^{\zeta_4\zeta_3\msba}_{5,0}\cdot F^{\msba}_{1,0}
+2G^{\zeta_4\msba}_{3,0}\cdot G^{\zeta_3\msba}_{3,0}
\end{align}
It is clear from the fact that the $G$ quantities have a scheme-independent functional form, and the form of the relation between $g^{\MOp}$ and $g^{\msbar}$ in Eq.~\eqref{MS10}, that we may again replace $G^{\msbar}$ by $G^{\MOp}$; so we obtain
\begin{align}
\frac32F^{(\zeta_3)^2\msbar}_{6,1}=&G^{\zeta_4\zeta_3\MOp}_{6,0}(g^{\msbar})+G^{\zeta_4\zeta_3\msba}_{5,0}\cdot F^{\msba}_{1,0}
+2G^{\zeta_4\msba}_{3,0}\cdot G^{\zeta_3\msbar}_{3,0}\nn
=&G^{\zeta_4\zeta_3\MOp}_{6,0}(g^{\msbar})+G^{\zeta_4\zeta_3\msba}_{5,0}\cdot F^{\msba}_{1,0}
+G^{\zeta_4\msba}_{3,0}\cdot G^{\zeta_3\msba}_{3,0}\nn
&+G^{\zeta_3\msba}_{3,1}\cdot G^{\zeta_4\msba}_{3,-1}\nn
=&F^{\zeta_4\zeta_3\MOp}_{6,0}(g^{\msbar}),
\end{align}
where we have used Eq.~\eqref{Grels:a} in the penultimate step. We therefore have
\be
F^{\zeta_4\zeta_3\MOp}_{6,0}(g^{\msbar})=\frac32F^{(\zeta_3)^2\msbar}_{6,1}=\frac14\beta^{(\zeta_3)^2\msbar}_6.
\label{MS26}
\ee

We may derive in a similar way starting from Eq.~\eqref{Grelsa:b}
\begin{align}
\frac38F^{\zeta_4\zeta_3\msbar}_{6,1}=&G^{(\zeta_4)^2\MOp}_{6,0}(g^{\msbar})+G^{(\zeta_4)^2\msba}_{5,0}\cdot F^{\msba}_{1,0}
+\frac12G^{\zeta_4\msba}_{3,0}\cdot G^{\zeta_4\msba}_{3,0}\nn
=&F^{(\zeta_4)^2\MOp}_{6,0}(g^{\msbar})-\frac12G^{\zeta_4\msba}_{3,0}\cdot G^{\zeta_4\msba}_{3,0}.
\label{MS27}
\end{align}

Now extending Eqs.~\eqref{MS9}, \eqref{MS10} to include $(\zeta_4)^2$ terms, and including second order effects as mentioned after Eq.~\eqref{MS10}, we obtain
\be
g^{\msbar}=g^{\MOp}+F^{\zeta_4\mop}_{3,0}\zeta_4+F^{\zeta_4\MOp}_{4,0}\zeta_4+F^{\zeta_4\MOp}_{5,0}\zeta_4+F^{(\zeta_4)^2\MOp}_{6,0}(\zeta_4)^2+\ldots
\ee
which again may readily be rewritten as
\begin{align}
g^{\MOp}=&g^{\msbar}-F^{\zeta_4\msba}_{3,0}\zeta_4-F^{\zeta_4\MOp}_{4,0}(g^{\msbar})\zeta_4-F^{\zeta_4\MOp}_{5,0}(g^{\msbar})\zeta_4\nn
&-\left(F^{(\zeta_4)^2\MOp}_{6,0}(g^{\msbar})
-G^{\zeta_4\msba}_{3,0}\cdot G^{\zeta_4\msba}_{3,0}\right)(\zeta_4)^2\ldots
\label{MS28}
\end{align}
This result is required in Eq.~\eqref{delz8} in the main text, during the discussion of $\beta^{\zeta_8}_{7}$.

\end{document}